\documentclass[reqno,a4paper,12pt,makeidx,english]{amsbook}
\usepackage{amssymb,amsmath,graphics}
\usepackage{tensind}
\usepackage{doublespace}

 \numberwithin{equation}{chapter}
\tensordelimiter{?} \tensorformat{none}

\numberwithin{figure}{chapter}
\numberwithin{section}{chapter}

\usepackage{hyperref}

\newtheoremstyle{springer_thm}
  {1em}
  {1.5em}
  {\normalfont}
  {}
  {\bfseries}
  {}
  {\newline}
  {}

\newtheoremstyle{springer_rem}
  {1em}
  {3pt}
  {\normalfont}
  {}
  {\itshape}
  {.}
  {12pt}
  {}

\theoremstyle{springer_rem}
\newtheorem*{example}{Example}
\newtheorem*{remark}{Remark}

\theoremstyle{springer_thm}
\newtheorem{Thm}{Theorem}[chapter]

\newtheorem{lemma}[Thm]{Lemma}
\newtheorem{obs}[Thm]{Observation}
\newtheorem{co}[Thm]{Conjecture}
\newtheorem{definition}[Thm]{Definition}
\newtheorem{defi}[Thm]{Definition}

\def\a{\alpha}
\def\be{\beta}
\def\A{\Lambda}
\def\w{\omega}
\def\Vac{\Omega}

\def\e{\epsilon}

\def\s{\sigma}
\def\g{\gamma}
\def\G{\Gamma}
\def\de{\delta}
\def\De{\Delta}
\def\m{\mu}
\def\n{\nu}
\def\vp{\varphi}
\def\la{\lambda}
\def\La{\Lambda}
\def\p{\rho}

\def\d{\partial}
\def\D{\nabla}

\def\OpD{\mathcal{D}}

\def\v#1{{\mathbf{#1}}}
\def\h#1{#1^*}
\def\con#1{#1^\dagger}
\def\b#1{\overline{#1}}
\def\T#1{\tilde{#1}}
\def\*{\cdot }
\def\nin{\in \hspace{-3.5mm} /}
\def\dir#1{#1 \hspace{-2.6mm} \slash}
\def\del#1{\widetilde{#1}}
\def\ra{\rightarrow}
\def\rest{\upharpoonright}

\def\H{\mathcal{H}}
\def\K{\mathcal{K}}
\def\Fk{\Fou_\K}
\def\C{\mathcal{C}}
\def\Dom{\mathcal{D}}

\def\Fou{\mathcal{F}}
\def\A{\mathcal{A}}
\def\W{\mathcal{W}}
\def\Lag{\mathcal{L}}
\def\Alg{\hbox{$O\hspace{-2.2pt}\iota
\hspace{-2.2pt}^{\prime}\hspace{-3.1pt}\acute{}\hspace{3pt}$}}

\def\O{\mathcal{O}}

\def\ess_sp{\s_{ess}}
\def\F1{\Fou^{(1)}}
\def\PDO{$\Psi$DO}

\def\R{\mathbb{R}}
\def\Z{\mathbb{Z}}
\def\N{\mathbb{N}}
\def\U{\mathbb{U}}
\def\S{\mathbb{S}}
\def\id{\mathbf{1}} 

\def\dom{\C_0^\infty}
\def\smooth{\C^\infty}
\def\lsq{L^2(\R^3,d^3x)}

\DeclareMathOperator{\supp}{supp} \DeclareMathOperator{\sgn}{sgn}
 
\DeclareMathOperator{\tr}{tr} 
\def\t#1{\mbox{#1}}

\def\sp{\vspace{1em}}

\makeindex 

\begin{document}

\title{Quantum Electrodynamics in external fields}
\author{Piotr Marecki}

\begin{titlepage}
\addtocounter{page}{1} \mbox{} \vspace{1.5cm}
\begin{center}
\Huge

\huge\mbox{\bf Quantum Electrodynamics}\\
\mbox{\bf on Background External Fields}\\
\large \vspace{6cm}
{Dissertation}\\
{zur Erlangung des Doktorgrades}\\
{des Fachbereichs Physik}\\
{der Universit\"at Hamburg}\\

\vspace{7em}

{Piotr Marecki}\\

\mbox{}\\
\vfill Hamburg\\ 2003
\end{center}
\end{titlepage}

\
 \vfill

\begin{center}
{\bf Abstract}
\end{center}
The quantum electrodynamics in the presence of background external fields is
developed. Modern methods of local quantum physics allow to formulate the
theory on arbitrarily strong possibly time-dependent external fields.
Non-linear observables  which depend only locally on the external field are
constructed. The tools necessary for this formulation,  the parametrices of
the Dirac operator, are investigated.

\vfill

\begin{center}
{\bf Zusammenfassung}
\end{center}
In dieser Arbeit wird die Quantenelektrodynamik in \"au\ss eren
elektromagnetischen Fel\-dern entwickelt. Die modernen Methoden der lokalen
Quantenphysik erm\"oglichen es, die Theorie so zu formulieren, dass die
\"au\ss eren Felder weder statisch noch schwach sein m\"ussen. Es werden
nicht-lineare Observable konstruiert, die nur lokal von den
Hintergrundfeldern abh\"angen. Die dazu ben\"otigten Werkzeuge, die
Parametrizes des Dirac\-operators, werden untersucht.

 \vfill

\setcounter{tocdepth}{2} \tableofcontents \setcounter{chapter}{0}


\chapter{Introduction}\label{chapter:intro}

\section{Formulation of the problem}
In this work Quantum Electrodynamics will be developed in which the Dirac
field propagates on an external field background. Perhaps the best way to
explain precisely what theory we have in mind is to look at its action.
Suppose\footnote{In the units $\hbar=1=c$; the $cgs$-Gauss units are
restored in appendix \ref{units}.}:
\begin{equation*}
      S=\int \t{d$^4$x } \left\{i\ \b \psi\g^a\d_a \psi-m \b\psi\psi
  +e \b\psi\g^a\psi\  A_a-
  \left[\frac{1}{16\pi }F_{ab}F^{ab}\right]+J^a A_a \right\}.
\end{equation*}
Here $J^a(t,\v x)$ denotes some external electromagnetic current which is a
fixed function of time and space; $\psi$ denotes the Dirac field and $A_a$
the electromagnetic field. We divide $A_a$ into two parts,
\begin{equation}\label{A_class_B}
    A_a=A^{class}_a+\A_a,
\end{equation}
where $A^{class}$ is a solution of the inhomogeneous Maxwell equations
\begin{equation}\label{A_Maxwell}
    \d^bF_{ab}^{class}=4\pi J_{a}.
\end{equation}
When substituted into the action $S$, the splitting \eqref{A_class_B}
leads to an action of which the only dynamical variables are $\psi$ and
$\A^a$:
\begin{equation*}
      S=\int \t{d$^4$x } \left\{\b \psi\ (i\g^a\d_a +e\g^a A_a^{class}-m)\ \psi
  +e \b\psi\g^a\psi \A_a-
  \left[\frac{1}{16\pi }F_{ab}F^{ab}\right]+J^a A_a \right\}.
\end{equation*}
The variation with respect to $\b \psi$ and $\A_a$ leads to the
Euler-Lagrange equations:
\begin{align*}
  \left(i\g^a\d_a+eA_a^{class}- m\right) \psi &=-e\A_a\psi,\\
  \d^b F^\A_{ab}+\d^b F^{class}_{ab}&=4\pi\ (\b \psi\g_a\psi+J_a).
\end{align*}
Taking into account \eqref{A_Maxwell}, we get the following system
\begin{align*}
        \left(i\g^a\d_a+eA_a^{class}- m\right) \psi &=-e\A_a\psi,\\
         \d^b F^\A_{ab}&=4\pi\ \b \psi\g_a\psi.
\end{align*}
That was the classical field theory. \emph{Quantum} electrodynamics on
external field backgrounds is the quantum field theory of the interacting
Dirac and Maxwell fields. We first quantize the free fields, which obey
the differential equations
\begin{align}
        \left(i\g^a\d_a+eA_a^{class}- m\right) \psi &=0,\\
         \d^b F^\A_{ab}&=0,
\end{align}
and then investigate their interaction following the steps of the causal
perturbation theory\footnote{The interaction Lagrangian of the perturbation
theory is $\Lag_I=e:\b\psi(x)\g^a\psi(x):\A_a$.}. We note that the division
\eqref{A_class_B} is unique only up to the solutions of the homogeneous
Maxwell equations, which thus can be included either as $A^{class}$ or as
$\A$. The classical, external current $J^a(t,\v x)$ is produced by some
external sources (for instance by a heavy nucleus or by charged electrodes)
and, by assumption, \emph{is not influenced by the (charged) quantized Dirac
field $\psi$}.

\section{General motivation}

There are good reasons to investigate external field QED. The most
important of which\footnote{Apart from the fact that the external field
QED provides the best currently accepted explanation of such a fundamental
phenomenon as the spectrum of the hydrogen atom.}, in our opinion, is the
fact that this theory has much in common with the more difficult theory of
 quantum electrodynamics on a background curved spacetime (i.e. in
 the presence of gravitation). The problems posed by the latter theory are
tremendous, yet nobody doubts it touches the central problem of theoretical
physics which is to understand the relation between gravitation and quantum
phenomena. Perhaps the most striking similarity between external field QED
and QED on a curved space-time is the lack of a preferred vacuum state for
the Dirac field. In the absence of a distinguished state many traditional
concepts require (at least) a redefinition; to name some of them: the normal
ordering of the field quantities or the concept of particles. Normal
ordering is crucial if anything else than operators linear in the fields are
to be considered\footnote{For instance, one would like to investigate the
currents, the definition of which requires however the normal ordering.}.
The presence of particles in general causes certain  characteristic
responses of various detector arrangements. Particles are quasi-local
excitations. However, if no vacuum is distinguished, it is impossible to say
which configuration describes "excitations". Different basis states (the
analogues of the vacuum) will give rise to different detector responses none
of which can be distinguished as "preferred". There is no way to calibrate
our detectors.

The definition of non-linear quantities and the understanding of the
association between detector responses and the presence of particles are not
the only important issues which, when resolved in the external field QED,
may help in the development of  QED on a curved spacetime. After all, the
external field theories are by no means fundamental theories. It is natural
to expect the external field approximation to break down in certain regimes.
The expectation is that the back reaction effects are to be regarded as a
test if a given external field theory is a reliable approximation or not.
The back-reaction in the context of external field QED means the additional
(apart from $J_a$ which is the source of $A^{class}$) electromagnetic field
produced dynamically by the quantum Dirac field $\psi$. To say that the
external field approximation is justified means to regard the quantum fields
propagating in it as \emph {test fields}. Sometimes the
 back-reaction effects are naturally small as is for instance the reaction of an
 electron on the field produced by a macroscopic magnet. In other cases
  the back reaction is essential as for example in the free electron laser
  (FEL), where the synchrotron radiation emitted by a bunch of electrons interacts
 with this bunch and alters its dynamics\footnote{This and other
 main phenomena which occur at the FEL are reported eg. in the paper by
 S.V.Milton et al. \cite{FEL}.}. In the external field approximation
 it is possible that every state produces some back-reaction effects,
 even "the vacuum"\footnote{In quotation marks because there rarely  exists a
 privileged state.}. More importantly, in QED on a curved spacetime
it would be interesting to know what is the energy-momentum content
$\{T_{\m\n}(x)\}_\Vac$ of a certain "vacuum" state $\Vac$ in the process of
a collapse of a heavy star or, equally dramatically, does the black hole
evaporate due to Hawking radiation. None of the above fundamental questions
can reliably be addressed at the moment, partially because the evolution
equations for the gravitational fields are highly complicated. We write
partially, because there is another fundamental problem: what exactly is the
back-reaction current/energy-momentum tensor, if no vacuum is
distinguished\footnote{This question is not trivial, even if a certain vacuum
is distinguished - as in the no-external field case. Just that there is a
unique quantity to subtract from the infinite expectation value does not
mean that what remains is indeed the source of
gravitation/electromagnetism.}? Thus - partially - the back-reaction question
can be investigated more easily in external field QED, as the effect would
add up to the given external field (Maxwell equations are linear).

\section{Relation to other formulations of external-field QED}

    The development of external field QED commenced almost simultaneously
with the development of QED, in part due to the urge to describe atomic
systems. The early investigations consisted almost exclusively of a double
expansion: in $A^{class}_a$ and in $\A_a$. More precisely, the free fields
were supposed to fulfill the equations\footnote{This approximation can also
be recognized by the usage of free Dirac field propagators in the
calculations.}
\begin{align}
    \left(i\g^a\d_a- m\right) \psi &=0,\\
    \d^b F^\A_{ab}&=0,
\end{align}
and the perturbation theory was developed with the external field as well
as the quantum electromagnetic field on the same footing:
\begin{equation*}
\Lag_I=e:\b\psi(x)\g^a\psi(x): \A_a+e:\b\psi(x)\g^a\psi(x): A^{class}_a.
\end{equation*}
In such a way many processes of great physical importance have been
explained, among others bremsstrahlung and $e^+e^-$ pair production in the
field of a nucleus \cite{Ber_QED,akh_ber}. Although physically one has
learned a lot from those investigations, they implicitly assume that the
external field is weak.
    Indeed a more profound theory has also been developed called the
Furry picture or strong field QED \cite{mohr,Ber_QED}. This theory is very
similar to the one developed in this paper. The quantized free fields are
supposed to fulfill the system of equations
\begin{align*}
         \left(i\g^a\d_a+eA_a^{class}- m\right) \psi &=0,\\
         \d^b F^\A_{ab}&=0,
\end{align*}
which is the same as ours, and the interaction is \emph{formally} the same,
\begin{equation*}
    \Lag_I=:\b\psi \g^a\psi:\A_a,
\end{equation*}
though the Wick product in the Furry picture QED means the normal ordering
which can be written as
\begin{equation*}
    :\b \psi \psi:=\b \psi \psi-(\Vac,\b\psi\psi\ \Vac),
\end{equation*}
where $\Vac$ is the vacuum (defined in a certain way).

We aim at a better understanding of the quantum electrodynamics than the
Furry picture QED gives. It is therefore necessary to put forward the
weaknesses of the latter. In our opinion the main unsatisfactory features
of this theory which are common to all of its formulations  are:
\begin{enumerate}
\item In the definition of quantities nonlinear in the Dirac field (such
as, for instance, the normal ordering required in the first order interaction
processes) non-local objects are employed. This non-locality (elaborated upon
in chapter \ref{non_linear}) manifests itself in a delicate way, namely, the
observables defined as they are in the Furry picture QED do depend on the
external field not only in the region of their support. For instance, a
detector sensitive to the electric charge placed in a region $\C$,
\begin{equation*}
    D(f)=\int d^3x\ :\h \psi(\v x)\psi(\v x): f(\v x) \quad \t{with } \supp
    f=\C,
\end{equation*}
would be local if as an operator it depended at most on the external field
in $\C$. However, if $::$ means what it does in the Furry picture of QED,
then
\begin{equation*}
\frac{\delta D\ \ }{\delta A^{class}_a}\neq 0,
\end{equation*}
even if the support of the variation $\delta A^{class}_a$ does not intersect
with $\C$. We emphasize the need for \emph{local observables}. The states of
the quantum field carry non local information, and that is a characteristic
feature of relativistic quantum field theory. Locality means that at least
observables should be free of acausal influences\footnote{The precise
formulation of this new type of locality has been given in \cite{BFV}, see
also chapter \ref{non_linear}.}.

\item Almost all of the literature on external field QED assumes the
external fields to be static. This unnecessary assumption carries with
itself a false feeling of uniqueness of the vacuum representation which is
employed. While it is true that the ground state on a static background is
privileged as the state of lowest energy, we stress that not all external
fields are eternally static. Some external fields are\footnote{For instance,
the trapping potentials in the ion traps.} turned on in the distant past of
the experiment. It is highly likely that in such situations the state of the
Dirac field at later times is not the ground state of the static potential.
Also concepts like "adiabatic switching" of the external field require time
dependence of the external field.
\end{enumerate}
\sp
 We regard the drawbacks named above as very important, and we will not follow
the Furry picture of QED any further. On the other hand, these drawbacks do
not preclude the authors from deriving physically observable properties of
matter, which are later compared with experimental results and yield a
reasonable agreement. It is one of the remaining dilemmas whether the same
or similar results can be derived from the improved foundations which we
develop in this thesis.

In a separate development the theory of quantum fields on curved spacetime
has recently acquired a very satisfactory status. Indeed the works of many
authors over the past decade resulted in an almost complete picture of the
(interacting) electrodynamics on curved spacetime\footnote{To our great
regret the various results have never been gathered together in a single
reference. The physical (Dirac, Maxwell) fields are investigated by some
authors, but the interacting theory (a version of causal perturbation theory)
is only done for scalar fields.} \cite{Wald,BF,HW1,HW2,BFV}. A very modern
approach allowed to remedy all the drawbacks similar to those named above.
The renormalization theory in that scheme uses the language of distribution
theory. One speaks of distributions, their extension to coinciding points
and of the uniqueness of this procedure. This contrasts sharply with the
language of divergent integrals and tricky extractions of the finite parts
from them which are so common in the literature on quantum electrodynamics.
Although in the no-external-field context all these formulations of the
renormalization lead to the same results the mathematical transparency of
causal perturbation theory is encouraging \cite{Scharf}. It seems that
certain problems of uniqueness of the renormalization of the causal
perturbation theory on external field backgrounds have not even been
realized in the Furry picture QED.

 Our work thus attempts to achieve the following:
\begin{enumerate}
\item To formulate Quantum Electrodynamics on
external field backgrounds in a modern way, using the methods of QFT on
curved spacetimes together with the causal approach to the (perturbative)
construction of interacting field theories.

\item To construct the theory with a local
dependence on the external background.

\item To construct the theory on all possible external field backgrounds,
even time-dependent ones.
\end{enumerate}

\section{Structure of the paper}

The thesis contains seven chapters and five appendices. Here we shall
briefly summarize their content.

\vspace{4mm} The second chapter is where our investigations begin. It deals
with the quantization of the Dirac field in the presence of external field
backgrounds. The first section of this chapter recalls standard properties
of the classical Dirac field on external, possibly time-dependent
potentials. Results on the selfadjointness and the type of the essential
spectrum are gathered there. In the second section we attempt to remove one
of the main unsatisfactory features of the current formulations of the
external field QED \cite{shabajev,mohr}. This feature is the restriction to
one particular representation\footnote{In the static case this is the ground
state based representation.} of the free Dirac field algebra. We remove this
unnecessary restriction with the standard methods and results of the
algebraic approach to quantum field theory \cite{Haag}. In this apparently
new application of these methods we rely upon quantum field theory on curved
spacetimes, where such an application already proved to be useful. It is
enlightening to realize that the global equivalence of states at all times,
previously insisted on by many authors, is not necessary for the development
of quantum electrodynamics. Although some observables, for instance the
number operator or the total-energy operator, are lost in this way, we are
still able to describe the response of localized detectors which in our
opinion link the theoretical description with experimental setups.

We formulate the theory for a class of locally equivalent states - the
Hadamard states. We allow all possible, non-singular external
fields\footnote{On external gravitational backgrounds the Hadamard property
as a spectrum condition rules out spacetimes with closed time-like curves -
see \cite{RKW}. The case of non-smooth external fields requires a separate
investigation.}. The concrete predictions can be obtained in any
representation based upon \emph{an arbitrary quasi-free Hadamard state}.
Such states can be found on time-dependent environments. In particular it is
relatively easy to construct Hadamard states, if the external field is static
for some (possibly short) time interval. In the third section of the second
chapter we recall the standard construction of the ground state
representation. Mostly known results are gathered there.

\vspace{4mm} The third chapter  deals with the quantization of the free
electromagnetic field, which is the other basic field of quantum
electrodynamics. In our theory,  the free electromagnetic field $\A$ fulfills
the standard Maxwell equations, and so the quantization procedure is
standard  (the Gupta-Bleuler method).

\vspace{4mm} In the fourth chapter, which is rather technical, we develop
tools which enable us (in later chapters) to remove the other main
unsatisfactory feature of the standard approaches to QED. This feature is
the non-local dependence on the external field of these theories. The tools
we develop are parametrices of the Dirac operator. To our knowledge they
have not been extensively studied in the literature. Although the
coefficients of those parametrices are written down in \cite{dosch_mueller},
we have found it valuable to present our own derivation of them. It helps us
later to study directly their short-distance limit, their scaling,
uniqueness, dependence on the external field and their gauge covariance.
Additionally, we expand the parametrix (which is a distribution of two
variables) in a power series in the distance of its arguments. This
straightforward computation allows us to see important things. For instance,
we can foretell that the instantaneous ground states (employed by some
authors in the context of time-dependent external fields) are not Hadamard
states which is a drawback of such states.

\vspace{4mm} The fifth chapter deals with the very important concept of the
Hadamard property. It describes the short-distance singularity structure of
the allowed class of states. In this chapter we gather important theorems
which assure that a broad class of states shares this property. We also
recall the connection between two possible ways to define Hadamard states,
namely, in terms of their short-distance singularity expansion (the Hadamard
series) of the two-point function and in terms of the wave front set of this
two-point function. The equivalence of both definitions, first realized by
M.Radzikowski \cite{Rad} for scalar fields and proven by S.Hollands
\cite{hollands_dirac} and K.Kratzert \cite{KK} for the Dirac field, is also
reported here as it joins together various important parts of this thesis.

\vspace{4mm} The sixth chapter is in many ways the central one. It deals
with the construction of non-linear field observables. These are the
pointwise products of field operators smeared with test functions. There are
at least two contexts for which non-linear observables are of fundamental
importance. The first is the investigation of the current density and the
energy-momentum density of the free quantum Dirac field. The other is the
perturbative construction of interacting quantum electrodynamics. Our
intention is to address both of these contexts.

In the first section we recall the inductive construction of perturbative
quantum electrodynamics. We use the framework of causal perturbation theory,
which on the one hand is one of many formulations of the no-external-field
quantum electrodynamics \cite{Scharf}, and on the other hand is flexible
enough to be applied to the construction of interacting quantum field
theories on background spacetime manifolds \cite{BF}. The purpose of our
investigations is to construct the building blocks of causal perturbation
theory (the time-ordered products) in the lowest orders. In the second
section we do a step in this direction by defining the algebra $\W$ of Wick
polynomials of fermionic field operators. This algebra will also contain the
time-ordered products which describe the interacting evolution in a finite
order of the perturbation.

 The third section defines the most important
concept of this thesis which is the local dependence of the observables on
the external field. All of our important results are consequences of it. We
motivate this requirement physically by showing it to be closely related to
one of the foundations of general relativity. This foundation, the local
position invariance, is well-tested experimentally and intuitively clear in
content. Much of our subsequent work is a deduction from this very natural
assumption\footnote{The investigation of the dependence of quantum processes
on the background fields on which they take place clarifies to a certain
extent the meaning of local position invariance - cf. the remark in the
section \ref{subsec:LPI}.}. In the later sections we show, by means of simple
examples, that both the normal ordering prescription and the renormalization
subtraction scheme  employed in known formulations of the external-field
quantum electrodynamics are not local. Having established this, we proceed
constructively and build the local Wick and time-ordered products in the
lowest two orders of  perturbation theory.

In another development in the seventh section  we discuss the definition of
the current operator for the free quantized Dirac field. This is of prime
importance for the investigations of the back-reaction process. The
requirement of locality allows for the first time to reduce the huge
ambiguity of its definition to a finite number of constants. Previously only
differences of current densities of two states could be defined; here, we can
define the absolute charge density of a given state.

\vspace{4mm} In the seventh chapter we begin to analyze the consequences of
the local construction of quantum electrodynamics. Here, we only show how
various ingredients are combined together in calculations of the probability
amplitudes of physically important processes in the presence of static
external fields. A fair amount of work still has to be invested in order to
derive concrete predictions of the theory. Specifically the construction of
states in concrete situations is particularly cumbersome. The purpose of the
seventh chapter is to outline the way in which concrete predictions can be
obtained.

\vspace{4mm}
 The five appendices vary in importance and content.
The first one deals with the electromechanical units which are employed in
this paper. This issue can have important consequences, as for instance the
dimensional analysis alone (together with the postulate of locality) reduces
the ambiguity in the definition of the current density to three arbitrary
numbers.

The second appendix contains a brief exposition of the main theorems of
micro local analysis which find their application in the chapter on the
Hadamard form.

The third appendix presents the vacuum representation of the Dirac field in
the absence of external potentials which may help the reader not familiar
with the external-field QED to recognize the familiar expressions in their
generalization presented in chapter \ref{chapter:quantization}.

The fourth appendix discusses our model of the atomic spontaneous emission
of light. Although this model is only partially related to the main theme of
this thesis, we  believe it to give an important insight into the dynamics of
the interacting theory. Here, on the basis of the Weisskopf-Wigner approach
to the spontaneous-emission problem, we construct a model in which a system
of two (non-relativistic) bound states of the electron interacts with the
radiation field restricted to the vacuum and the one-photon sector. Instead
of using the perturbation theory, we derive an equation (which is an integral
equation) for the excited state's amplitude. In contrast to the perturbation
theory, this equation is reliable also for large times. More importantly the
calculation shows that the spontaneous emission is influenced directly by
the two-point function of the radiation field. Due to the nature of this
equation, it is straightforward to investigate various two-point functions of
the radiation field, not only the vacuum one. We can, for instance,
investigate the modification of the emission process on physical spacetimes
(eg. Robertson-Walker) or in the presence of boundaries (eg. Casimir-like
geometry).

The fifth appendix discusses the peculiarities of the construction of states
of fermionic systems (i.e. representations of the CAR algebra). In a simple
example we show what happens if the GNS construction is performed with a
mixed "basis" state. The phenomena which occur are symptomatic of the
problems which might occur in the general construction of representations of
the CAR algebra in the presence of external backgrounds.

\chapter{Quantization of the free Dirac field}\label{chapter:quantization}

This chapter deals with the quantization of the Dirac field in the presence
of external field backgrounds. It begins with a section on properties of the
classical Dirac field and of the Dirac operator on various external field
backgrounds. Most of the results are standard; we recall those which are
particularly important for the further development of the theory.

\section{Classical Dirac field}
In the following, let the Hilbert space be
\begin{equation*}
  \H=\lsq^4.
\end{equation*} The Dirac equation governs the time evolution of the
vectors $\psi\in\H$:
\begin{equation*}
  i\d_t \psi(t)=H(t) \psi(t),
\end{equation*}
where
\begin{equation*}
  H(t) =\a^i[-i\d_i+e A_i(t,\v{x})]-e A_0(t,\v{x}) +m \g^0
\end{equation*}
is the Hamiltonian, $e$ denotes the electronic charge and $m$ is the mass
of the electron. The symbols $\a^i, \ \g^0$ stand for the Dirac gamma
matrices\footnote{In the spinor (Weyl) representation the gamma matrices
are expressed in a simple manner by the 2$\times$2 Pauli matrices:
\begin{equation*}
  \a^i=\g^0\g^i=\begin{pmatrix}
    0 & -\s^i \\
    \s^i & 0
  \end{pmatrix}, \qquad \g^0=\begin{pmatrix}
    0 & \s^0 \\
    \s^0 & 0
  \end{pmatrix}.
\end{equation*}
}. The classical external electromagnetic field $A_\m(t,\v x)$ is assumed
to be such that the Hamiltonian $H(t)$ at each instant of time is
self-adjoint on a suitable domain $\OpD(H)\subset\H$.

\vspace{1em}
\subsection{Theorems on properties of the Dirac operator}
The Dirac operator in the presence of external fields can be split
according to
\begin{equation*}
  H=H_0+V(A),
\end{equation*}
where
\begin{equation}\label{free_H}
  H_0=-i\a^i\d_i+m\g^0
\end{equation}
is the free part (independent of the external field $A_\m(t,\v x)$); and
\begin{equation*}
  V(A)=e[\a^iA_i(t,\v x)-A_0(t,\v x)]
\end{equation*}
is the potential matrix, which is a multiplication operator. The matrix
elements of $V$ will be denoted by $V_{ij}$, $i,j=1,\ldots,4$. In the sequel
we shall specify the domain of definition of $H$, $\Dom(H)$, and recall some
results on its properties depending on the potential $V(A)$.

\vspace{1em}

 If $V(A)$ is time-independent and Hermitian, then the following
 theorems hold true:

\begin{Thm}[Theorem 4.3 of \cite{Thaller}]
If each matrix element of $V$ is a smooth function of $\v x$,
\begin{equation*}
  V_{ij}\in \smooth(\R^3),
\end{equation*}
then $H$ is essentially self-adjoint\index{Dirac
operator!self-adjointness} on $\dom(\R^3)^4$.
\end{Thm}

\begin{remark}
The above theorem covers quite a substantial area of physical situations -
the non-differentiable potentials are often only convenient approximations
which are physically smoothed out at short distances. Even the Coulomb
potential of an atomic nucleus is typically smoothed out inside the
nucleus; a notable counterexample, where the singularity is not smoothed
out, is the Coulomb field of an electron which appears not to be modified
at any distance at all.
\end{remark}

The situation of external fields which possess a Coulomb-like singularity is
covered by the following

\begin{Thm}[Theorem 4.2 of \cite{Thaller}] If all the elements of
the matrix $V_{ij}$ are majorized by Coulomb-like terms,
\begin{equation*}
  |V_{ij}|\leq \frac{a}{2|\v x|} +b \quad \t{with } \v x\in\R^3\backslash \{0\},\ b>0, \ a<1
\end{equation*}
then $H$ is essentially self-adjoint on $\dom (\R^3\backslash\{0\})$;
moreover, $H$ is self-adjoint\index{Dirac operator!self-adjointness} on the
Sobolev space\footnote{The first Sobolev space $H^1$ is the space of
$L^2$-functions whose first derivatives are also square-integrable.}
$H^1(\R^3)^4$.
\end{Thm}

\begin{remark}The above theorem is sensitive to the constant which
multiplies the potential. In the proof the theorem of Kato-Rellich
\cite{RS2} is utilized. In case of the Coulomb field \[A^0(\v x)=eZ/|\v
x|,\] after restoration of units, the theorem guarantees essential
self-adjointness up to $Z=68$. For the Coulomb potential this is still not
the maximal charge for which the essential self-adjointness property holds
because of
\end{remark}

\begin{Thm}[Theorem 4.4 of \cite{Thaller}]
If the external field is electrostatic, i.e.  $A^i=0$, $A^0=A^0(\v x)$, and
singular with the singularity not stronger than ${a}/{|\v x|}$, more
precisely
\begin{equation*}
    \sup_{\v x\in \R^3\setminus \{0\}}|\ |\v x| \ A^0(\v x)|<a,
\end{equation*}
then the corresponding Dirac Hamiltonian is essentially self-adjoint on
$\dom (\R^3\backslash\{0\})$, if $a<c\sqrt{3}/2=118.6$ in atomic units (see
appendix \ref{app:spontaneous}). For larger values of $a$ not greater than
$c=137$ there exists a unique selfadjoint extension of $H$ whose domain is
contained in the domain of $V(A)$.
\end{Thm}

If the potential is time-dependent and bounded, the evolution is described
in terms of the Dyson series. In the interaction representation the
evolution propagator is given by
\begin{equation*}
    \T U(t,s)=\sum_0^\infty \frac{(-i)^n}{n!} \int_s^t d\tau_1\ldots d\tau_n\
    T\left[\T V(\tau_1)\ldots \T V(\tau_n)\right],
\end{equation*}
where $T$ denotes the chronological order of the $\T V$ operators and
\begin{equation*}
    \T U(t,s)=e^{iH_0 t}\ U(t,s)\ e^{-iH_0s},\qquad
    \T V(t)=e^{iH_0 t}\ V(t)\ e^{-iH_0t}.
\end{equation*}
The unitary propagator $U(t,s)$ fulfills the strong operator equations,
\begin{align*}
    i\frac{\d}{\d t}\ U(t,s)\psi&=H(t) U(t,s) \psi,\\
    -i\frac{\d}{\d s}\ U(t,s)\psi&=-U(t,s) H(t) \psi,
\end{align*}
for all $\psi$ in the domain of $H$, only if the family of interaction
operators $\T V(t)$ is strongly continuous in time\footnote{This is the
case, if the commutator $[H_0,V(t)]$ is strongly continuous, cf. \cite{RS2}
chapter X.12.}, for otherwise only weak, distributional solutions can be
expected.

In the quantization of the Dirac field on static external fields the
question of the type of the spectrum of the Dirac operator is of interest.
It therefore appears appropriate to recall general results which settle the
question of the essential\index{Essential spectrum} spectrum\footnote{The
essential spectrum $\ess_sp$ is the set of all accumulation points and
infinite-degenerate eigenvalues.} $\ess_sp$ of the Dirac operator $H$.

The essential spectrum of the free Dirac operator $H_0$ is
\begin{equation*}
  \ess_sp(H_0)=(-\infty,-mc^2]\cup[mc^2,\infty).
\end{equation*}
This property is stable under the addition of static potentials decaying at
infinity.

\begin{Thm} Let $H=H_0+V(\v x)$ be self-adjoint on a certain domain $D(H)$, and let $V(\v x)$
be decaying at infinity,
\begin{equation}\label{potential_cond_spectrum}
  \lim_{|\v x|\rightarrow\infty} V_{ij}(\v x)=0,
\end{equation}
then the Dirac operator $H$ possesses the same essential spectrum as the
free Dirac operator:
\begin{equation*}
  \ess_sp(H)=\ess_sp(H_0).
\end{equation*}
\end{Thm}

\begin{remark} As a consequence, it is appropriate to have in mind
a picture of the spectrum of $H$ consisting of two continua (free Dirac
operator) and a ladder of bound states which can have $\{-mc^2,mc^2\}$ as
the only accumulation points, as this will be the case in interesting
applications. More on the essential spectrum of the Dirac field can be found
in section 4.3.4 of \cite{Thaller}.
\end{remark}

\sp
\section{Construction of states on general external field backgrounds}
The purpose of this section is to remove one of the most unsatisfactory
features of the recent constructions of the external field QED, which is the
fact that they are founded upon a certain "vacuum" representation of the
canonical anticommutation relations. This not only introduces an unnecessary
assumption that the external field is static but also carries an unjustified
claim that such a construction is unique and necessary. On the other hand, in
the constructive approach to the quantum field theory the existence of many,
even unitarily inequivalent representations of the canonical anticommutation
relations is well-known. The temptation of applying the modern methods of
local quantum physics \cite{Haag} to the external field problem of quantum
electrodynamics has resulted in the section that follows.

This section discusses in detail how to find representations of the CAR
algebra on a Cauchy surface. This is the algebra of fields $\psi(f),\ \h
\psi(f)$, smeared on a Cauchy surface with complex functions
$f\in\dom(\R^3)$, together with their polynomials. The fields  fulfill the
Canonical Anticommutation Relations (CAR):
\begin{align*}
    \{\psi(f),\h \psi(g) \}&=(f,g)\\
    \{\psi(f), \psi(g) \}&=0.\\
\end{align*}

If the external fields are static, the construction of a (vacuum)
representation poses no particular difficulty and is described in many
textbooks on QED. All that is needed in order to define such a vacuum state
is a projection operator $P_+$ which describes the splitting of the
underlying Hilbert space $\H$ into the electron/positron subspaces $\H_\pm$.
Such a projection on a static background is provided by\footnote{The
representation produced by such a projection describes what is usually
called "the Dirac sea". Here we assume that the Hamiltonian has an empty
kernel.} \[P_+=(1+\sgn H)/2\] and is distinguished as it leads to the vacuum
which is a ground state.

 Static external fields comprise, however, a narrow family of
allowed environments. After all, hardly any field available in experiments
is static for all times. With an important exception of the Coulomb field of
an eternal charge, all external fields that are static during an experiment
are rather generated earlier from the no-external-field environment,
stabilized for the duration of the experiment and later turned off. It is
important to realize that the ground state of the Dirac field in the static
external field configuration is different from the state which was a ground
state before the experiment and evolved in time while the fields were being
turned on. A byproduct of this fact is the observation of G.Scharf
\cite{Scharf} that the adiabatically turned on Coulomb field should be
modified on short distances by vacuum-back-reaction currents. Such an effect
does not, however, occur, if the field is strictly static.

What follows is the adaptation to the external field problem of the
methods presented in \cite{hollands_dirac}. We shall also make use of
various results of \cite{pow_stroe}.

\subsection{Introduction}
Suppose the external field $A^\m(x)$ is time-dependent. The classical
Dirac equation
\begin{equation*}
    [i\d_t -H(t)]\ \psi(t,x)=0
\end{equation*}
with
\begin{equation*}
    H(t)=-i\g^0\g^i\d_i+m\g^0+e[\a^iA_i(t,\v x)-A_0(t,\v x)]
\end{equation*}
will be investigated. Suppose we restrict ourselves to the Cauchy surface
$t=t_0$. The Hamiltonian at that instant of time, $H(t_0)$, is an
essentially self-adjoint operator on some dense domain in $\H$.
 In what follows we shall make use of the spectral properties of $H(t_0)$;
they are\footnote{In most cases the index $\v k$ will be continuous if the
eigenvalue corresponding to $\psi^\sharp_\v k$ belongs to the essential
spectrum of $H(t_0)$ and discrete if $\psi^\sharp_\v k$ is a bound state of
$H(t_0)$. The upper index which denotes the sign of the eigenvalue will in
the following be a variable. The summation rule with respect to this index
will assumed.}
\begin{itemize}
  \item the spectral measure $d\m(\v k)$,
  \item the (possibly generalized) eigenfunctions of positive $\psi^+_\v k(\v
  x)$ and negative $\psi^-_\v k(\v x)$ frequency.
\end{itemize}

The smeared two-point function of a state
\begin{equation*}
    \w(\psi(f)\psi^*(g))
\end{equation*}
restricted to the Cauchy surface under consideration may be parameterized
with the help of a positive, bounded operator $0\leq B\leq 1$. Let us
prescribe the action of this operator with the help of an integral kernel
$B(\v x,\v y)$
\begin{equation*}
  (B\ g)^C(\v x)=\int d^3y\ B(\v x,\v y)^{CA}\ g_A(\v y),
\end{equation*}
for $g\in \H$. Using the generalized eigenfunctions of $H(t_0)$, we define
\begin{equation}\label{def_B}
 B^{nm}(\v k,\v p)=\int d^3x\ d^3y\ \left(\psi^n_\v k\right)^\dag_A(\v x)\
 B(\v x,\v y)^{AC}\ \left(\psi_\v p^m\right)_C (\v y),
\end{equation}
where the dagger denotes the conjugation in $\H$ (i.e. complex adjunction
supplemented by a transposition) and the indices $m,n$ were introduced and
summed over in order to keep track of the positive/negative frequencies. The
action of $B$ can now be expressed in terms of $B^{nm}(\v k,\v p)$:
\begin{equation}\label{decomposition_B}
(B\ g)_C(\v x)=\int d\mu(\v k) d\mu(\v p) \   \left(\psi^{n}_{\v
k}\right)_C(\v x)\ B_{nm}(\v k,\v p)\ \ (\psi^m_{\v p},g),
\end{equation}
where $(.,.)$ on the RHS denotes the scalar product in $\H$.  As the field
operator is smeared with the test functions which are elements of the
underlying Hilbert space $\H$:
\begin{equation*}
    \psi(f)=\int d^3x\ {f(\v x)}^\dag_A \psi^A(\v x),
\end{equation*}
so that
\begin{equation*}
      \w(\psi(f)\psi^*(g))=\int d^3x\ d^3y\ f^\dag_A(\v x)\ \w^{AC}(\v x,\v
      y)\ g_C(\v y),
\end{equation*}
we infer that the integral kernel of the operator $B$, $B^{AC}(\v x,\v y)$,
coincides with the two-point distribution $\w^{AC}(\v x,\v y)$. The
two-point function of the free Dirac field is thus defined as
\begin{equation*}
    \w(\psi(f)\psi^*(g))=(f,B\ g).
\end{equation*}
It is the goal of the whole section to find a representation of the CAR
algebra $\Alg$ on certain Hilbert space $\Fou$. In such a representation
there exists typically a certain "base state" $\Vac\in\Fou$ from which all
other states (excitations) are generated with the help of (if they exist)
creation operators. If we assume this "base state" to be
quasi-free\footnote{The quasi-free state has the property that all n-point
functions can be expressed in terms of the two-point function. Only those
n-point functions do not vanish which are the expectation values of an
equal number of field operators and their adjoints, that is:
\begin{equation*}
    \w\bigl(\psi(f_1)\ldots\psi(f_n)\h\psi(g_m)\ldots\psi(g_1)\bigl)=
    \de_{n,m}\ \det[(f_i,B g_j)].
\end{equation*}
What appears on the right hand side is the determinant of a matrix whose
$i,j$-th elements are $(f_i,B g_j)$. The quasi-free property is thought in
the literature to be a modest one, because many known states of quantum
fields can be expressed by density operators in the representation based
upon a  quasi-free state.} then a construction named after Gelfand,
Naimark and Segal tells us how to find the representation $\pi$, the
Hilbert space $\Fou$ and the "base vector" $\Vac$:

\subsection{GNS construction}\index{GNS construction}

The GNS construction starts from an observation that the elements of the CAR
algebra $\Alg$ (that is sums of products of field operators and their
adjoints) may be regarded as vectors in a Hilbert space. Indeed, the
quantized Dirac field operator $\psi(f)$, $f\in\H$, on static backgrounds is
an element of $\Alg$. This operator (cf. section \ref{section_static})
corresponds to a vector in the Fock space $\Fou$ via
\begin{equation*}
\psi(f)=a(P_+f)+\h b(P_- f)\ \longleftrightarrow\ \psi(f)\Vac=\h
b(P_-f)\Vac,
\end{equation*}
which is a one-positron state with wave packet $P_-f$. The idea of GNS is to
construct the whole Fock space by applying products of field operators to
the vacuum $\Vac$. At the beginning, however, we have only $\Alg$ and the
state $\w$; what is needed is:
\begin{itemize}
\item the Hilbert space structure (the scalar product),

\item the creation operators which are elements of $\Alg$ with appropriate
smearing functions (in the example above $\h b(P_-f)=\psi(P_-f)$, $f\in \H$).
\end{itemize}
The scalar product of $A,C\in\Alg$ is provided by the state $\w$:
\begin{equation*}
    (A\Vac,C\Vac)=\w(\h AC).
\end{equation*}
It is semi-definite, because for some elements of $\Alg$ it may happen that
\begin{equation*}
    \w(\h A A)=0.
\end{equation*}
Those elements are precisely the annihilation operators which are the
adjoints of what is sought in order to construct the whole representation
space $\Fou$. The search for the null space of $\w$ is by far the most
non-trivial step of the GNS construction. This null space forms a left
ideal\footnote{Algebraically this follows from the Schwarz inequality
\begin{equation*}
    0\leq|\w\bigl(\h{(CA)}CA\bigr)|^2\leq\w(\h A A)\w(\h A\h C C \h C C A)=0
\end{equation*}}
(named after Gelfand) of $\Alg$ in agreement with the fact that $C A \Vac$
is always a null vector for all $C\in\Alg$, if $A$ is an annihilation
operator.

If $B$ which describes $\w$ is a projection operator on $\H$, then it is
relatively easy to find the annihilators in the one-operator
subspace\footnote{The annihilators in higher subspaces are a rather trivial
addition, for they only enforce Pauli's principle (the appropriate
statistics) in $\Fou$.} of $\Alg$. If $A=\psi(f)+\h \psi(g)\in \Alg$, then
\begin{equation*}
    \w(\h A A)=0\Leftrightarrow \ |P_- f|^2+|P_+ g|^2=0.
\end{equation*}
Therefore $\psi(f_+)=a(f_+)$ and $\h \psi(f_-)=b(f_-)$, $f_\pm=P_\pm f\in
\H$, are the annihilation operators present in the algebra $\Alg$. Products
of their adjoints applied to $\Vac$ generate the electron-positron Fock
space. From the commutation relations of the field operators it follows
\begin{align*}
    \{a,a\}&=0, & \{b,b\}&=0,\\
    \{a,b\}&=0, & \{a,\h b\}&=0,\\
    \{a(f),\h a(g)\}&=(f, P_+ g),& \{\h b(f), b(g)\}&=(f, P_- g).
\end{align*}
The representation obtained in such a way (the standard vacuum
representation) is irreducible. What happens, if $B$ is not a projection, is
studied in appendix \ref{Thermo_field}, as it leads away from the case of
highest interest.

\sp
\subsection{Time evolution, local and global quasi-equivalence}

\index{Quasi-equivalence} There are two ways to incorporate the time
evolution into the theory. One of them is to view the time evolution as an
automorphism of the CAR  algebra $\Alg$, which was the algebra of field
operators on a Cauchy surface. The field operators were operator-valued
distributions smeared with elements of the classical Hilbert space $\H$. The
time evolution can be described by requiring the test functions to evolve in
time. More precisely, the map
\begin{equation*}
    \psi(f)\ra \psi\left(U^*(t,t_0) f\right)
\end{equation*}
preserves the algebraic relations and is reversible. Thus, it describes an
automorphism of the CAR (which will be denoted by $\a_U$).

 The other way to look at the time evolution is to investigate the
two-point function of the state $\w$. Quite frequently it can be decomposed,
as in \eqref{decomposition_B}, with the help of some generalized
eigenfunctions of some selfadjoint operator. The eigenfunctions $\psi^r_{\v
k}(\v x)$ will evolve in time just as classical solutions of the Dirac
equation; the unitary propagator $U(t,t_0)$ describes this evolution. The
two-point function at a later time $t$, $\w_t(\v x,\v y)$, can be found from
the two-point function at the initial time $t_0$, $\w_{t_0}(\v x,\v y)$, with
$\psi^r_{\v k}(\v x)$ replaced by $U(t,t_0)\psi^r_{\v k}(\v x)$. We get
\begin{equation}\label{eq:automorphism}
    \w_t(f,g)=\w(U^*_tf,U^*_tg)=[\w\circ\a_t] (f,g),
\end{equation}
where $\a_t$ is the automorphism of the algebra described earlier. Both
descriptions are equivalent; they correspond to the Heisenberg and
Schr\"odinger pictures of quantum mechanics, respectively.

In the literature there was a strong tendency to relate all the quantities
to the no-external-field (call it Minkowski) situation, partially because one
has regarded the unique Minkowski vacuum as an anchor to experimentally
observable phenomena like the existence of particles\footnote{In view of
local quantum physics, particularly with the examples from quantum field
theory on curved spacetimes (where there are no reasons to relate anything
to the Minkowski vacuum), such a view must be regarded as obsolete.}. The
following question has been addressed: under what conditions on the time
dependence of the external potential is the time evolution unitarily
implementable in the GNS Fock space of the initial state? In other words,
can the state $\w_t$ for all times $t$ be expressed as a vector state in the
initial Hilbert space, or even in the Minkowski Fock space? The restriction
of the allowed types of time-dependent backgrounds obtained in this way was
quite dramatic \cite{Scharf,Thaller,Ru}. The time evolution can be
implemented in the Fock space based upon the "Minkowski vacuum" only for
some\footnote{In \cite{Ru}, chapter 3D, which is our main reference for this
result, the potential $A_0$ is assumed to be a smooth function with rapid
decay.} external fields of electric type, $A_0=A_0(t,\v x)$, $A_i=0$. One
way out of this trouble is to use the asymptotic notions. The state in the
far future (the initial state that evolved in the presence of external
background) can be expressed in the initial Hilbert space under much weaker
assumptions on the time dependent potential. In order to enlighten the way
in which such questions are investigated we present the the
Shale-Stinespring criterion (theorem \ref{Shale}) which settles the issue of
quantum field theoretical implementability of classical unitary
transformations.

\vspace{2em}

{\bf Implementation of unitary transformations.} Consider a representation
of the CAR defined in the Hilbert space of some pure "base state", which is
uniquely described by its projection $P_+$. Let $U:\H\rightarrow\H$ denote a
unitary transformation of the classical Dirac field\footnote{$U$ may denote
the unitary propagator $U(t,t_0)$ or the scattering operator $S$.}. It can
be decomposed as follows:
\begin{equation*}
  U= U_++U_{+-}+U_{-+}+U_-
\end{equation*}
with
\begin{align*}
  U_+&=P_+U P_+, &  U_-&=P_-U P_-,\\
  U_{+-} &=P_+U P_-, & U_{-+}&=P_-U P_+.
\end{align*}
On the algebraic level $U$ can be promoted to an automorphism of the
algebra, because the operators
\begin{equation*}
    \T \psi(f)=\psi(U f)
\end{equation*}
also fulfill the CAR. Suppose $\psi(f)$ denotes already the representative of
the field operator on a Hilbert space. The question arises, whether the field
operator $\T \psi(f)$ can be represented as a unitary (Bogoliubov)
transformation \index{Bogoliubov transformation} of $\psi(f)$,
\begin{equation*}
  \T \psi(f)=\psi(U f)=\U \psi(f) \U^*,
\end{equation*}
for a given $U$. In this case, the original representation and the tilde
representation are \emph{globally equivalent}. The answer when this is the
case is provided by the criterion of D.Shale and W.F.Stinespring \cite{SS}:

\begin{Thm}[Shale-Stinespring criterion\index{Shale-Stinespring criterion}]
\label{Shale} If the operators $U_{+-},\ U_{-+}$ are Hilbert-Schmidt, then
$U$ is unitarily implementable. The state $\w\circ \a_U$ can be expressed
in the Fock space of the state $\w$ as a vector state:
\begin{equation*}
    \T \Vac=\U \Vac,
\end{equation*}
where $\U$ is the implementation of $U$.
\end{Thm}

\begin{remark}
The content of the Shale-Stinespring criterion is the following. We can
always define the tilde representation of the CAR,
\begin{equation*}
  \T \psi(f)=\psi(Uf),
\end{equation*}
where $\psi(f)$ is the base representation with the vacuum $\Vac$. We
could also define the creation/annihilation operators $\T a,\ \T b$ which
would fulfill the same commutation relations as $a$ and $b$:
\begin{equation*}
\T a(f)\equiv \T \psi(P_+f),\qquad \T b(f)\equiv \h{\T\psi}(P_-f).
\end{equation*}
If there existed a vector $\T\Vac$ in the Fock space based upon $\Vac$,
annihilated by $\T a,\ \T b$, then the existence of the unitary
implementation would be automatic, namely the implementation of $U$ could be
defined on a dense subspace of $\Fou$ via
\begin{equation*}
\U\left[\psi(f_1)\ldots\psi(f_n)\h \psi(g_1)\ldots\psi(g_m)\Vac\right]=
\psi(U f_1)\ldots\psi(U f_n)\h \psi(U g_1)\ldots\psi(U g_m)\T \Vac.
\end{equation*}
The Shale-Stinespring criterion tells us when the vector $\T \Vac$ can be
found.
\end{remark}

\sp

We note that the existence of a unitary operator $\U$ which intertwines
between two representations of the CAR algebra is usually referred to as
\emph{global equivalence} of these representations. The Shale-Stinespring
criterion tells us when two representations constructed upon pure "base
states" whose projections are connected by a classical unitary operation
are globally equivalent.

\sp

{\bf Local equivalence.}

In local quantum physics global equivalence of states is a very strong
criterion. There is absolutely no physical reason why all the allowed states
should be globally equivalent to the, for instance,  Minkowski vacuum.
Indeed, phenomena like the thermal radiation present in the future of a
gravitational collapse indicate that the condition of global equivalence is
too strong. For instance, the Minkowski vacuum and the KMS (thermal
equilibrium) state are not globally equivalent; if, due to some interaction
with external sources (for instance cosmic background radiation), the Dirac
field were to thermalize, then it would leave the folium of the Minkowski
vacuum.

In QFT one should, however, always restrict the folium\footnote{Folium means
some reference state $\Vac$ and all the states obtained as density operators
in the representation connected with $\Vac$.} of allowed states. Instead of
insisting on  global equivalence one can investigate \emph{local
equivalence}. It says that in the bounded open regions the states should be
normal to one another. In other words, there would always exist a density
operator which expresses one of the states in the Fock (GNS) space of the
other. Why to postulate local equivalence?  Throughout this paper in order
to define the non-linear observables it is necessary to restrict the
investigations to the class of Hadamard states. Each two Hadamard states are
locally normal\footnote{This has been proven in the context of curved
spacetimes by R.Verch \cite{verch} for scalar fields and for Dirac fields by
C.D'Antoni and S.Hollands in \cite{DAH}.}. In other words: only for locally
normal states we are able do define (relative) non-linear observables like
the current operator or Wick products or time-ordered products.

The issues of global equivalence of states on the CAR algebra have been
studied by R.Powers and E.St\"ormer in \cite{pow_stroe} and also by H.Araki
\cite{araki_CAR}. They have considered the CAR on a Cauchy surface and have
used their characterization with the help of a selfadjoint operator $B$,
$0\leq B\leq 1$ the way we have outlined it above. Among other things they
have obtained the following result:

\begin{Thm}
Two states on the CAR algebra which are parameterized by the operators
$B_1$ and $B_2$ (not necessarily projections) are globally equivalent if
\begin{align*}
    \|\sqrt B_1-\sqrt B_2\|_{H.S.}&<\infty, &
    \|\sqrt {1-B_1}-\sqrt {1-B_2}\|_{H.S.}&<\infty.
\end{align*}
\end{Thm}

\begin{remark}
If the operators $B$ are projections then global equivalence reproduces the
Shale-Stinespring criterion. To see this, let $P_+$ denote one of the
projections and let $B_2=S P_+ S^*$, i.e. the second projection is connected
to the first by means of a unitary transformation (Bogoliubov
transformation). The Shale-Stinespring criterion says, that there exists a
unitary transformation $\S$ on the whole Fock space related to $P_+$ which
implements $S$, if and only if
\begin{equation*}
    \|S_{+-}\|^2_{H.S.}=\tr [(P_+ S P_-)^*(P_+ S P_-)]< \infty.
\end{equation*}
Using the properties of the trace one verifies that the above criterion is
equivalent to
\begin{equation*}
    \tr(P_+-S^* P_+ S P_+)<\infty.
\end{equation*}
On the other hand the criterion of global equivalence, in case the $B$'s
are projections as above, is
\begin{equation*}
    \tr[(P_+-SP_+S^*)(P_+-SP_+S^*)]=\tr(P_+-P_+SP_+S^*-P_+SP_+S^*+P_+)<\infty,
\end{equation*}
and so it is the same as the Shale-Stinespring criterion.
\end{remark}

Let us now look at the local equivalence of two states $\w_1$ and $\w_2$. We
restrict those states to an open ball $\C$ on a certain Cauchy surface
$\Sigma_t$. The restriction of a state to $\C$ means that we multiply its
projection $P_{1,2}$ by a characteristic function of the region $\C$ from
both sides. We also restrict the underlying Hilbert space to
$\H\rest_\C=L^2(\C,d^3x)^4$. The states are locally equivalent, if the
representations induced by the restrictions of the states to $\C$ are
unitarily equivalent\footnote{For all possible $\C$ on all Cauchy surfaces.}.

\begin{remark}
Even if the state $\w$ is pure, so that it is characterized by a projection
$B$ the restriction to $\C$, $\w\rest_{\C}$, will not be a pure state, i.e.
will not be given by a projection. On the other hand it is possible do
define a pure state on a given region $\C$ and extend it to the full algebra.
Such a state, however, will not be locally equivalent to any Hadamard
state\footnote{Statements of this sort, although without a proof here, are a
common lesson of local quantum physics \cite{Haag,fre_super}, and are related
to the fact, that local algebras $\Alg(\O)$ are factors (possess a trivial
center) of the type $III_1$. This ensures that there are no pure states on
$\Alg(\O)$ locally normal to any quasi-free Hadamard state.}.
\end{remark}

The following theorem of \cite{pow_stroe} gives a condition for local
equivalence of states

\begin{Thm}\label{local_normality}
Two states on the CAR algebra which are parameterized by the operators $B_1$
and $B_2$ are locally normal, if
\begin{equation*}
    \|\sqrt{B_1}-\sqrt{B_2}\|_{H.S.}<\infty.
\end{equation*}
The Hilbert-Schmidt norm is evaluated on the restricted Hilbert space
$\H\rest_\C$; the above condition should hold for any open bounded region
$\C$.
\end{Thm}

On the mathematical level it is clear that local equivalence is much
weaker than the global one. As an example of locally equivalent states
which fail to be globally equivalent let us consider the following

\begin{example} Let us take for simplicity two pure states described
by projection operators. Moreover, let us assume both of them can be
expressed as functions of some selfadjoint operator which possesses a
continuous spectrum. The eigenfunctions of this operator will be denoted by
$\psi_p(\v x)$. The second two-point function $\w_2$ be built out of the
first, $\w_1$ by subtracting a projection operator $\De$ which is smaller
than $\w_1$. More precisely, the projection will be characterized by a
characteristic function\footnote{The possible values of $\chi(\v p)$ are $0$
and $1$, because we want it to be a projection, moreover, we assume $\chi$ is
such that the operator $\w_2$ is positive (i.e. $\De<\w_1$).} $\chi(\v p)$
of compact support. We set\footnote{This must be done in such a way, that
$\w_2$ as an operator remains positive.}
\begin{equation*}
\w_2(\v x,\v y)=\w_1(\v x,\v y)-\int d^3p\ \chi(\v p) \psi_p(\v
x)\psi^*_p(\v y)\equiv \w_1(\v x,\v y)-\De(\v x,\v y).
\end{equation*}
Global equivalence would imply
\begin{equation*}
    \|B_1-B_2\|_{H.S.}<\infty,
\end{equation*}
where $B$'s denote the projections which characterize both states. Clearly
the integral kernel of the difference $B_1-B_2$ is just $\De(\v x,\v y)$.
Its Hilbert-Schmidt norm is
\begin{equation*}
    \|B_1-B_2\|_{H.S.}=\int d^3x\ d^3y \ |\De(\v x,\v y)|^2=
    \int  d^3p\ d^3k\ \h{\chi(\v k)}\chi(\v p)\
    \left(\psi_k,\psi_p\right) \left( \psi_p, \psi_k\right),
\end{equation*}
where $(\psi_p,\psi_k)$ is the scalar product, from $\H$, of two scattering
states, which is equal to $\de(\v p-\v k)$. Thus the above Hilbert-Schmidt
norm is infinite.

On the other hand, the question of local equivalence, may be answered with
the help of the inequality proved by Powers and St\"ormer:
\begin{equation*}
    \|\sqrt{B_1}-\sqrt{B_2}\|^2_{H.S.}\leq \|B_1-B_2\|_{tr}.
\end{equation*}
Let $g(\v x)$ denote the characteristic function of the region $\C$. We
estimate the RHS:
\begin{equation*}
\|B_1-B_2\|_{tr,\C}=\|g\circ \Delta\circ g\|_{tr}=\|\De\circ g\|_{H.S.}^2,
\end{equation*}
because $\De$ is a projection. The Hilbert-Schmidt norm of $\De\circ g$
can be evaluated:
\begin{align*}
\|\ldots\|_{H.S.}^2&=\int d^3p\ d^3k\ \chi(\v p)\chi(\v k)\
|(\psi_p,g\psi_k)|^2\leq\\ &\leq \int d^3p\ d^3k\ \chi(\v p)\chi(\v k)\
\|g\psi_p\|^2 \|g\psi_k\|^2=\left[\int d^3p\ \chi(\v p)\
\|g\psi_p\|^2\right]^2,
\end{align*}
by the Schwarz inequality. As the norm of the scattering states $\psi_p$
restricted to the region $\C$ is a bounded function of $\v p$ (at least if
the external fields are not singular) and $\chi(\v p)$ is a function of
compact support, we infer that the states $\w_1$ and $\w_2$ are locally
equivalent in every bounded open region $\C$.
\end{example}

The preceding discussion has made it clear that in the external-field
electrodynamics there is an enormous freedom of choosing "a basis state" for
the construction of the representation.  In what follows we shall
investigate an example, where we compare two natural candidates for such
"basis states", namely the state which evolved out of a preferred state in
the past (when for instance the external fields were absent) and the state
termed "instantaneous ground state" which is prescribed by a projection on
the positive-frequency subspace of the instantaneous Hamiltonian.

\subsection{An example (equivalence of states, instantaneous vacua)}
\begin{example} We are going to consider a model of the Dirac field
on such external field background that there would be positive- and
negative-frequency bound states\footnote{We think of this model as of
fermions bound by gravity. Indeed the model will resemble the construction
of representations of the CAR given for Robertson-Walker spacetimes by
S.Hollands \cite{hollands_dirac}.}. In order to simplify the computational
side we assume the energies of the $n$-th electronic bound state to be equal
in absolute to the energies of positrons in the bound state of the same
quantum number. Additionally, there would be a time dependence of the
Hamiltonian. Again, for simplicity, we assume that the interaction couples
only the bound states of the same $n$ (but positive and negative subspaces
are allowed to interact). The Hilbert space of the system, $\H$, is
\begin{equation*}
    \H=\C^2\otimes\ell_2.
\end{equation*}
Then a convenient orthonormal basis of this space is formed by the vectors
\begin{align*}
 \psi^+_n&=\begin{pmatrix}
      1 \\
      0 \\
    \end{pmatrix}\otimes |n\rangle,
 &
 \psi^-_n&=\begin{pmatrix}
      0 \\
      1 \\
    \end{pmatrix}\otimes |n\rangle,
\end{align*}
where $n\in \N$. The time evolution of the model is governed by the
Hamiltonian
\begin{equation*}
    H(t)=\sum_n  \left(E_n \ \s_3+
    \frac{f(t)}{n}\ \sigma_2\right)\otimes  |n\rangle\langle n|,
\end{equation*}
where $\s$'s are Pauli matrices and $f(t)$ denotes an arbitrary function
of time.

Let the first state, $\w_1$, be described by a projection on the positive
frequencies at $t=0$. Suppose that the function $f$ \emph{vanished} at that
surface. Thus at that surface we have
\begin{equation*}
    \w_1=\sum_{n} \psi^r_n \ B_1^{rs} \ \psi^{*s}_n,
\end{equation*}
where \[B_1=\begin{pmatrix}
  1 & 0 \\
  0 & 0 \\
\end{pmatrix}\]
and\footnote{Here we introduce a new index $r$ for brevity. Its possible
values are $+$ and $-$ which correspond to the upper and lower vector
components respectively.} $\psi^r_n$ is the orthonormal basis introduced
earlier which is also the basis of the eigenvectors of the Hamiltonian
with $f=0$.

The goal is to compare the state $\w_1(t)$, evolved from $0$ to $t$
according to the time evolution governed by the Hamiltonian, with some
instantaneous ground state defined via projection on what at $t$ appears to
be the positive-frequency subspace. Thus, the second state is defined as
\begin{equation*}
    \w_2^{t}=\sum_{n} \psi^r_{t,n} \ B_2^{rs} \ \psi^{*s}_{t,n},
\end{equation*}
where
\[
B_2=\begin{pmatrix}
  1 & 0 \\
  0 & 0 \\
\end{pmatrix}
\]
and $\psi^r_{t,n}$ are instantaneous eigenvectors of the Hamiltonian at time
$t$:
\begin{align*}
 \psi^+_{t,n}&=\begin{pmatrix}
      \cos(\a/2) \\
      i \sin(\a/2) \\
    \end{pmatrix}\otimes |n\rangle,
 &
 \psi^-_{t,n}&=\begin{pmatrix}
     \sin(\a/2) \\
      -i\cos(\a/2) \\
    \end{pmatrix}\otimes |n\rangle,
\end{align*}
where $\a=\a(n,t)$ and
\begin{align*}
    \sin(\a/2)&=\sqrt{\frac{1}{2}\left(1-\frac{E_n}{\sqrt{E_n^2+f^2(t)}}\right)},\\
    \cos(\a/2)&=\sqrt{\frac{1}{2}\left(1+\frac{E_n}{\sqrt{E_n^2+f^2(t)}}\right)}.\\
\end{align*}

What is needed in order to check the local equivalence\footnote{Locality
means that we assume finiteness of the norm of each $\psi_n^r$.} of both
states is to express the (evolved) $\psi_n^r(t)$ in the new basis
$\psi^r_{t,n}$. As a matter of fact, it is generally impossible to find the
unitary time evolution operator explicitly -  we have to solve the
Schr\"odinger equation:
\begin{equation*}
    [i\d_t-H(t)]\psi_n(t)=0,
\end{equation*}
which in the interaction picture
\begin{equation*}
    \psi_n(t)=e^{-iE_n\s_3 t}\vp_n(t)
\end{equation*}
leads to the equation
\begin{equation*}
    i\dot \vp_n=\frac{f(t)}{n}\ e^{iE_n\s_3t}\ \s_2\ e^{-iE_n\s_3t}\ \vp_n,
\end{equation*}
that is
\begin{equation*}
i\dot \vp_n=\frac{f(t)}{n}\*[\cos(2E_nt)\s_2+\sin(2E_nt)\s_1]\ \vp_n.
\end{equation*}
As the operator in square brackets depends on time\footnote{Thus, the
generator of the evolution cannot be written as a certain function of time
times some time-independent operator.}, it is not possible to write down an
explicit solution of this equation generally, i.e. without investigating the
concrete structure of $f(t)$ and $E_n$.

Due to the above observation, we are forced to consider special cases in
hope of gaining some insight into the construction of various states on
time-dependent external fields.

Suppose the state is defined by a projection operator $B(0)$ and that the
interaction term vanishes, $f=0$. Then at a later time the state is defined
via a time-evolved projection which may be written as
\begin{equation*}
B(t)=\sum_n\ e^{-iE_nt\s_3}\ \psi_n^r\ B_{rs} \ \psi^{*s}_n\
e^{iE_nt\s_3}.
\end{equation*}
In other words, $B(t)$ restricted to the subspace $n$ is
\begin{equation*}
B(t)=\begin{pmatrix} B_{+}& B_{+-} e^{2iE_nt}\\
B_{-+}e^{-2iE_nt}& B_{-}\end{pmatrix}.
\end{equation*}

Is $B(t)$ equivalent to $B(0)$? As both operators, $B(t)$ and $B(0)$, are
projections the condition on the equivalence of the respective
representations,
\begin{equation*}
\|\sqrt{B(0)}-\sqrt{B(t)}\|^2_{H.S.}<\infty,
\end{equation*}
may be replaced by
\begin{equation*}
\|B(0)-B(t)\|^2_{H.S.}<\infty.
\end{equation*}
The difference inside the norm is easily seen to be equal to
\begin{equation*}
\De B=\begin{pmatrix} 0& B_{+-}( e^{2iE_nt}-1)\\
B_{-+}(e^{-2iE_nt}-1)& 0\end{pmatrix}
\end{equation*}
with $B_{+-}=B_{-+}^*$ (they are numbers). We find
\begin{equation*}
(\De B)^* (\De B)=4 \sin^2(E_n t)
\end{equation*}
on the subspace $n$. The Hilbert-Schmidt norm may now be evaluated, say,
w.r.t. the basis $\psi_n^r$ of the Hilbert space $\H$. We find
\begin{equation*}
\|.\|^2_{H.S.}=\tr \left(\sum_{n=0} 4 \sin^2(E_n t) |B_{+-}|^2
|n\rangle\langle n|\right).
\end{equation*}
The trace is then calculated, and the condition for the equivalence of $B(t)$
and $B(0)$ becomes
\begin{equation*}
\sum_n \sin^2(E_n t)\ \sin^2[2\a(n)]<\infty,
\end{equation*}
where we have employed the explicit parametrization of $B$ at the subspace
$n$, namely:
\begin{equation*}
B(0)=
B(t)=\begin{pmatrix} B_{+}& B_{+-} \\
B_{-+}& B_{-}\end{pmatrix} \equiv
\begin{pmatrix} \cos^2[\a(n)]& \frac{1}{2}\sin[2\a(n)]\\
\frac{1}{2}\sin[2\a(n)]& \sin^2[\a(n)]\end{pmatrix}.
\end{equation*}
where $\a$ is an angle which may vary with $n$.

We observe the following: not every projection operator is compatible even
with the free time evolution. Suppose the spectrum of the Hamiltonian is
arbitrary, then the angles $\a(n)$ have to be chosen in such a way that the
operator $B_{+-}$ is Hilbert Schmidt which means that they have to decrease
for large $n$ quicker than $1/n$.

On the other hand, if the energies $E_n$ accumulate around $0$ for large $n$
quickly (similarly, $E_n$ has to decrease quicker than $1/n$) than the
equivalence of $B(t)$ and $B(0)$ will be guaranteed irrespective of $B(0)$.

We see, therefore, that both concrete ingredients, the projection operator
$B(0)$ and the spectrum of the free Hamiltonian, may play a role in the
equivalence of states at different times.

\end{example}

\subsection{Physical meaning of local and global equivalence}

Physically the notions discussed in this section mean the following: global
equivalence means that the vacuum corresponding to the state $\w_2$, when
expressed in the Fock space of $\w_1$, contains finitely many "particles";
local equivalence means that the relative energy density (defined by a
point-splitting procedure) is finite.

\newpage
\section{Quantization on the static external field backgrounds}
\label{section_static} This section contains an important special case of
the general construction of states, namely, the construction of a ground
state of the Dirac field on static external backgrounds. An even greater
specialization, namely the case of absence of any  external backgrounds has
been included in appendix \ref{swobodne pole diraca} for comparison. We
recall the construction of this representation, because the static external
fields provide the simplest non-trivial example of the external-field
problem.

\subsection{Negative and positive frequency subspaces of $\H$}\quad
If the external fields are time-independent, it is possible to define a
splitting of  $\H$ into positive- and negative-frequency parts which will be
invariant under the time evolution. More precisely, let us define the
operator
\begin{equation}\label{involution_signum}
  \tau=\sgn H=\frac{H}{|H|},
\end{equation}
if the Hamiltonian $H$ has a vanishing kernel. With the help of $\tau$ we
define the projections
\begin{equation}\label{spinorowe projektory}
  P_\pm=\frac{1}{2}(1\pm \tau),
\end{equation}
and the splitting of the Hilbert space $\H$ into the positive/negative
frequency subspaces
\begin{align*}
  P_\pm \H_\pm&=\H_\pm\\
  \H &=\H_+\oplus\H_-.
\end{align*}
If the zero belongs to the spectrum of $H$, it is necessary to settle the
splitting of $\ker H$ by a separate prescription. The splitting presented
above is invariant under the time evolution, because $[P_+,H]=0$ and the
Hilbert spaces $P_+ \H$ defined at different times are the same.

\begin{remark}
The operator $\tau$ does not preserve the localization; namely if one takes
$\psi\in\H$ to have a compact support within a certain region $N\subset\R^3$,
then in general $\psi_+=P_+\psi\in\H_+$ does not have a compact support.
Consequently, the term "relativistic quantum mechanics" has little sense, as
the electron which should be described as a positive-frequency solution of
the Dirac equation propagates acausally.
\end{remark}

On the  Hilbert space $\H$ we may also investigate the charge
conjugation\index{Charge conjugation} operator which is defined (in the
spinor representation) by
\begin{equation*}
  C\psi=i\g^2 \psi^\dag,
\end{equation*}
for $\psi\in \H$ ($\psi^\dag$ denotes the Hermitian conjugation of $\psi$,
and $\g^2$ the second Dirac matrix). The operator $C$ possesses the following
properties:
\begin{itemize}
\item it is anti-unitary: $(C\psi,C\chi)=(\chi,\psi)$;

\item double application of $C$ is the identity\footnote{In the spinor representation
this is due to $(i\g^2)^*=i\g^2$.}: $C(C\psi)=\psi$

\item $C$ "conjugates the
charge", that is
\begin{equation*}
C H(e) C^{-1}=-H(-e).
\end{equation*}
\end{itemize}

\sp
\subsection{Representation of the CAR algebra}\quad

In the following we shall recall the construction of the Fock representation
which is based upon the splitting prescribed by $P_+$. The Hilbert space on
which the operators will be represented, the Fock space, is constructed with
the help of creation operators. Let $f_i$ denote the orthonormal basis
vectors of the space $\H_+$ and $g_i$ the orthonormal basis vectors of
$\H_-$. One defines a Hilbert space $\Fou^{(n,m)}$ as a space spanned by the
vectors of the form
\begin{equation*}
  \h a(f_1)\ldots\h a(f_n)\h b(g_1)\ldots\h b(g_m)\Vac.
\end{equation*}
The space $\Fou^{(n,m)}$ is equipped with the scalar product, whose form
follows from the classical scalar product $(.,.)$ and the anti-commutation
relations:
\begin{align*}
  \{a(f_i),\h a(f_j))\}&=(f_i,f_j), & \{b(g_i),\h b(g_j))\}&=(g_j,g_i), \\
  \{a(f_i),\h b(g_j))\}&=0, & \{a(f_i), b(g_j))\}&=0, \\
  \{a(f_i),a(f_j))\}&=0, & \{b(g_i), b(g_j))\}&=0. \\
\end{align*}
The Fock space\index{Fock space} is a direct sum of all the spaces
$\Fou^{(m,n)}$ for all $m,n\in\mathbb N$
\begin{equation*}
  \Fou=\bigoplus_{n,m=0}^\infty \Fou^{(n,m)},
\end{equation*}
where the case $m,n=0$ denotes the vacuum.

The representation of the field operators can now be given:

\begin{definition}\label{def:dirac_field}
The field operator \index{Dirac field operator} is an operator valued
distribution which takes as a test function all the 4-component $\dom(\R^3)$
functions on the surface of constant time (such functions are dense in
$\H$). If $h\in \dom(\R^3)$ then we express the field operator in terms of
the creation/annihilation operators:
\begin{subequations}\label{op_pola}
    \begin{equation}
        \psi(h)=a(P_+h)+b^*(P_-h).
    \end{equation}
The (Fock space) adjoint of the field operator is given by
    \begin{equation}
        \psi^*(h)=a^*(P_+h)+b(P_-h).
    \end{equation}
\end{subequations}
\end{definition}

The field operators are bounded, antilinear and fulfill the canonical
anti-commutation relations (CAR)\index{Canonical anti-commutation relations},
\begin{equation}\label{CAR_t}
  \{\psi(f),\h \psi(g)\}=(f,g),
\end{equation}
where $(.,.)$ denotes the usual scalar product in $\H$, i.e.
\begin{equation*}
  (f,g)=\int d^3x\ \b{ f(x)} \g^0 g(x)=\int d^3x \ f(x)^\dag\ g(x).
\end{equation*}

The definition of the field operators completes the construction of the
vacuum representation of the CAR algebra \eqref{CAR_t}\footnote{The
abstract symbols $\psi(f)$ have acquired a concrete realization as
operators on a Hilbert (Fock) space.}.

\vspace{1em}
\subsection{Implementability of the unitary evolution}\quad

The next question we are going to discuss is whether the free time evolution
which is given on the  classical level by a unitary group $U_t$ can be
implemented as a unitary group on the Fock space. This question is important
because we would like to relate the field operator at the time $t$, $\psi_t$
to the filed operator at the time $t=0$, $\psi$ by an action of a unitary
operator:
\begin{equation*}
    \psi_t(f)=\U_t \psi(f) \h \U_t,
\end{equation*}
which is just the Bogoliubov transformation. We may define
\begin{equation*}
  \psi_t(f)=\psi_0(\h U_t\ f),
\end{equation*}
where $U_t$ denotes the classical unitary evolution operator, and the star
denotes the $\H$-adjoint operation. In order to answer the question under
consideration we just have to apply the Shale-Stinespring criterion (theorem
\ref{Shale}). In the case of time-inde\-pen\-dent potentials, if the
splitting of the Hilbert space $\H$ is given by some function of the
Hamiltonian, for instance
\begin{equation*}
  P_\pm=\frac{1}{2}(1 \pm \sgn H),
\end{equation*}
it is particularly easy to implement the time evolution. The classical
evolution is given by the one-parameter unitary group
\begin{equation*}
  U(t)=e^{-itH},
\end{equation*}
and, consequently (because $P_+$ commutes with $H$)
\begin{equation*}
  U_{+-}=0=U_{-+}.
\end{equation*}
The Shale-Stinespring tells us that the classical evolution in the case
under consideration  is \emph{always} unitarily implementable.

\subsection{Ground state}\label{diff_inv}\quad
Let us  consider the situation of a vacuum state in static potentials
vanishing at infinity. For such potentials  there are finitely many points
of the spectrum of the Dirac-Hamiltonian $H$ in the vicinity of $E=0$.

If $P_+$ denotes the projection operator on the subspace of positive
eigenvalues of $H$, then, according to the preceding construction, we may
obtain a vacuum state. However, there is a freedom in the choice of the
splitting point (here $E=0$). In what follows we will argue that different
splittings lead to a globally equivalent representation (Bogoliubov
transformation) and that only $E=0$ leads to the state of minimal energy
which will be called the ground state\index{Ground state}.

\begin{lemma}
Let $\H=\ell_2(\Z)$ be the Hilbert space spanned by the orthonormal basis
$f_i$, $i\in\Z$. Let the splitting $\H=\H_+\oplus\H_-$ be prescribed by
the projection operator on the vectors with positive indices:
\begin{equation*}
    P_+ f_i=f_i\quad \forall i>0,
\end{equation*}
both $\H_+$ and $\H_-$ are assumed to be infinite-dimensional. A splitting of
this sort may correspond, for instance, to a splitting into positive- and
negative-frequency subspaces of a Hamiltonian $H$. Suppose, furthermore, that
the system evolved in time and the classical evolution is given by the
unitary operator
\begin{equation*}
    U f_i=f_{i+1} \quad \forall i\in\Z.
\end{equation*}

Then the unitary transformation $U$ can be implemented in the original Fock
space. In other words the state $\w$ corresponding to $P_+$ and the
state\footnote{See equation \eqref{eq:automorphism}.} $\T \w=\w\circ \a_t$
are globally equivalent.
\end{lemma}

\begin{proof}
The vacuum $\Omega$ of the representation associated with $\w$ is defined
by the relations
\begin{align*}
  a(P_+f)\Omega &=0, &  b(P_-f) \Omega&=0
\end{align*}
for all  $f\in\H$. The new representation of the CAR is defined via
\begin{equation*}
    \T \psi(f)=\psi(U f)
\end{equation*}
for all $f\in\H$. Specifically we obtain the following Bogoliubov mapping
of the creation/annihilaiton operators:
\begin{align*}
    \T a(P_+ f)&=a(U_{+}f)+\h b(U_{-+}f),\\
    \T b(P_- f)&=\h a(U_{+-}f)+b(U_-f),
\end{align*}
which are the Bogoliubov relations. From the definition of $U$ we infer
\begin{align*}
    U_{+-}&=P_+ U P_-=P_0,\\
    U_{-+}&=P_- U P_+=0,
\end{align*}
where $P_0$ is the projection onto $f_0$. We may summarize this as follows:
\begin{align*}
    \T a(P_+ f)&=a(P_{+}f),\\
    \T b(P_- f)&=\h a(P_0 f)+b(P_-f).
\end{align*}
Thus the only nontrivial Bogoliubov transformation is
\begin{equation*}
    \tilde b(f_0)=a^*(f_0).
\end{equation*}
In this way we obtain the mapping of the two representations. What remains
to be shown is that the state $\T\w$ can be expressed as a vector  in the
Fock space constructed upon the vacuum $\Vac$, which is annihilated by $\T
a $ and $\T b$:
\begin{align*}
  \T a(P_+f)\T \Omega &=0, &  \T b(P_-f) \T \Omega&=0.
\end{align*}
That this is the case may easily be verified, namely
\begin{equation*}
  \tilde \Omega=a^*(f_0) \Omega,
\end{equation*}
the operator $\T b(f_0)$ annihilates $\T\Vac$, because it is equal to $\h
a(f_0)$ and we have Pauli's exclusion principle.
\end{proof}

If the states with different splitting of $\H$ are expressed in the same
Fock space (as in the preceding lemma) we may ask: which of them has the
minimal energy? The energy operator (the self-adjoint generator of the time
evolution unitary group) up to an arbitrary constant equals
\begin{equation*}
    \mathbb H=\sum_n\  [E_+(n)\ \h a(f_n) a(f_n)-
    E_-(-n)\ \h b(f_{-n}) b(f_{-n})]+E_+(0)\ \h a(f_0) a(f_0),
\end{equation*}
where $E_\pm(n)$ denote the eigenvalues of the time-independent Hamiltonian
$H$ corresponding to the eigenvectors $f_n$, $n\in \N$. We assume that
$E_+(n)$  are \emph{ positive} numbers in contrast to $E_-(-n)$. If we
compare the expectation value of $\mathbb H$ in different states (regarded,
due to the global equivalence, as different vectors in the same Fock space),
we arrive at

\begin{lemma}Of all possible states characterized by  projection operators
on the subspaces of frequencies greater then $E$ the state with $E=0$ has
the lowest energy.
\end{lemma}
\begin{proof}In the case, where the splitting is done with respect to $E=0$,
 the expectation value of $\mathbb H$ evidently vanishes. Suppose that
 we investigate $\T \Vac$ where one positive-frequency vector $f_0$ is included in $\T P_-$. We calculate
\begin{equation*}
    (\T\Vac,\mathbb H\ \T\Vac)=E(f_0)\ \bigl(
    \h a(f_0)\ \Vac,\h a(f_0) a(f_0)\ \h a(f_0)\ \Vac\bigr)=E(f_0)
\end{equation*}
which is greater than zero.
\end{proof}

\begin{remark}The above lemma tells us that it is not at all arbitrary
where to put the cut into positive/negative frequencies. Indeed, any choice
other than $E=0$ leads to states of the Dirac field which are not the states
of lowest energy in the Fock space. All the states of lower energy, however,
differ in charge from $\T\Vac$ so that it would be difficult to imagine an
electrodynamic process which could extract this energy.
\end{remark}

\subsection{Time-dependent external fields}
\quad Suppose that we have external fields which vanish in the far future and
past. In such a case, instead of trying to implement the (classical) unitary
propagator $U(t,s)$ in the free (no external field) Fock space, which is
rarely possible, one might try to implement the scattering matrix $S$ only.
The Shale-Stinespring criterion (theorem \ref{Shale}), when applied to the
scattering matrix, provides a (rather weak) tool to decide for which external
fields the unitary implementation $\S$ of $S$ exists:

\begin{Thm}[Theorem  8.25 of \cite{Thaller} and 5.1 of \cite{Scharf}]
Let $W_n(t,\v x)$ denote the strong derivatives of the potential \[V(t,\v
x)=e\g^0[\g^aA_a(t,\v x)]\] with $n=0,1,2$, that is $W_n(t,\v x)=dV^n/dt^n$.
Let the family $W_n(t,\v x)$ be strongly continuous in $t$. Furthermore, let
the Fourier transform of $W_n(t,\v x)$ satisfy
\begin{equation*}
  \int_{-\infty}^{\infty} \|\hat W(t,.)\|^k\ dt<\infty
\end{equation*}
for each $n=0,1,2$ and all $k=1,2$. Then $S_{+-}$ and $S_{-+}$ are
Hilbert-Schmidt operators, and thus the scattering operator $S$ is unitarily
implementable.
\end{Thm}

\begin{remark}In the course of the proof of the above criterion it is easily seen that,
 given the assumptions of the theorem, the first-order operators
\begin{align*}
  S^{(1)}_{+-} &=-ieP_+\int ds \ e^{iH_0s} \g^0 \g\*A(s,\v x)\ e^{-iH_0s}P_-,\\
  S^{(1)}_{-+} &=-ieP_-\int ds \ e^{iH_0s} \g^0 \g\*A(s,\v x)\ e^{-iH_0s}P_+,
\end{align*}
are also of the Hilbert-Schmidt type.
\end{remark}

\chapter{Quantization of the electromagnetic field}\label{chapter_EM}

The theory of the free, quantum radiation field (i.e. electromagnetic field
without sources) will be presented in this chapter. We shall follow the
standard method\index{Gupta-Bleuler method} of Gupta and Bleuler
\cite{gupta,bleuler} explained in a brief manner in \cite{itzykson_zuber}.

\section{Quantization of the vector potential}

The vector potential $\A_\mu (x)$ is the quantity that will be
quantized.
We introduce a sesquilinear form $\langle.,.\rangle$ on the space of
functions of fast decrease (Schwartz space)
\begin{equation*}
    \langle f,g\rangle= -\frac{1}{(2\pi)^3}\int\frac{d^3p}{2p^0} \ \b{f_\mu(\v p)}
    g_\nu(\v p)\eta^{\mu\nu}.
\end{equation*}
This form is not positive-definite. The space of four-component Schwartz
functions together with the sesquilinear form $\langle.,.\rangle$ will be
called the Krein space\index{Krein space} $\K$.

The standard decomposition of the four-vector electromagnetic potential
(classical), which is a basis of the subsequent quantization, reads
\begin{equation}\label{operator_pola_radiacyjnego}
  \A_\mu(x)=\frac{1}{\sqrt{2\pi}^3} \int \frac{d^3k}{\sqrt{2k^0}}
  \  e^\a_\m(\v k)
  \left\{ a^*_\a (\v k)\ e^{ikx}+a_\a (\v k)\ e^{-ikx}\right\},
\end{equation}
where $k_0=|\v k|$, and $ e^\a(\v k)$ denote four real polarization
vectors\index{Polarization vectors} which are supposed to be orthogonal to
one another and normalized so that three of them are spatial, namely $ e^1,
e^2, e^3$, and $ e^0$ is timelike. Moreover, as all spatial polarization
vectors will be $\v k$-dependent, we choose $e^1,  e^2$ to be of the form
$e^{1,2}=(0,\v e^{1,2})$ and to be orthogonal to  $(0,\v k)$:
\begin{align*}
    \v e^\a (\v  k)\  \v e^\be (\v k)&= \de^{\a\be},\\
    \v e^\a (\v k)\   \v k &=0,
\end{align*}
for $\a,\be=1,2$. Altogether $(e^\a)_\mu$ form a four-tetrad:
\begin{equation*}
\eta_{\a\be}\ (e^\a)_\mu (e^\a)_\nu=\eta_{\m\n}.
\end{equation*}

We can now quantize the theory turning $a^\a(\v k)$ into
operators\footnote{Those symbols will be operators when smeared with
appropriate test functions.} which are supposed to fulfill the commutation
relations
\begin{equation}\label{komutacje_radiation}
[a_\a(\v p), a_\be^*(\v k)]=-\eta_{\a\be}\ \delta(\v p-\v k)
\end{equation}
and to act on the Fock space constructed out of the Krein\footnote{Here we
can understand why the indefinite product space $\K$ has been introduced. In
quantum theory we always search for a representation of the observable
quantities as selfadjoint operators on a \emph{Hilbert space}. Here, however,
such a representation cannot be constructed covariantly (i.e. to respect the
covariant CCR \eqref{commutation_AA}). The reason for this is that the
relation $[a_0,\h a_0]=-1$ cannot be fulfilled by selfadjoint operators on a
positive product (Hilbert) space.} space $\K$. Such a Fock-Krein
\index{Fock-Krein space} space $\Fk$ is constructed in a standard manner.
Its n-particle subspace is the symmetrized tensor product
\begin{equation*}
    \Fk^{(n)}=S_n \ \K\otimes\ldots\otimes\K.
\end{equation*}
The operators $a_\a$, $\h a_\a$ act on $\Fk$ as standard creation
operators:
\begin{align}
    a^*(f)\ S_{n}  h_1\otimes \ldots \otimes h_n
    &=\sqrt{n+1}\ S_{n+1} f\otimes h_1\otimes \ldots \otimes h_n,\\
    a(f)\ S_{n}  h_1\otimes \ldots \otimes h_n
    &=\frac{1}{\sqrt{n}}\sum_{i=1}^n \langle f,h_i\rangle\
    S_{n-1} h_1\otimes \ldots \otimes\tilde h_i\otimes\ldots \otimes h_n,
\end{align}
where the tilde denotes the omission of $h_i$ in the symmetrized product.
The vacuum is defined via
\begin{equation*}
    a_\a(\v k)\Vac=0.
\end{equation*}

 This is the standard Gupta-Bleuler
representation of the field operators\footnote{The Fock-Krein space is
sometimes called Gupta-Bleuler space.}. Note that the adjoint $^*$ means the
Krein-adjoint\index{Krein adjoint}. With such notations the operator
$\A_0(f)$,
\begin{equation*}
    \A_0(f)=\int d^4x \ f(x) \A_0(x),
\end{equation*}
is Krein symmetric:
\begin{equation*}
    \langle \A_0(f) F,G\rangle=\langle F, \A_0(f) G\rangle\qquad \forall
    F,G\in\Fk.
\end{equation*}

\section{The Lorentz condition and the physical Hilbert space}
The Krein product on $\K$ is not positive-definite\footnote{Which means there
are vectors $F$ for which
\begin{equation*}
    \langle F,F\rangle<0.
\end{equation*}}. However, up to now the vector potential
does not fulfill the Maxwell equations. It does fulfill
\begin{equation*}
    \Box \A_\mu (x)=0
\end{equation*}
which however follows from
\begin{equation*}
    \d_\n F^{\m\n}=0
\end{equation*}
with
\begin{equation*}
    F_{\m\n}=\d_\m \A_\n-\d_\n \A_\m,
\end{equation*}
only if the Lorentz condition is imposed
\begin{equation*}
    \d_\m \A^\m=0.
\end{equation*}
It is very fortunate that we can remedy two shortcomings in one stroke,
namely, a way to impose the Lorentz condition as a strong operator equation
at the same time deletes the negative "norm" states from the Krein space.
More specifically, the physical Hilbert space $H_{p}$ is defined with the
help of
\begin{equation*}
    \chi(x)=  \d^\m \A_\mu(x)^{(+)}=\frac{1}{\sqrt{2\pi}^3}
    \int \frac{d^3k}{\sqrt{2k^0}}\  e^\a_\m(\v k)k^\m\
  a_\a (k) e^{-ikx}
\end{equation*}
via\footnote{This condition is (as $\chi(x)$) Lorentz-invariant.}
\begin{equation*}
    H_p=\{F\in\K:\ \chi(x) F=0\}.
\end{equation*}
The condition\footnote{As we have already said this is a strong operator
equation. It is weaker than an operator equation (which would assert $\d^\m
\A_\mu(x)^{(+)}=0$ on an algebraic level). It is, however, stronger than the
weak operator equation $\langle F, \d^\m \A_\mu(x)^{(+)}\ G\rangle$ for all
$F,G\in\Fk$.}
\begin{equation*}
\d^\m \A_\mu(x)^{(+)} \ \Vac=0
\end{equation*}
with our choice of the polarization vectors means
\begin{equation*}
    (a_0-a_3)\ F=0\qquad \forall F\in\H_p.
\end{equation*}
A way to see the structure of $\H_p$ is to introduce a new set of
creation/annihilation operators:
\begin{align*}
    b&=\frac{1}{\sqrt{2}}(a_3-a_0),\\
    c&=\frac{1}{\sqrt{2}}(a_3+a_0),
\end{align*}
which fulfill\footnote{Here we omit the arguments of the
creation/annihilation operators as we only intend to illustrate the
structure which appears. }
\begin{equation*}
    [b,\h b]=0, \qquad [c,\h c]=0,
\end{equation*}
\begin{equation*}
[b,\h c]=1.
\end{equation*}
The Lorentz gauge condition is a strong operator equation $b F=0$ for all
$F\in\H_p$. Clearly the states from $\H_p$ may still contain longitudinal and
scalar excitations in coherent pairs as
\begin{equation*}
    b \ (\h b)^n\ \Vac=0.
\end{equation*}
On the other hand no excitations created by $\h c$ can be present, due to
\begin{equation*}
    b\ (\h c)^m\  \Vac\neq 0.
\end{equation*}
Therefore the physical Hilbert space $H_p$ is a Fock space generated by $\h
a_1$, $\h a_2$, $\h b$. There are still zero norm vectors in $\H_p$. In fact
all vectors which contain scalar-longitudinal excitations have zero norm,
which follows from
\begin{equation*}
    \langle{\h b}^n\ \Vac,{\h b}^n \Vac \rangle=\langle\Vac,b^n{\h b}^n\
    \Vac\rangle=0.
\end{equation*}
Those states can be factored out, which means that we may assume them to be
equal to the zero vector. In that way we obtain the \index{Hilbert
space!transversal} transversal Hilbert space $H_{tr}$ in which only the
physical, transversal photon polarizations are present. The Krein product,
when restricted to $H_{tr}$, is positive-definite. Henceforth, we may work in
a standard Fock-Hilbert space of transversal photons. However, one should not
forget the $H_{tr}$ is a factor space: a Lorentz transformation may lead to
the appearance of zero-norm components which, due to factorization, are
equivalent to the zero vector.

The other important reason not to forget the non-physical polarizations is
the following: the commutation relations \eqref{komutacje_radiation} lead
to the standard commutation relations for the potentials:
\begin{equation}\label{commutation_AA}
    [\A_\mu(x),\A_\nu(y)]=-\eta_{\m\n} \Delta(x,y),
\end{equation}
only if we do not omit $e_0$ and $e_3$ in the expressions for the quantized
potential; here $\Delta(x,y)$ is the massless Pauli-Jordan distribution. The
same happens with the two-point function,
\begin{equation}\label{two_point_AA}
    (\Vac,\A_\mu(x)\A_\nu(y)\ \Vac)=-\eta_{\m\n} \Delta_+(x,y),
\end{equation}
which contains not only the transversal part:
\begin{align*}
\Delta_T^{ij}(x,y)&=\frac{1}{(2\pi)^3}\int\frac{d^3p}{2p}\
e^{-ip(x-y)}\left(\de^{ij}-\frac{\v p^i\v p^j}{|\v p|^2}\right),\\
\Delta_T^{\m 0}(x,y)&=0,
\end{align*}
but the full scalar two-point function:
\begin{equation}\label{eq:omega_0}
\Delta_+(x,y)=\frac{1}{(2\pi)^3} \int\frac{d^3p}{2p}\ e^{ip(x-y)}.
\end{equation}
On the other hand, if the expectation values of gauge-invariant quantities
are investigated, then the non-physical polarizations make no contribution.
For instance,
\begin{equation*}
    \langle E^i(x) E^j(y)\rangle_\Vac=\d_x^0\d_y^0\ \langle
    \A^i(x)\A^j(y)\rangle_\Vac+\d_x^i\d_y^j\ \langle
    \A^0(x)\A^0(y)\rangle_\Vac=\d_x^0\d_y^0 \Delta_T^{ij}(x,y),
\end{equation*}
even if we make use of \eqref{two_point_AA}.

\chapter{Parametrices of the Dirac equation on external field backgrounds}

\label{chapter:Local_Solutions}

This chapter contains an exposition of a general method of finding
parametrices of partial differential operators\footnote{Many formulas will
be present in this chapter, some of which are easy to derive. Because of
that we have decided to emphasize some of the more important equations with
a box in order to distinguish them from what may serve as a reminder of
their derivation.}. In particular, we shall consider the scalar Klein-Gordon
operator and the Dirac operator.

Parametrices are distributions which, when acted upon by the wave operator,
vanish or produce a smooth function. They are useful tools in the
construction of fundamental solutions\footnote{Such as retarded/advanced
solutions of the d'Alembert equation.} of partial differential operators. It
is therefore natural that much of this chapter is based upon the standard
textbooks on partial differential equations
\cite{hilbert_courant2,friedlander}. It is the aim of this chapter to
investigate the structure of singular solutions of the Dirac equation. The
two-point functions of the quantum Dirac field are examples of such
solutions. Being distributions in two variables $x$ and $z$, they are always
singular at the coincidence point $x=z$. We shall present a way to deduce
the singular part of such solutions which will be local, that is, it will
depend only on the external field backgrounds in a neighborhood containing
$x$ and $z$.

\begin{figure}[h]\centering
\includegraphics{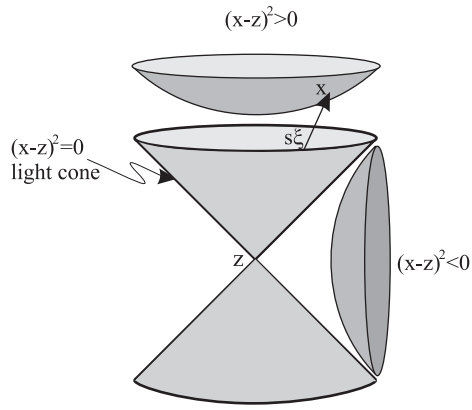}
\caption{Geometrical setting. The point $z$ is fixed, and there the initial
condition is specified. The parametrix fulfills the wave equation in the
variable $x$.} \label{obraz_geometryczny}
\end{figure}

Before we begin, let us introduce the geometrical setting (figure
\ref{obraz_geometryczny}). The point $z$ is fixed, it is the origin of the
construction of the parametrix. The point $x$ varies, and the parametrix
fulfills the wave equation with respect to this point by construction. There
is a unique geodesic curve which joins $x$ and $z$, which is just the
straight line. The tangent vector of this line is denoted by $\xi$. This
vector is normalized to $\pm 1$, if $x$ and $z$ are not null-related.
Moreover, the (positive-valued) geodesic distance of $x$ and $z$ is denoted
by $s$. The square of the Lorentzian geodesic distance is denoted by
\begin{equation*}
  \G=(x-z)^2=(x_0-z_0)^2 -(\v x-\v z)^2.
\end{equation*}
 We have
\begin{equation*}
  x^a=z^a+s \xi^a.
\end{equation*}

\section{Scalar field case}

Before we concentrate on the Dirac field, let us look for the fundamental
solutions of the scalar wave operator
\begin{equation}\label{wave_equation}
\left[\Box^x + B^a(x)\d^x_a +c(x) \right] \vp(x,z)=\OpD_x \ \vp(x,z)=0.
\end{equation}
If the functions  $B^a(x),\ c(x)$ exhibit little or no symmetry, then it is
difficult to find the fundamental solutions explicitly. We shall present
here a method of Hadamard which allows us to find approximate fundamental
solutions of the wave equations. They are approximate in the sense that
\begin{itemize}
  \item they differ from the exact fundamental solutions by a residual
  term which is differentiable up to an arbitrarily high order,
  \item the differential operator $\OpD_x$ acting on the residual term
  of order $N$ gives a distribution which vanishes as $\G^N$ with $\G\ra 0$.
\end{itemize}

\subsection{Progressing wave expansion}
In order to find approximate fundamental solutions we use the method called
'the progressing wave expansion'\footnote{See \cite{friedlander} section
3.6.; \cite{hilbert_courant2}.}. The solutions of order $N$ will be
parameterized as follows:
\begin{equation*}
    \vp(x)=\sum_{n=0}^{N} U_n(x) \* S_n[\chi(x)],
\end{equation*}
where $U_n(x)$ are $\C^\infty$ \emph{amplitudes}, $S_n$ are distributions of
one variable called \emph{the wave forms}, and $\chi$ is a smooth function,
\emph{the phase}. Any discontinuity of $\vp$ comes from the distributions
$S_n$, and thus surfaces of constant phase are possible surfaces of
discontinuity.

It is clear from the outset that the progressing wave expansion introduces a
considerable redundancy. Not only it is possible to describe different types
of singular solutions of the wave equations (with different surfaces of
singularity), but also the same solutions may be expressed in different
ways\footnote{We may, for instance, change the distributions $S_n$ away from
their singular support and absorb this change into a redefinition of the
$U_n$'s.}.

We use this redundancy in order to specify the concrete situation (we look
for a singularity on the light cone $\G=0$) and to simplify the search for
the amplitudes.  In the construction we fix from the beginning the wave
forms $S_n$ and the phase function and ask, whether it is possible to choose
the amplitudes so that the distribution $\vp$ fulfills the wave equation.

The wave operator $\OpD$ has the same principal symbol as $\Box_x$;
because of that it will turn out advantageous to require the wave forms to
fulfill
\begin{equation*}
   \frac{d S_n}{d\G}=S_{n-1}.
\end{equation*}
Moreover, if we choose the phase function to be equal to the square geodesic
distance\footnote{This is only possible after a regularization of $\G$, see
the next section.} between $x$ and some fixed basis point $z$, an additional
equation between the wave forms must be fulfilled:
\begin{equation}\label{eq:fundamental_cond}
\G S_{-2}(\G)+2S_{-1}(\G)=0,
\end{equation}
for otherwise no choice of amplitudes can lead to a solution of the wave
equation.

If we choose a set of wave forms $S_n$ which fulfill both equations and if
we find for this set the amplitudes $U_n(x)$ (which we will do in the
subsequent sections), then the problem of finding $\vp$ is solved by means
of a series (of progressing waves).

The equation \eqref{eq:fundamental_cond} prescribes the degree of homogeneity
of $S_{-1}$. Formally, the following distributions fulfill this requirement:
\begin{align*}
    S_{0}&=\frac{1}{\G} &\hspace{-2cm} \t{or} \qquad S_{0}&=\de(\G).
\end{align*}
Irrespective of this choice and also of the choice of the regularization of
$\G$, the method of progressing waves will yield \emph{the same} amplitude
functions. For instance, the retarded and advanced solutions as well as the
Feynman parametrix have the same progressing wave amplitudes and differ only
by the wave forms.

From now on we shall concentrate on
\begin{align*}
S_{0}(\G)&=1/\G,\\ S_{1}(\G)&=\ln \G.
\end{align*}
In this case the progressing wave expansion coincides with the ansatz of
\cite{DeWittBrehme} which will be later recognized as the universal form of
the singularity of the two-point function of the quantum Dirac field (the
Hadamard form):
\begin{equation}\label{ansatz}
  \vp(x,z)=\frac{u(x,z)}{\G}+v(x,z)\ln(\G) + w(x,z),
\end{equation}
where $v$ and $w$ are sums of higher order wave forms and their
coefficients:
\begin{align*}
v&=v_0(x,z)+v_1(x,z) \G +v_2(x,z) \G^2+\ldots,\\
w&=w_0(x,z)+w_1(x,z) \G +w_2(x,z) \G^2+\ldots.
\end{align*}
The amplitude functions $u,v_n,w_n$ will be constructed locally (in the
smallest causal normal neighborhood containing $x$ and $z$) and will fulfill
a set of (first-order differential) transport equations.

\subsection{Regularization of the phase function}
In the method of progressing waves the wave forms were distributions of one
variable taking as an argument the value of the phase function $\chi(x)$.
Such a composition should be a distribution on the space of spacetime-valued
test functions. Its action on $g\in \dom(\R^4)$ can be obtained from
\begin{equation*}
  \langle S[\chi],g\rangle=\int d\tau\ S(\tau) \de[\tau-\chi(x)]\ g(x)\ d^4
  x.
\end{equation*}
The operation is justified, only if the integration over $d^4x$ yields a
smooth function of compact support in $\tau$. This point is in general
non-trivial. A detailed investigation\footnote{Notably the theorem 2.9.1.}
reported in \cite{friedlander} section 2.9 assures that, as long as $\chi(x)$
is smooth and the gradient of $\chi$ is non-vanishing, the function
\begin{equation*}
    f(\tau)=\int \de[\chi(x)-\tau] \phi(x) \ d^4x
\end{equation*}
is indeed a $\dom(\R)$ function for all test functions $\phi\in\dom(\R^4)$
so that the composite distributions $S=S[\chi(x)]$ are well-defined. The
phase function $\chi(x)=\G$, however,  possesses a vanishing gradient at the
origin\footnote{A similar difficulty arises in the attempts to define
distributions of the radial coordinate $r=|\v x|$ in $\R^3$. If $x=0$
belongs to the singular support of (the one dimensional distribution) $S(x)$,
then $S$ cannot be extended to a composite distribution of the radial
coordinate $S[r(\v x)]$. Indeed, expressions like $\de(r)$ have an obscure
meaning.} $x=z$ (for $\G=0$), and so it is not directly suitable for the
method of progressing waves, although we know from simple examples that the
fundamental solutions of the wave equations are precisely singular solutions
whose singularities lie on the light cone $\G=0$. One finds them with the
help of the Fourier transform. There the transformed fundamental solution
(eg. of the d'Alembert operator) $\hat H (p)$ fulfills
\begin{equation*}
  \OpD(p) \hat H(p)=e^{ipy}.
\end{equation*}
The inverse of $\OpD(p)$ has a nontrivial manifold of zeros which makes
$H(p)$ not locally integrable. Such expressions have to be regularized.
There are different possibilities of defining the inverse, and they lead to
different distributions; as we know, there are different fundamental
solutions of the d'Alembert equation.

In the case at hand we regularize the phase function $\chi=\G$ by means of a
weak limit $i\e\rightarrow 0$. Specifically, we define
\begin{align*}
  \G_\e &=(x_0-z_0+i\e)^2-(\v x-\v z)^2, \\
  \G_F &=(x_0-z_0)^2-(\v x-\v z)^2+i\e.
\end{align*}
Both the above regularizations of $\chi(x,z)=\G$ and the distributions of
them coincide, if the test functions with which they are integrated are not
supported at the origin.

As is the case with momentum space regularizations of $1/p^2$ which have
very different properties in the coordinate space (eg. retarded vs. advanced
solution), different regularizations of $\G$ have different properties in
momentum space. Regarded as distributions of both arguments, $x$ and $z$,
they have different wave front sets.

The regularization of the distribution at the origin does not alter its
homogeneity and therefore all regularizations of $\de(\G)$ and $1/\G$
fulfill \eqref{eq:fundamental_cond}. All their progressing wave amplitudes
are therefore the same.

\subsection{Construction of the parametrix; the transport equations}\quad

Consider the wave equation $\OpD_x \vp(x,z)=0$ together with the
particular progressing wave expansion (taken with some regularization of
$\G$):\index{Parametrix!scalar field}
\begin{equation}\label{parametrix}
  \vp(x,z)=\frac{u(x,z)}{\G}+v(x,z) \ln \G +w(x,z),
\end{equation}
where the smooth functions $v$, $w$ are given by the power series:
\begin{align*}
    v(x,z)&=\sum_{n=0}^\infty v_n(x,z)\ \G^n, \\
    w(x,z)&=\sum_{n=1}^\infty w_n(x,z)\ \G^n.
\end{align*}
The method of progressing waves is a way to find fundamental solutions
which are singular. There always exists a possibility of adding a smooth
solution of the wave equation to $\vp$. This freedom is expressed here as
a freedom to choose an arbitrary, smooth amplitude $w_0(x,z)$ which (as
we shall see in the moment) influences all the $w_n(x,z)$, $n>0$.

With
\begin{align*}
  \d^x_a \G & = 2 (x-z)_a=s \xi_a, \\
  \d^a\G\d_a\G&=4\G,\\
  \Box^x \G&=8,
\end{align*}
we find
\begin{multline*}
  \d_a \vp= -\frac{u}{\G^2}\ \d_a \G + \frac{1}{\G} \ \d_a u + \d_a\G \sum^\infty_0 v_n \G^{n-1} +
  \ln \G \left(\d_a\G\sum^\infty_1 n \ v_n \G^{n-1}  + \sum^\infty_0 \d_av_n \ \G^n\right)\\
  +\sum^\infty_1\left(\d_a w_n \ \G^n + \d_a \G \ n\ w_n  \G^{n-1}  \right),
\end{multline*}
where all the differentiations are executed with respect to $x$, the variable
$z$ being merely a parameter. We also find
\begin{multline*}
  \Box \vp=\frac{1}{\G^2}\ (-2\ \d_a\G\d^au )+ \Box u\frac{1}{\G} + 8 v_0 \frac{1}{\G} +
  8 \sum^\infty_1 v_n\G^{n-1} +  \d_a\G \d^a v_0 \frac{1}{\G}+ \d_a \G \sum^\infty_1 \d^av_n
  \G^{n-1}+\\
  -4  v_0 \frac{1}{\G} + 4\sum^\infty_1 (n-1) \ v_n \G^{n-1} +  \d^a\G\d_a v_0 \frac{1}{\G}
  +\d^a\G\sum^\infty_1 \d_av_n\ \G^{n-1}  +4\sum^\infty_1n\ v_n\G^{n-1} +\\+
  \ln \G\left[8\sum^\infty_1n\ v_n \G^{n-1}+\d_a\G \sum^\infty_1 n\ \d^av_n\ \G^{n-1}+4\sum^\infty_1 n(n-1)\ v_n\G^{n-1}+
  \sum^\infty_0 \Box v_n \ \G^n+\right.\\ +\left.\d^a\G\sum^\infty_1n\ \d_av_n\ \G^{n-1}\right]+
  \sum^\infty_1 \left[\Box w_n \ \G^n+ 2n\ \d^a\G\d_aw_n\ \G^{n-1}+4n(n-1)\ w_n\G^{n-1}+8n\
  w_n\G^{n-1}\right].
\end{multline*}
Ordering the above expression so that appropriate progressing waves appear
together, we find:
\begin{multline*}
  \Box \vp=\frac{1}{\G^2} (-2 \d^a\G\d_a u) + \frac{1}{\G}\left(\Box u
  +4v_0+2\d^a\G\d_av_0\right)+\\
  +\sum^\infty_0\ln \G \ \G^n
\left[2(n+1)\ \d^a\G\d_av_{n+1}+4(n+1)(n+2)\ v_{n+1}+\Box v_n\right]+\\
+\sum^\infty_0\G^n \left[2\ \d^a\G\d_av_{n+1}+4(2n+3)\ v_{n+1}+2(n+1)\
\d^a\G\d_aw_{n+1}+ 4(n+1)(n+2)\ w_{n+1} +\Box w_n\right],
\end{multline*}
(we have set $w_0=0$). Finally, by setting all the coefficients in front of
the factors $\G^n$ and $\G^n \ln \G$ to zero, we obtain the following system
of differential equations of first order for $u, v_0$ and $v_n, w_n$:
\begin{subequations}
\begin{center}
\framebox{\parbox{6.0in}{
\begin{align}
  2\ \d^a\G\d_a u+uB^a\d_a\G &=0, \\
  2\ \d^a\G\d_av_0+v_0(4+B^a\d_a\G) &=-\OpD_x u \label{rown_v0},\\
  2\ \d^a\G\d_av_n+v_n\left[4(n+1)+B^a\d_a\G\right]&=
  -\frac{1}{n}\OpD_x v_{n-1}\label{rown_vn},
\end{align}
}}
\end{center}
and
\begin{center}
\framebox{\parbox{6.0in}{
\begin{multline}\label{rown_wn}
  2\ \d^a\G\d_aw_n+w_n[4(n+1)+B^a\d_a\G]=\\-\frac{1}{n}
  \left\{v_n[4(2n+1)+B^a\d_a\G]+2\d^a\G\d_av_n+\OpD_x w_{n-1}\right\}.
\end{multline}
}}
\end{center}
\end{subequations}
\vspace{1em}

This is a system of partial differential equations. If we recognize that the
contractions of the partial derivatives  on the left-hand side can be written
as derivatives along the geodesic, $\tfrac{d}{ds}$,
\[\d^a\G\d_a f=2s\xi^a\d_a f=2s\frac{df}{ds},\]
we realize that we are dealing with a recursive system of ordinary
differential equations. They will be solved in the following order: first
$u$, then $v_n$ with growing $n$, and finally $w_n$.  At each step we have an
ordinary differential equation to solve, yet the whole system is a system of
partial differential equations, because the knowledge of all the previous
amplitudes in \emph{the neighborhood} of the geodesic (there are normal
derivatives in $\OpD$) is necessary in order to establish the equation for
the next amplitude.

 Apart from the ambiguity of choosing $w_0$, the system has a
unique smooth solution with initial condition $u(z,z)=1$. Setting the
function $u$ equal to $1$ on the diagonal $x=z$ and demanding the smoothness
of $v_0,v_n,w_n$ determines these functions uniquely. The proof of the above
statement will be apparent from the construction which follows. The role of
smoothness as an initial condition is the following: the derivatives with
respect to $s$ are all multiplied by $s$ and therefore have to vanish at
$s=0$, if we want the solutions to be smooth. The transport equations
themselves taken at $s=0$ give us the initial conditions for $v_0, v_n$ and
$w_n$ in a recursive manner.

\subsection{Solution of the transport equations}\quad\\

The system of differential equations from the previous section will now be
solved constructively.

\vspace{1em}
{\bf Equation for $u(z,x)$:}\\
This equation allows us to find a smooth coefficient $u(z,x)$,  i.e. the
coefficient which appears in front of the most singular term in the
parametrix \eqref{parametrix} with appropriate initial condition $u(z,z)=1$.
Due to $\d_a\G=2s\xi_a$, we find
\begin{equation*}
  u=\exp\left[-\frac{1}{2}\int_0^sB^a\xi_a\ d\tau\right],
\end{equation*}
where only $B^a$ is $\tau$-dependent. It is important to realize that the
above solution is smooth in the limit $s\rightarrow 0$.

\vspace{1em}
{\bf Equation for $v_0(z,x)$:}\\
The equation \eqref{rown_v0} for $v_0$ may be integrated as soon as the
function  $u(z,x)$ is already known. We find
\begin{equation*}
  v_0=-\frac{u}{4s}\int_0^s \frac{\OpD_x u}{u}[\tau] \ d\tau
\end{equation*}
which is a smooth function if $u(x,z)$ is smooth.

\vspace{1em}
{\bf Equations for $v_n(z,x)$:}\\
In order to find smooth solutions of  \eqref{rown_vn} one proceeds
recursively. Here we shall present the typical method of finding such
solutions in a compact form. Noting that
\begin{equation*}
  B^a\d_a\G =-\frac{1}{u}2\ \d^a\G\d_a u=-4s\ \frac{1}{u}\frac{d u}{ds},
\end{equation*}
it is possible to transform \eqref{rown_vn} into
\begin{equation*}
  4s \frac{d}{ds}\left(\frac{v_n}{u}\right)+4(n+1) \frac{v_n}{u}=-\frac{\OpD_x v_{n-1}}{n \*
  u}
\end{equation*}
and then to obtain straightforwardly,
\begin{equation*}
  v_n=-\frac{u}{4 s^{n+1}}\int_0^s
  \tau^n \ \frac{\OpD_x v_{n-1}}{n \* u} \ d\tau.
\end{equation*}
The function $v_n(x,z)$ is smooth, if $v_{n-1}$ is also smooth.

\vspace{1em}
{\bf Equations for $w_n(z,x)$:}\\
The equations \eqref{rown_wn} can be solved similarly. According to
\eqref{rown_vn} the RHS of \eqref{rown_wn} can be transformed into
\begin{equation*}
  RHS_{\text{\eqref{rown_wn}}}=
  \frac{\OpD_x v_{n-1}}{n^2}-4v_n -\frac{1}{n} \OpD_x w_{n-1}.
\end{equation*}
Analogously to the $v_n$-case one finds:
\begin{equation*}
  w_n=-\frac{u}{4s^{n+1}}\int_0^s\tau^n\ \left(-\frac{\OpD_x v_{n-1}}{n^2}+
  \frac{\OpD_x w_{n-1}}{n}+4v_n\right)\ d\tau.
\end{equation*}

\begin{remark}It is perhaps valuable to note that all the coefficients,
$u(x,z),\ v_n(x,z),\ w_n(x,z)$, are functionals of the external field.
Moreover, what counts is the external field in an infinitesimal neighborhood
of the geodesic (straight line) which connects\footnote{Not only the
external field \emph{on the geodesic}, because there are partial derivatives
of $A_a(y)$ in all directions involved.} $x$ and $z$. Due to the powers of
$s$ in the denominator, however, those functionals do not vanish in the
limit $s\ra 0$; they rather have a form of certain averages of the external
field on the line $x-z$.
\end{remark}

\section{Dirac field}

Now we shall consider the case of a free Dirac field propagating in a given,
fixed electromagnetic environment. Our goal is to find the parametrix
exactly the way we did in the last section in the scalar case. By doing so,
we shall also investigate the transport equations for the expansion
coefficients in much greater detail, in part because the Dirac parametrix
will be relied on this thesis.

We consider the Dirac equation:

\begin{equation}\label{dirac_local}
i\g^a[\d_a-ieA_a(x)]H(x,z)-mH(x,z)\equiv ?D_xA^C? H_{CB}(x,z)=0,
\end{equation}
where we have emphasized the bi-spinorial character of $H(x,z)$. We define
also an auxiliary  differential operator,
\begin{equation}\label{hat_D}
  \hat D_x =i\g^a\d_a+e\g^aA_a(x)+m,
\end{equation}
and make use of a standard ansatz, namely, the parametrix of the
Dirac\index{Parametrix!Dirac field} equation \eqref{dirac_local} is given by
\begin{equation}\label{ansatz_dirac}
  H_{AB}(x,z)=   \hat{D}?\,_{xA}^C?\ \phi_{CB}(x,z),
\end{equation}
where $\phi(x,z)$ is the (bi-spinorial) parametrix fulfilling the
following second order partial differential equation:
\begin{equation*}
  \OpD \phi \equiv D \hat D \phi=[i(\dir \d -ie \dir A)-m]
  [i(\dir \d -ie \dir A)+m]\phi=0
\end{equation*}
modulo smooth terms\footnote{There will always remain a freedom of
choosing an arbitrary smooth function $w_0(x,z)$.}; here
\begin{equation}\label{kwadrat_Dirac}
\OpD=[\Box-2ie\ A^a\d_a+m^2+e^2A^2]\cdot
?\de^A_B?-\frac{ie}{2}(\g^a\g^b-\g^b\g^a)?\, ^A_B?\d_aA_b
\end{equation}
is the spinorial differential operator which acts on the $x$-variable (and
its spinorial index "$A$"). From now on the external electromagnetic field is
considered to be in the Lorentz gauge $\d A=0$. Making use of the definition
$\tfrac{1}{2}[\g^a,\g^b]=\s^{ab}$ we obtain
\begin{equation*}
\OpD=[\Box-2ie\ A^a\d_a+m^2+e^2A^2]\cdot ?\de^A_B?-\frac{e}{2}\
?\s^{ab}^A_B?\ F_{ab}.
\end{equation*}
Now we introduce certain abbreviations which will make the following
considerations much more apparent\footnote{The introduction of $B_a$ serves
only as a link to the scalar field case of the previous sections. We shall
not use it frequently, as it tends to obscure the issue of gauge
invariance.}:
\begin{equation}\label{dirac_opd}
  \OpD=[\Box+B^a(x)\d_a + c(x)]\cdot ?\de^A_B? -\frac{e}{2}?\s^abA_B?\ F_{ab}(x),
\end{equation}
where
\begin{align*}
  B^a(x) &=-2ie A^a(x), \\
  c(x) & =m^2+e^2A^a(x)A_a(x),
\end{align*}
and so the Lorentz gauge condition reads: $\d_aB^a=0$.

Clearly the operator \eqref{dirac_opd} differs from the scalar operator
considered in the previous section. Although the first part of it is
multiplied with a spinorial delta $?\de^A_B?$, i.e. effectively acts as a
scalar operator , the second part  (which does not contain any
differentiation) contains a term $?\s^ab^A_B?$ which mixes spinorial
indices. In other words, it is only because of the second term
 that we have to consider the spinorial character of the expansion
 coefficients.

The bi-spinorial\footnote{Each spinor index is attached to a different
spacetime point.} parametrix will be found with the help of the progressing
waves. The issue to be investigated throughout this section is the gauge
covariance of the system of the transport equations. They are:
\begin{center}
\framebox{\parbox{6.0in}{
\begin{align*}
    s\ \xi^a(\d_a -ieA_a )_x\ u(x,z)&=0,\\
    s\ \xi^a(\d_a  -ieA_a )_x\ v_0(x,z)+v_0(x,z)&=-\frac{1}{4} \OpD_x u(x,z),\\
    s\ \xi^a(\d_a -ieA_a )_x\ v_n(x,z)+(n+1)v_n(x,z)&=-\frac{\OpD_x
    v_{n-1}(x,z)}{4n},
\end{align*}
}}
\end{center}
and
\begin{center}
\framebox{\parbox{6.0in}{
\begin{multline*}
    s\ \xi^a(\d_a -ieA_a)_x\ w_n(x,z)
    +(n+1)w_n(x,z)=\\-\frac{1}{n}\left[s\ \xi^a(\d_a -ieA_a )_x\ v_n(x,z)
    +(2n+1)v_n(x,z)+\frac{\OpD_x}{4}\  w_{n-1}(x,z)\right].
\end{multline*}
}}
\end{center}
\sp

 By writing the transport equations in this way, we have illuminated
their covariance w.r.t. the gauge transformations, namely, if the external
field is changed by
\begin{equation*}
    A_a'=A_a+\d_a\La,
\end{equation*}
then the primed coefficients of the Hadamard expansion are related to the
initial ones by
\begin{equation*}
u'(x)=e^{ie\La(x)}u(x)
\end{equation*}
and fulfill the (primed) transport equations.

A couple of general remarks about the system of differential equations for
the parametrix coefficients are in order. Firstly, they are supplied with
only one boundary condition $u(z,z)=1$ (times the identity matrix). All the
boundary conditions for $v_n$ are to be derived recursively from the
condition that they remain finite as $s\ra 0$. This is only possible because
$s=0$ is the singular point of each of those equations. Secondly, there is an
impression that we deal with ordinary differential equations. Indeed, as the
geodesic tangent vector $\xi$ is fixed, at least the homogeneous (i.e. LHS)
part of those equations contains only derivatives w.r.t. the geodesic
distance $s$. However, the RHS, in particular the operator $\OpD_x$ contains
also derivatives in directions orthogonal to $\xi$. As a consequence the
expansion coefficients $v_n(x,z)$ depend not only on the external field on
the geodesic which joins $x$ and $z$, but also on its values in an
arbitrarily small neighborhood of this geodesic. The third important issue
is the dependence of the coefficients on $z$. Up to now $z$ was regarded as
"origin" of the expansion and played no active role. However, we may pose the
boundary value problem at a different point, say $\T z$, and ask to compare
the expansion coefficients $v(x,z)$ and $v(x, \T z)$. We shall investigate
this issue in a separate section (\ref{left_right}).

The equation for $?u^A_B?(x,z)$ reads
\begin{equation*}
  \frac{d}{ds} ?u^A_B?=ieA^a\xi_a \ ?u^A_B?
\end{equation*}
with the initial condition $?u^A_B?(z,z)=?\de^A_B?$. As the operator
multiplying $?u^A_B?$ on the RHS of the above equation is a scalar, the
solution of the above transport equation must have form of a spinorial delta
$?\de^A_B?$ multiplying a scalar, smooth function $\tilde u$, that is
\begin{equation*}
  u=?\de^A_B?\ \tilde u(z,x).
\end{equation*}
We find, as in the scalar case\index{Parametrix!coefficients}
\begin{equation}\label{rozw_u_dirac}\boxed{
  \tilde u(z,x)=\exp\left[ie (x-z)^a \int_0^1 A_a(y)\ dr\right],}
\end{equation}
where $y=z+r(x-z)$. Now the gauge covariance of this coefficient may be fully
understood: although we see that the potential along the whole geodesic
seems to count for $\T u$, the gauge term $\d_a \La$ influences the value of
$\T u$ only at the boundary points, that is only the $\La(x)$ and $\La(z)$
are important. To see this we recall that
\begin{equation*}
    \xi^a\d_a F(x)=\frac{d}{dr} F(x),
\end{equation*}
for any function $F(x)$, where $r$ denotes the distance along the geodesic.
We have
\begin{equation*}
    \T u'(x,z)=\T u(x,z)\ \exp\left[ie\int_0^s \frac{d}{dr}\La(y)\
    dr\right],
\end{equation*}
and so
\begin{equation*}
\T u'(x,z)=e^{ie\La(x)} \T u(x,z) e^{-ie\La(z)},
\end{equation*}
which is the covariant transformation of a bi-scalar.

 The properties of the scalar distribution $\T u$ are all
that is needed to establish an equation for  $?v_0^A_B?$:
\begin{equation*}
  2\ \d^a\G\d_av_0+v_0(4+B^a\d_a\G) =-\OpD_x u. \\
\end{equation*}
Using the properties of $\T u$ we find
\begin{equation*}
  2\d^a\G\left[\d_a?v_0^A_B?-?v_0^A_B?\frac{1}{\T u} \d_a\T u\right]+4?v_0^A_B?=-\OpD_x ?u^A_B?
\end{equation*}
and thus
\begin{equation}\label{rozw_v0_dirac}\boxed{
  ?v_0^A_B?(x,z)=-\frac{\tilde u(x,z)}{4s}
  \int_0^s \frac{\OpD_y u(y,z)^A_B}{\tilde u(y,z)}\
  d\tau,}
\end{equation}
where $y(\tau)=z+\xi \tau$. We may also write
\begin{equation*}
  ?v_0^A_B?(x,z)=-\frac{\tilde u(x,z)}{4}\int_0^1 \frac{\OpD_y u(y,z)^A_B}{\tilde u(y,z)}\
  d\tau,
\end{equation*}
where $y=z+(x-z)\tau$. All the remaining $v's$ can be determined
recursively:
\begin{equation}\label{rozw_vn}\boxed{
  ?v_n^A_B?=-\frac{\T u(x,z)}{4 }\int_0^1 \tau^n \
  \frac{\OpD_y ?v_{n-1}^A_B?(y,z)}{n\ \T u(y,z)} \ d\tau.}
\end{equation}

 Now that the formulae for the coefficients of the progressing
wave expansion are explicitly written, we may argue, that the following
theorem holds true:

\begin{Thm}
If the external field is smooth then all the amplitudes of the progressing
wave expansion are smooth functions for all their arguments.
\end{Thm}

Before proving the theorem let us first denote
\begin{equation*}
    \chi(x,z)=-\frac{1}{2}\xi_a\int_0^s B^a(z+\tau \xi) \ d\tau,
\end{equation*}
so that
\begin{equation*}
\OpD \ \T u=\left[(\Box +B^a\d_a)\chi+(\d
\chi)^2+c-\frac{e}{2}\s^{ab}F_{ab}\right]\T u,
\end{equation*}
and prove:

\begin{lemma} If the external field $A_\mu(x)$ is real and smooth then the
function $\chi(x,z)$ is also smooth.
\end{lemma}
\begin{proof}The function $\chi$ can be rewritten in the form
\begin{equation*}
    \chi(x,z)=-\frac{1}{2}(x^a-z^a)\int_0^1 B_a[z+r(x-z)] \ dr,
\end{equation*}
from which the conclusion is evident, as $B_a$ is proportional to the smooth
$A_a$.
\end{proof}

\section{Explicit form of the singularity of the Dirac parametrix}

The parametrix of the Dirac equation \eqref{ansatz_dirac} is
given by $\phi(x,z)$ which is acted upon by $\hat D_x$:
\begin{equation*}
  H_{AB}(x,z)=(e\dir A+m)\phi(x,z)+i\dir \d\, \phi(x,z).
\end{equation*}
Suppose that some quantity, non-linear in the Dirac field, must be
regularized in order to make it an operator-valued distribution. If the
quantity of interest does not involve differentiations of the field
operators, then in the point-splitting limit only finitely many terms of the
parametrix will be important (because e.g. $\G \ln \G \ra 0$). In what
follows we shall find all those terms explicitly for a good purpose, namely,
in the rich literature on QED in external fields there have been many
attempts to regularize such nonlinear quantities; all of them use some
subtractions. The singular part of the parametrix gives the \emph{universal}
singularity structure of all such subtractions. In other words: what does
not match the singularity behavior which will be presented below certainly
does not lead to operator valued-distributions (at least if we consider the
representation of the CAR in some Hadamard state).

If we let $\hat D$ act on the scalar field parametrix
\begin{equation*}
  \phi=\frac{u}{\G}+v_0 \ln \G+\sum^\infty_1 v_n\G^n \ln \G +\sum^\infty_0 w_0\G^n,
\end{equation*}
with the abbreviation $\a=i\dir\d\,\G=i\g^a\ 2(x_a-z_a)$, we will find
\begin{multline}\label{Dirac_parametrix}
H_{AB}(x,z)=-\frac{\a u}{\G^2}+\frac{\hat D u+\a
v_0}{\G}+\sum^\infty_0\left[\hat D v_n+\a(n+1)v_{n+1}\right]\G^n\ln
\G+\\+\sum^\infty_0 \left[\hat D w_n+\a(n+1)w_{n+1}+\a v_{n+1}\right]\G^n.
\end{multline}
We will characterize explicitly the part of the Dirac parametrix which is
singular in the coincidence limit. Evidently,
\begin{equation}\label{minimal_singularity}
\boxed{H^{sing}_{AB}(x,z)=-\frac{\a u}{\G^2}+\frac{\hat D u+\a
v_0}{\G}+\left[\hat D v_0+\a v_{1}\right]\ln \G.}
\end{equation}
In what follows we will deal with the coefficient of the $\ln \G$-term,
which is the least singular one. Its computation, however, requires the
knowledge of $v_1$ which is the third coefficient in the recursive hierarchy
of the progressing wave expansion and thus, by calculating it, we will
simultaneously illustrate the practical side of the progressing wave
expansion. The appropriate expansions of the other coefficients can also be
obtained, but since their derivation is less illuminating we shall not
present them here.

 We recall that
\begin{align*}
    u(x,z)&=\exp[ie \ \chi(x,z)], & \chi&=(x-z)^a\int_0^1 A_a(y)\ d\tau,
\end{align*}
where as usual $y=z+\tau(x-z)$ is the point which lies on the geodesic from
$z$ to $x$ at the distance $\tau$ away from $z$. The action of $\hat D$,
\[\hat D_x=i\g^a\d_a+e\g^aA_a(x)+m\], leads to

\[
  \boxed{  f_a\equiv \d_a \chi=\int_0^1A_a(y) \ d\tau +
    (x-z)^b\int_0^1\d^y_aA_b(y)\ \tau d\tau},
\]
\begin{align*}
    \d_a u&=ie\ f_a,\\
    \hat D u&=[m+e(\dir A-\dir f)],
\end{align*}
where from now on we will keep all the partial differentiations under the
integral to be with respect to $y$. The formula for $v_0(x,z)$,
\eqref{rozw_v0_dirac},
\begin{equation*}
    v_0=-\frac{u}{4}\int_0^1\frac{\OpD_y u}{u}\ d\tau,
\end{equation*}
requires the knowledge of $\OpD
u=[\Box-2ieA\*\d+m^2+e^2A^2-\frac{e}{2}\s\*F]u$. With the help of the
Lorentz gauge condition $\d A=0$ we find
\begin{equation*}
    \Box u=\left\{ie\left[(x-z)^a\int_0^1 \Box A_a\ \tau^2 d\tau\right]
    -e^2f_af^a\right\}u,
\end{equation*}
\begin{align*}
    \OpD_y u(y,z)&=\left\{ie\left[(y-z)^a\int_0^1 \Box A_a\ \tau^2 d\tau\right]
    -e^2f_af^a\right\} u(y,z)+\\&+(m^2+e^2A^2-\frac{e}{2}\s\* F)\ u(y,z)+2e^2A^af_a
    \ u(y,z).
\end{align*}
The above expression divided by $u(x,z)$, that is without $u$ at the end of
it, will be called $\psi(y,z)$ for brevity\footnote{Again we stress that we
do not divide by a bi-spinor $?u^A_B?=?\de^A_B?\ \T u$ but rather by a
scalar $\T u$.}:
\begin{equation*}\boxed{
\psi(y,z)= e^2 \left\{2A^af_a+A^aA_a
-f_af^a\right\}+ie\left[(y-z)^a\int_0^1 \Box A_a\ \tau^2
d\tau\right]+(m^2-\frac{e}{2}\s\* F).}
\end{equation*}
With that abbreviation we have
\begin{equation}\label{nice_v0}
\boxed{ v_0(y,z)=-\frac{u(y,z)}{4}\int_0^1 \psi(w,z)\ ds}
\end{equation}
with $w=z+s(y-z)$.

The difficult task which lies ahead is to determine the least singular term
of the parametrix, the one which comes with the $\ln \G$. What needs to be
computed is  $\hat D v_0(x,z)$ and even $v_1(x,z)$. The task will be
significantly simplified, as we are only interested in the point-splitting
limit $x\ra z$. If we expand the coefficient of $\ln \G$ into a Taylor series
in $(x-z)$, then only the zeroth term, i.e. $\left.\left[v_1(z,z)+\hat D
v_0(x,z)\right]\right|_{x=z}$, is important in the point-splitting
limit\footnote{Appropriately, the first five terms of the Taylor expansion of
the coefficient of $1/\G^2$, \eqref{minimal_singularity}, are important in
the point-splitting limit and also the first three terms of the coefficient
of $1/\G$.}. In other words, if two different smooth functions $v_1(x,z)$ and
$\T v_1(x,z)$ are investigated, then the point-splitting limit vanishes,
\begin{equation*}
    \lim_{x\ra z}\ln\G\ [v_1(x,z)-\T v_1(x,z)]=0,
\end{equation*}
as long as $v_1(z,z)=\T v_1(z,z)$ irrespective of the direction of the limit
(space- ,time- or light-like).

Before we proceed further, it is appropriate to note that
\begin{equation*}
\d^a f_c=\int_0^1 (\d^aA_c+\d_cA^a)\ \tau d\tau+(x-z)^b\int_0^1 \d_c \d^a
A_b \ \tau^2 d\tau,
\end{equation*}
\begin{equation*}
\Box f_c=\int_0^1 \Box A_c \ \tau^2d\tau+(x-z)^b\int_0^1\d_c\Box A_b\
\tau^3d\tau.
\end{equation*}

The calculation of $v_1(z,z)+\hat D v_0(x,z)|_{x=z}$ will proceed much
easier, if we first give the coincidence limit of various expressions:
\begin{align*}
\lim f_a&=A_a,   &   \lim \d_af_b=\tfrac{1}{2}(\d_aA_b+\d_bA_a),\\
\lim \Box f_a&=\tfrac{1}{3}\Box A_a, &&
\end{align*}
\begin{align*}
    \lim \psi&=e^2A^aA_a+m^2-\tfrac{e}{2}\s\* F,\\
    \lim \d^b\psi&=4e^2 A_a\d^bA^a+\tfrac{ie}{3}\ \Box A_b-\tfrac{e}{2}\ \s^{cd}\d^b
    F_{cd},\\
    \lim \Box \psi&=e^2\left(3\ \d_aA_b\ \d^aA^b+\d_aA_b\ \d^bA^a+4A^b\Box
    A_b\right)-\tfrac{e}{2}\s \Box F,
\end{align*}
where all quantities on the right-hand side of the equations are taken at
the point $z$.

We can now evaluate the coincidence limit of  $\hat D v_0(x,z)$ (cf.
\eqref{nice_v0}):
\[
\lim_{x\ra z} -4\ \hat D v_0(x,z)=(m+e\dir A)\psi+i\g^a[(ie)
A_a\psi+\tfrac{1}{2}\d_a\psi]=m\psi+\tfrac{i}{2}\dir \d \psi,
\]
and so
\[\boxed{
\hat D v_0(x,z)=-\frac{1}{4}\left\{m\psi(z)-\frac{i}{2}\g^a\ \d_a
\psi(z)\right\}}
\]
As far as $v_1$ is concerned, from  \eqref{rozw_vn} we have
\begin{equation*}
    4\ v_1(x,z)=-u(x,z)\int_0^1\tau d\tau\ \frac{\OpD_y v_0(y,z)}{u(y,z)},
\end{equation*}
thus in the coincidence limit
\begin{equation*}
    32\ v_1(z,z)=\OpD_y\left[ u(y,z)\int_0^1d\tau\ \psi(w,z)\right]_{y=z}.
\end{equation*}
A calculation shows that in the coincidence limit
\begin{equation*}
    \lim_{x\ra z} (\Box-2ieA^a\d_a)\left(u(x,z)\*\int_0^1 \psi(y,z) d\tau\right)=
    e^2A^2(z)\psi(z)+\tfrac{1}{3}\Box\psi(z),
\end{equation*}
and so finally
\begin{equation*}\boxed{
     v_1(z,z)=\frac{1}{32}\left\{\frac{1}{3}\Box \psi+\left(2e^2A^2+m^2
     -\frac{e}{2}\s\*
    F\right)\psi(z)\right\}.}
\end{equation*}

\vspace{1em}
\section{Left/right parametrices of the Dirac operator}\label{left_right}
In the following we make a comment on the issue, whether what we have found,
namely, the left parametrix of the Dirac operator is also a right parametrix
of some other differential operator. As later on the parametrix we construct
will be used to regularize the Wick square of the Dirac field operators
$:\psi(x)\b\psi(z):$, this problem is of great importance. Let us recall that
in the case of the scalar Klein-Gordon operator the left parametrix is also
the right parametrix of the adjoint Klein-Gordon operator. This means that
from
\begin{equation*}
 \left[\Box^x + B^a(x)\d^x_a +c(x) \right] \vp(x,z)=\OpD_x \ \vp(x,z)=0
 \t{ modulo } \C^\infty,
\end{equation*}
it follows
\begin{equation*}
 \left[\Box^z + B^a(z)\d^z_a +c(z) \right] \vp(x,z)=\vp(x,z)\
 \overleftarrow{\OpD_z}=0
 \t{ modulo } \C^\infty;
\end{equation*}
here the Klein-Gordon operator is self-adjoint in the generalized sense
\begin{equation*}
    \langle\OpD_x \vp,f\rangle=\langle\vp,\OpD_x f\rangle.
\end{equation*}

However, the Dirac-operator parametrix is formed from the Klein-Gordon
parametrix with an action of the $\hat D_x$ on the left variable of
$\vp(x,z)$. The adjoint Dirac operator is given by
\begin{equation*}
    \overleftarrow{D^T_z}=  -i\overleftarrow{\dir \d_z}+e\dir A-m,
\end{equation*}
so that for a classical solution $\phi(x)$ of the Dirac equation there also
holds
\begin{equation*}
      \b\phi(z)\ \overleftarrow{D^T_z} =0,
\end{equation*}
where the bar denotes the Dirac conjugation. The question therefore is:

\sp \emph{Does the Dirac-operator parametrix $H(x,z)$ (cf.
\eqref{Dirac_parametrix}) fulfill the adjoint Dirac equation
\begin{equation*}
    H(x,z)\ \overleftarrow{D^T_z}=0 \t{ modulo }\C^\infty \ \t{?}
\end{equation*}
} \sp

We do not possess, at present, any answer to this important question. As a
consequence we do not know whether $H(x,z)$ should be used in section
\ref{sec:local_Wick} as a regularizing distribution of $:\psi(x)\b\psi(z):$
or $:\psi(x)\h\psi(z):$ with both quantities differing by a right
multiplication with a $\g^0$-matrix\footnote{For undifferentiated Wick
products this issue can be attacked with laborious "brute force" calculation
of the right action of $D^T_z$ on the $H(x,z)$ defined with the help of only
few terms of the Hadamard expansion.}.

\chapter{Hadamard form}\label{chapter_hadamard}
The singularity structure of the two-point function of the Dirac quantum
field on an external background has a central role in the theory under
development. On the one hand it provides a selection criterion for the
allowed class of states (the Hadamard states), on the other hand the
particular form of this singularity allows for a pointwise multiplication of
the distributions which posses it and in turn allows for the development of
the causal perturbation theory.

    There are two possible formulations of the Hadamard property, one
which uses the Hadamard series elaborated upon in chapter
\ref{chapter:Local_Solutions}, and the other which employs the notions of
microlocal analysis\footnote{See appendix
\ref{appendix:microlocal_analysis}.} and characterizes the Hadamard
distributions in terms of their wave front sets.

    The chapter which follows reflects our struggle to translate the
results obtained in the context of the quantum field theory on curved
spacetimes to the external-field case. In particular, we shall attempt to
translate the following theorems/propositions:

\begin{enumerate}
\item The definition of the Hadamard form in terms of its wave front set
and in terms of the Hadamard series are equivalent, theorem
\ref{definition_equivalence}.

\item Every two Hadamard states are locally equivalent.

\item The ground state on a static background is a Hadamard state.

\item If $\w$ is a Hadamard state, then also $\w_t=\w\circ \a_t$, i.e. the
state composed with a unitary time evolution, is also a Hadamard state.

\end{enumerate}

All the above theorems provide the crucial intuition as to which states of
the Dirac field are allowed in our investigations. Indeed this class is much
broader than anything considered before in the context of the external-field
QED.

Before we proceed further, let us note the connection between the two-point
function of a state of the CAR on a certain Cauchy surface and the two-point
function of a state of the CAR on the spacetime as a whole. The latter
algebra is the algebra of symbols $\psi(f)$, $\h\psi(f)$ and their
polynomials, where $f\in\dom(\R^4)^4$ are complex, smooth and rapidly
decaying spinor functions. The algebraic relations are
\begin{align*}
\{\psi(f),\h \psi(g)\}&=S(f,g),\\
\{\psi(f), \psi(g)\}&=0,
\end{align*}
where $S(x,y)$ is the unique anti-commutator distribution. This distribution
solves the Dirac equation in the first variable and the adjoint Dirac
equation in the second variable. If $x^0=t=y^0$, then
\begin{equation*}
    S_{AB}(t,\v x,t,\v y)=\de(\v x-\v y)\ \de_{AB}.
\end{equation*}
We have an important
\begin{lemma}\label{expression:two_point}
The two-point function $\w^+(\v x,\v y)$ of a state of the Dirac field on a
Cauchy surface (of constant time, $t_0$) extends naturally to a two-point
function at different times,  $\w^+(t,\v x,s,\v y)$, via
\begin{equation*}
\w^+(t,\v x,s,\v y)=S(t,\v x, t_0,\v z)\circ \w^+(\v z,\v w)\circ S(t_0,\v
w,s,\v y),
\end{equation*}
where the composition means an integration with respect to $\v z$ and $\v w$
supplemented by an appropriate contraction of the spinor indices.  The
restriction of $\w_+(t,\v x,s,\v y)$ to $t=t_0=s$ gives $\w^+(\v x,\v y)$.
\end{lemma}
\begin{proof}We shall establish the above equalities instead of saying that
they follows trivially from the uniqueness property of the (weak) solutions
of the Cauchy problem for the Dirac equation. Let $\psi_p(\v x)$ denote the
(positive- and negative-frequency) generalized eigenfunctions of the Hamilton
operator at time $t=t_0$. They form a complete set:
\begin{equation*}
    \int d\m_p \ \psi_p(\v x) \h \psi_p(\v y)=\de(\v x-\v y),
\end{equation*}
where $d\m_p$ denotes the spectral measure of the Hamiltonian and we have
omitted the spinor indices for brevity. If  we replace in the integral the
eigenfunctions $\psi_p(\v x)$ by their time-evolved versions,
\begin{align*}
    \psi&\ra U(t,t_0) \psi,\\
    \h \psi&\ra [U(s,t_0)\psi]^*,
\end{align*}
where $U$ denotes the unitary propagator, then what results is a
bi-distribution which solves the Dirac equation in the first variable and
the adjoint equation in the second variable:
\begin{equation*}
    \int d\m_p \ \psi_p(t,\v x) \h \psi_p(s,\v y)=S(t,\v x,s,\v y).
\end{equation*}
This is the solution of the Cauchy problem, and it is unique, because the
unitary propagator $U(t,t_0)$ is unique. We can also write
\begin{equation*}
    \psi(t,\v x)=\int d^3y\ S(t,\v x,s,\v y) \psi(s,\v y)
\end{equation*}
for all solutions $\psi$ of the Dirac equation.

The two-point function on the Cauchy surface $t=t_0$, $\w^+$ can be
expressed in terms of the generalized eigenfunctions of the Hamiltonian:
\begin{equation*}
    \w^+(\v x,\v y)=\int d\m_p\ d\m_k\ \psi_p(\v x)\ B(\v p,\v k)\ \h
    \psi_k(\v y).
\end{equation*}
A composition with $S(t,t_0)$ on the left side and $S(t_0,s)$ on the right
side with the completeness relation for the $\psi_p$'s yields the desired
result.
\end{proof}

\section{Two definitions of Hadamard states}

Let a state be defined via its two-point function
\begin{equation*}
  \w^+_{AB}(x,z)=\w(\psi_A(x)\psi^*_B(z))=iG^+_{AC}(x,z)\ ?\g^{0 C}_B?
\end{equation*}
in the notation of chapter \ref{non_linear}, understood as a bi-distribution
on the double copy of the spacetime as a whole, not just the Cauchy surface.

\begin{defi}[Hadamard series definition]\label{def:hadamard}
A quasifree state of the Dirac field defined with the help of the two-point
function $\w^+_{AB}(x,z)$ is a Hadamard state, if for each $N$
\begin{equation*}
    \w^+_{AB}(x,z)-H^N_{AB}(x,z)
\end{equation*}
is a continuous function and all its derivatives up to the order $N$ are
also continuous. Here $H^N$ denotes the Hadamard parametrix
\eqref{Dirac_parametrix} with the series cut at $N$ and with the $w_0$-term
absent.
\end{defi}

The other definition of Hadamard states is:

\begin{defi}[Microlocal definition]\label{micro_definition} A quasifree
state of the Dirac field defined with the help of the two-point function
$\w^+_{AB}(x,z)$ is a Hadamard state, if the primed wave front set of $\w^+$
is
\begin{equation*}
    WF'(\w^+)=\{(x,\xi,y,\xi): x\sim y, \xi_0\geq 0\},
\end{equation*}
where $x \sim y$ means that the points $x$ and $y$ can be joined by the
lightlike geodesic (straight line) with tangent vector $\xi$.
\end{defi}

\begin{remark}Note that in the second definition the normalization of
$\w^+$ is important as $\a \w^+$ for an arbitrary constant $\a$ has the same
wave front set. Another important issue is that $\w^+_{AB}$ is a
bi-distribution, i.e. a 16-component matrix. The wave front set of such an
object is defined as the union:
\begin{equation*}
    WF(\w^+)=\bigcup_{AB}\ WF(\w^+_{AB}).
\end{equation*}
In the light of that, it is truly remarkable that the microlocal definition
is equivalent to the definition given in terms of the Hadamard series. The
situation is similar in quantum field theory on a curved
spacetime\footnote{See \cite{FV} section 2.3.}. With the identity of
Lichnerowicz\footnote{Here $\dir\D$ denote a covariant derivative which
contains an appropriate spin connection.},
\begin{equation}\label{KG_curved}
    (i\dir \D+m)(-i\dir\D+m)=\Box -\frac{1}{4}R+m^2,
\end{equation}
we may also write
\begin{equation*}
    \w^+_{AB}(x,z)=(-i\dir\D_x+m)\vp^+(x,z),
\end{equation*}
where $\vp^+(x,z)$ is a spinorial bi-distribution, the Hadamard property of
which needs to be investigated.
\end{remark}

The equivalence of both definitions, essential for the further development
of the theory, has been proved for the scalar field on a curved spacetime by
M.Radzikowski \cite{Rad}. For spinor fields the appropriate modification has
been given by S. Hollands \cite{hollands_dirac} and independently by
K.Kratzert \cite{KK}. The generalization to other vector-valued fields has
subsequently been given by H.Sahlmann and R.Verch in \cite{SV2}, where the
authors also remove a gap present in the previously cited works. We follow
the proof of S.Hollands in the sequel.

\begin{Thm}[Equivalence of both definitions]\label{definition_equivalence}
The definition of Hadamard states in terms of their wave front sets and the
definition with the help of the Hadamard series are equivalent.
\end{Thm}
\begin{proof}(Microlocal definition $\Rightarrow$ Hadamard series)\\
In the proof we shall make use of two plausible properties:
\begin{itemize}
\item The squared (spinorial) Dirac operator $\OpD_x$, eq.
\eqref{kwadrat_Dirac}, possesses four distinguished parametrices $\De_R$,
$\De_A$, $\De_F$, $\De_{\b F}$, which are called the retarded, advanced,
Feynman and anti-Feynman distributions. They are distinguished by their wave
front sets:
\begin{align*}
    WF'(\De_A)&=\{(x,\xi,y,\xi):\ x\sim y; x\in J^+(y)\},\\
    WF'(\De_R)&=\{(x,\xi,y,\xi):\ x\sim y; x\in J^-(y)\},\\
    WF'(\De_F)&=\{(x,\xi,y,\xi):\ x\sim y;\ [\xi_0>0 \t{ if }y\in J^+(x)],[
    \xi_0<0 \t{ if }y\in J^-(x)]\},\\
    WF'(\De_{\b F})&=\{(x,\xi,y,\xi):\ x\sim y;\ [\xi_0>0 \t{ if }y\in J^-(x)],[
    \xi_0<0 \t{ if }y\in J^+(x)]\}.
\end{align*}
Those parametrices are unique in the sense that, if a bi-distribution is a
solution of the inhomogeneous equation,
\begin{equation*}
    \OpD_x \De_.(x,y)=\de(x-y)+\t{smooth function},
\end{equation*}
 and possesses one of the wave front sets named above, then it must be
equal to the respective parametrix (modulo smooth function)\footnote{It is
important that we speak of inhomogeneous solutions, otherwise, if $\OpD
\De=0$ for some singular bi-spinorial distribution $\De$, then also, for
instance $\OpD (\De\*\g^0)=0$.}.

\item Up to smooth functions there holds
\begin{equation}\label{FF_AR}
    \De_F+\De_{\b F}=\De_A+\De_R.
\end{equation}
\end{itemize}
\sp
 Both of the above properties are plausible, because they hold for the
(squared) Dirac operator \eqref{KG_curved} on a curved spacetime and the
proof of this does not depend on the particular structure of what other terms
apart from the principal symbol appear in the differential operator. The
principal symbol is, however, the same on background external fields. Note
that the first property already exhibits almost the same strength as the
theorem does. It says that in each case the singularity structure of one of
the components already fixes (up to smooth terms) the whole 16-component
matrix.

Let us denote the (also distinguished) parametrices of the Dirac operator
by $S_A$, $S_R$, $S_F$, $S_{\b F}$:
\begin{equation*}
    S_\sharp(x,y)= \hat D_{x}\ \De_\sharp(x,y)=
    [i\g^a\d_a+e\g^aA_a(x)+m]\ \De_\sharp(x,y)
\end{equation*}
for all parametrices (the subscript $\sharp$ denotes either one of $A,R,F,\b
F $).

With those properties we may now follow the proof, the idea being to show
that $\w^+$ can be expressed (up to smooth terms) by $S_A$ and $S_F$. We
define
\begin{equation}\label{def_wF}
    \w_F=i\w^++S_A
\end{equation}
and show that
\begin{equation}\label{had_prop}
\w_F-S_F=\dom \t{ function}.
\end{equation}
Clearly, then
\begin{equation}\label{thm_conclusion}
    i\w^+=S_F-S_A,
\end{equation}
and what appears on the right-hand side is distinguished up to smooth
functions and therefore possesses the universal Hadamard expansion into
powers of $\G$.

In order to show \eqref{had_prop} we investigate the wave front set of
$\w_F$. We notice that, due to
\begin{equation*}
\w^++\w^-=S=S_A-S_R,
\end{equation*}
we also have
\begin{equation}\label{def_wF2}
    \w_F=-\w^-+S_R.
\end{equation}
For $x\nin J^-(y)$ the advanced solution $S_A$ vanishes and, therefore, for
such points $WF(\w_F)$ coincides with that of $WF(\w_+)$, which is known by
definition (cf. \ref{micro_definition}). On the other hand for $x\nin J^+(y)$
the retarded solution vanishes and from \eqref{def_wF2} and definition
\ref{micro_definition} we also know the wave front set. Hence we conclude
\begin{equation*}
       WF'(\w_F)=\{(x,\xi,y,\xi):\ x\sim y;\ [\xi_0>0 \t{ if }y\in J^+(x)],[
    \xi_0<0 \t{ if }y\in J^-(x)]\},
\end{equation*}
which is the same as the wave front set of $S_F$ \footnote{The application
of the differential operator $\hat D$ to the parametrix $\De_F$ does not
enlarge the wave front set, due to the pseudolocal property
\eqref{pseudolocal_property}.}. An analogous property holds for \[\w_{\b
F}=-\w^++S_R.\] We therefore conclude that $\w_{F,\b F}$ have the same wave
front sets as $S_{F,\b F}$. From  equation \eqref{FF_AR} and
$\w_F+\w_{\b{F}}=S_R+S_{\b{A}}$ we infer
\begin{equation*}
\w_F-S_F=\w_{\b F}-S_{\b F}
\end{equation*}
modulo smooth ingredients.  Thus
\begin{equation*}
    WF(S_F)\supset WF(\w_F-S_F)=WF(\w_{\b F}-S_{\b F})\subset WF(S_{\b F}).
\end{equation*}
But the sets $WF(S_F)$ and $WF(S_{\b F})$ are disjoint. Therefore $\w_F-S_F$
as well as $\w_{\b F}-S_{\b F}$ are smooth bi-spinors which is what we
intended to show. The Hadamard series is the necessary expansion of any
two-point function $\w^+$ which fulfills the microlocal definition because up
to smooth terms it can be expressed by the unique parametrices $S_F$ and
$S_R$, the expansion of which does have precisely the Hadamard form.
\end{proof}

With the equivalence at hand we obtain an immediate corollary:

\begin{Thm} Every two Hadamard states $\w_1$, $\w_2$ are locally equivalent.
\end{Thm}
\begin{proof}With theorem \ref{local_normality} we only need to show
that
\begin{equation*}
    \|\sqrt{B_1}-\sqrt{B_2}\|_{H.S.}<\infty,
\end{equation*}
where $B_1$, $B_2$ are the positive operators corresponding to $\w_1$,
$\w_2$ restricted to a bounded open region $\C$.  With the inequality of
Powers and St\"ormer \cite{pow_stroe},
\[
\|\sqrt{B_1}-\sqrt{B_2}\|^2_{H.S.}\leq \|B_1-B_2\|_{tr},
\]
 it suffices to show that the trace norm  $\|B_1-B_2\|_{tr}$ taken on $\C$ is finite.
 This, however, follows trivially, because, by definition \ref{def:hadamard} the difference
$B_1-B_2$ is an operator with smooth integral kernel. Such operators are
trace-class.
\end{proof}

\section{Time evolution preserves the Hadamard form}
Suppose that the external field was static for some time interval and that
the state of the Dirac field in this region coincided with the ground state.
It is very plausible that such a state is a Hadamard state (see the next
section). An important question arises whether the state remains Hadamard, if
we switch on in a smooth way some time-dependent external field. The answer
to this question is affirmative. More precisely, the following theorem has
been proven \cite{SV2} in a (much more complex) context of curved spacetimes:
\begin{Thm}
Let $\w(x,y)$ be a bi-solution (modulo smooth function) for the wave operator
$\OpD$ . Moreover, assume that there is a causal normal neighborhood $N$ of
a Cauchy  surface $\Sigma_t$  so that $\w$ is of Hadamard form for the wave
operator on $N$. Then, if $N'$ is a causal normal neighborhood of any other
Cauchy surface, $\Sigma_s$,  $\w$ is also of Hadamard form for the wave
operator $\OpD$ on $N'$.
\end{Thm}

\section{Static ground states and the Hadamard form}

In what follows we shall attempt to answer the question as to whether ground
states in time-independent external fields are Hadamard.

\begin{co} \label{had_form} Suppose that
the external field is time-independent and smooth. Then the
two-point functions of the ground state of the Dirac field are of Hadamard
form.
\end{co}

\begin{remark}The only ground state is the state for which the splitting of the
one-particle Hilbert space $\H$ selects functions of positive energy, that is
the corresponding projections \eqref{spinorowe projektory} are
\begin{equation*}
    P_{\pm}=(1\pm \sgn H_0)/2.
\end{equation*}
The above operator, as a function of the self-adjoint $H_0$, may be defined
in terms of the functional calculus. Indeed, in many references there appear
expressions of the form
\begin{equation*}
  G^{+}(\v x,\v y)=\int d\m(p)\ \theta(E_p) \ \b{\psi_p(\v x)} \psi_p(\v y)
\end{equation*}
where $\psi_p(\v x)$ denote the (possibly generalized) eigendistributions of
$H_0$ parameterized by $p$ and $d\m(p)$ denotes a measure\footnote{We
suppress the spinorial indices for brevity.} in $p$, roughly
\begin{equation*}
  d\m(p)=d^3p+d\m_{s}(p),
\end{equation*}
where $d\m_s$ is the singular part corresponding to the bound states. If the
integral kernel of the two-point function is written in that way, it may be
useful in some cases, but it \emph{obscures very well} the singularity
structure of $G^+$, in particular the leading divergent terms in the limit
$\v x\ra\v y$ are difficult to obtain.
\end{remark}

\begin{proof}[justification of the conjecture \ref{had_form}]
We will present an attempt to prove this conjecture. Its statement is
plausible, because the ground states (even all passive states) on the static
external gravitational background are known to be Hadamard states \cite{SV1}.

The approach we shall present makes use of the techniques of microlocal
analysis (see appendix \ref{appendix:microlocal_analysis}). We include it
here, because we believe this may lead to a complete proof in the future. In
overview the proof may proceed as follows:
\begin{enumerate}
\item Definition of the two-point functions of the ground states in terms
of the causal anti-commutator bi-distribution $S(x,y)$ and $P_+$, lemma
\ref{expression:two_point}.

\item Calculation of the wave front set of $G^{\pm}$ with the help of
microlocal techniques.

\item Equivalence of the description of Hadamard states by the Hadamard series
or the wave front set, theorem \ref{definition_equivalence}.
\end{enumerate}
Let us begin with the first point. The two-point function is a
bi-distribution which takes as arguments the four-component test functions,
i.e.
\begin{align*}
iG^+(x,z)&=(\Vac,\psi(x)\b\psi(z)\Vac),\\
iG^-(x,z)&=(\Vac,\b\psi(x)\psi(z)\Vac)
\end{align*}
act on ${\dom}^4$ in both of their variables. They are also solutions of
the Dirac and the adjoint-Dirac equations in the appropriate variables.

Due to lemma \ref{expression:two_point}, we have
\begin{equation*}
  G^+(x,z)=S(x,y)\circ P_+(y) \circ S(y,z).
\end{equation*}
In the following we shall investigate the wave front set of this
distribution. We shall begin with the investigation of $P_+\circ S$. As the
bi-distribution $S$ is not properly supported (and neither is the Schwarz
kernel of $P_+$ as we will see), we should explain, why the composition of
$P_+$ and $S$ is well-defined.

It is not difficult to see that the above problem causes no difficulties.
For a test function $f\in{\dom}^4$ the general composition of distributions
would require $S(x,f)$ to be a test function, which is not the case, because
it does not decay rapidly in the time direction. However, $P_+(y)$ is
supposed to act on the $L^2$ wave functions on the Cauchy surface $t=y^0$.
Certainly, $S(x,f)$ is such a function (as $f$ is smooth and $S:\dom\ra
\smooth$ and even compactly supported at $x^0=y^0$, due to the causal support
of S). Therefore, the composition $P_+\circ S$ is well-defined.

A standard fact, which we shall not elaborate on, is that the unique
anti-commutator distribution $S$ has the wave front set
\begin{equation*}
  WF'(S)=C=\{(x,\xi,y,\xi):\ x\sim y\},
\end{equation*}
where $x\sim y$ means that $x$ and $y$ are light-like related with $\xi$
being the tangent vector of the geodesic (null ray) which joins them. The
wave front set of $P_+\circ S$ will be found from the
propagation-of-singularities theorem which states that
\begin{equation*}
  WF[P_+(y) \circ S(y,z)]=C\cap[\s^{-1}(0)\setminus\{0\}\times \id],
\end{equation*}
$\s$ being the symbol of $P_+$.

There are two difficulties, however: the projection operator $P_+(y)$ which
may be written as $P_+(y)=\widehat{\theta(p_0)}$ is not a \PDO\
(pseudo-differential operator), because its "symbol" is not continuous at
$p_0=0$. The second pitfall is that the theorem we would like to invoke is
present in the literature only for properly supported \PDO, and $P_+$
certainly is not properly supported\footnote{Its Schwarz kernel is
proportional to $\frac{1}{x^0-y^0+i\e}$.}. Both of the above difficulties
will be cured in what follows.

The problem of the lack of differentiability at $p_0=0$ may be cured by
means of a redefinition of $P_+$ is such a way that it is smoothed out in a
compact vicinity of $p_0=0$, say on the interval $[-\e,\e]$. If zero does
not belong to the spectrum of the Hamilton operator (which we assume; if not
true, the splitting with respect to $P_+$ would result in peculiarities,
anyway), then the composition $P_+\circ S$ does not change as a
bi-distribution, as is easily seen from its Fourier transform w.r.t time.
This fact already settles the issue. In general, the definition of a
discontinuous symbol may be given in terms of a convergence of smooth
symbols, as indicated by exercise 6, p. 4 of \cite{tay2}.

As to the second issue, whether the conclusion of the
propagation-of-singularities theorem remains true even for the $P_+$, not
properly supported, we may proceed in the following manner: we can decompose
this \PDO\ into a singular part, which is properly supported, and a regular
part, which is not. The regular part possesses a $\smooth\times \smooth$
integral kernel. We have
\begin{equation*}
P_+(y)=P_+^s(y)+P_+^\infty(y).
\end{equation*}

At this point we stumble upon two difficulties. The first is that one would
need to argue that $P_+^\infty\circ S$ as a bi-distribution is actually a
smooth function in both variables.

If this is true  the whole wave front set of $P_+\circ S$ comes from
$P_+^s\circ S$. It is not difficult to show that the principal symbol of
$P_+^s$ is equal to that of $P_+$. It follows that the inverse of this
symbol, $\s^{-1}(x,p)$, is given by
\begin{equation*}
\s^{-1}(0)=\{(x,p):\ p_0\ge 0\}.
\end{equation*}

However, we would like to invoke the propagation-of-singularities theorem
\ref{thm:propagation}, which would give the desired result that the wave
front set of $G^+$ is:
\begin{equation*}
  WF'(P_+\circ S)=\{(x,\xi,y,\xi): x\sim y, \xi^2=0, \xi_0\geq 0\}
\end{equation*}
However, the second difficulty occurs, that for pseudo-differential operators
which are non-polynomial functions of \emph{only few momenta} this theorem
is not valid\footnote{We are grateful to W.Junker for pointing out this
important obstruction to us.}. The difficulty lies in the crucial decay
property of its symbol, which is not fulfilled by operators like
$\Theta(p_0)$ - see \cite{junker}.

\end{proof}

\chapter{Construction of local non-linear observables}
\label{non_linear}

\section{Causal perturbation theory - an overview}\label{sec:CPT}

The causal perturbation theory (CPT) is a mathematically precise
way to construct interacting quantum field theories. In this
section we shall give a brief overview of this method in order to
put what comes in later sections in a firm context. Our exposition
is based upon \cite{BF}.

Instead of starting from the (non-linear) Maxwell-Dirac field equations,
which are difficult to interpret as operator equations (at least at the
beginning), the CPT first attempts  to construct the evolution operator $\S$
as an operator-valued distribution with values in an algebra $\W$ to be
defined. This operator is a functional of $g$ and of the interaction
Lagrangian $\Lag$, symbolically given by the Dyson series
\begin{equation}\label{eq:dyson_series}
\S [\Lag,g ] = \sum_{n=0}^\infty \frac{\ i^n}{n !}\int d^4x_1\ldots \
d^4x_n \ g(x_1)\ldots g(x_n)\ T[\Lag(x_1)\ldots\Lag(x_n)],
\end{equation}
where
\begin{equation*}
\Lag(x)=e :\b \psi(x) \g^a \psi(x): \A_a(x)
\end{equation*}
is the interaction Lagrangian of the quantum electrodynamics. The functions
$g(x)$ are the test functions from a convenient test space; they are
intended to be equal to $1$ in the region where we wish the interaction to
take place\footnote{The results of the finite-order time dependent
perturbation theory of ordinary quantum mechanics are in general trustworthy
only for a limited time scope; the strong limit $|t|\ra \infty$ of the
finite-order interacting evolution operators $U_I(t,t_0)$, even if it exists,
only in special cases (like stationary perturbations) leads to results of
clear physical meaning. Thus, we shall think of $g(t,\v x)$ as having its
support on a finite (possibly small) time interval $t\in I$; the compactness
of the support in spatial dimensions will be less important, as we shall see
in concrete applications.} and to vanish outside of it. The series
\eqref{eq:dyson_series} should be understood as a formal power series in
$e$. The terms in the Dyson series\footnote{The time-ordered products can be
obtained from $\S[\Lag,g]$ by means of a functional derivative
\begin{equation*}
    T[\Lag(x_1)\ldots\Lag(x_n)]=
    \frac{\de^n}{i^n\ \de g (x_1) \ldots \de g(x_n)}\
    \S[\Lag,g]\Big|_{g=0}.
\end{equation*}
}
\begin{equation*}
    T[\Lag(x_1)\ldots\Lag(x_n)]
\end{equation*}
are called time-ordered products of $n$-th order. Obviously the whole
information about the QED is contained in the time-ordered products, and
their construction also contains  the renormalization difficulties.

The main idea of CPT is to obtain the time-ordered products by means of an
inductive procedure, namely, one attempts to construct $T_n$,  the $n$-th
order time-ordered product with the knowledge of all the time-ordered
products of lower orders, $T_{<n}$. This is possible with the assumption of
causality\index{Causality!of time-ordered products}.

\begin{figure}[h]\centering
\includegraphics{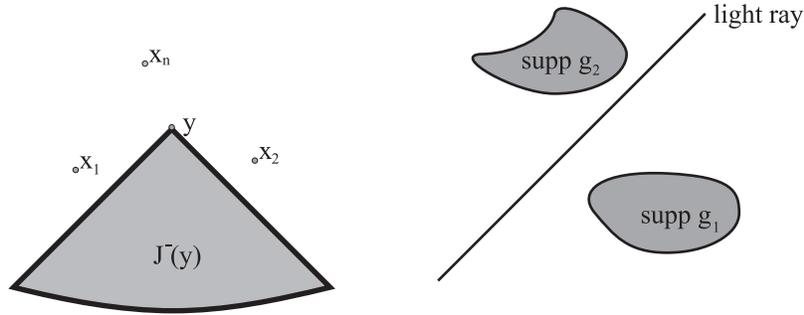}
\caption{Relative position of the point $y$ for the factorization of the TOP
and of the $\supp g_1$ for the factorization of $\S$.}\label{fig:causal_past}
\end{figure}

\begin{definition}[Causality of time-ordered products]
The hierarchy of time-ordered products satisfies the causality requirement,
if for any set of points, $x_1,\ldots,x_n,y_1\ldots,y_m$, such that the
points $x_i$ do not lie in the past of the points $y_i$ (cf. figure
\ref{fig:causal_past}) the factorization property holds:
\begin{equation*}
T[\Lag(x_1)\ldots \Lag(x_n)\Lag(y_1)\ldots\Lag(y_m)]=T[\Lag(x_1)\ldots
\Lag(x_n)]\ T[\Lag(y_1)\ldots\Lag(y_m)].
\end{equation*}
Here the LHS is a $(m+n)$-th order TOP, and the RHS is a product of $n$-th
and $m$-th order TOP. The factorization condition is equivalent to the (more
intuitive) condition on the evolution operators, namely
\begin{equation*}
    \S(g_1+g_2)=\S(g_2)\ \S(g_1),
\end{equation*}
whenever
\begin{equation*}
\supp g_2\nin J^-(\supp g_1).
\end{equation*}
\end{definition}

Given a set of points $x_1,\ldots, x_n$ we ask whether there is a partition
of it such that at least one of the points lies in the causal past of all
the others. Evidently, this is possible, only if the points do not all
coincide\footnote{The coincidence of all points will be called the
\emph{diagonal} of $(\R^4)^{\times n}$.} ($x_1=x_2=\ldots=x_n$). Therefore,
if at least one pair of points does not coincide, it is possible to construct
the $n$-th order TOP. The renormalization problem, in this scheme, manifests
itself as the problem of extending the $n$-th order TOP to the diagonal. The
TOP are, however, \emph{operator-valued distributions} with values in an
algebra $\W$ (to be defined later). Instead of searching for an extension of
OVDs, one reduces the problem to the extension of c-number-valued
\emph{distributions} with the help of the Wick expansion. \index{Wick
expansion}

\begin{Thm}[Wick expansion]\label{thm:Wick_expansion}
The product of Wick polynomials can be expressed as a sum of Wick
polynomials. Without loss of generality we have
\begin{equation*}
    :\h \psi(x_1)\psi(x_1): \ldots :\h \psi(x_n)\psi(x_n):\* :\h \psi(y)
    \psi(y):=\sum_{m=0}^{n+1} d_m(x_1,\ldots ,x_n,y)\ :\h\psi \psi^{(m)}:,
\end{equation*}
where $:\h\psi \psi^{(m)}:$ is the Wick product of $m$-th order which
contains $m$ field operators and $m$ adjoint field operators. The $d_m$
denote number-valued distributions\footnote{See \eqref{wick_2_1} for a
concrete example, where the product of two first-order Wick products has been
expanded.}.
\end{Thm}

Focusing attention on the TOP of the order $n$, $T_n$, we may factorize it
to become a product of lower order TOP's and, after Wick expansion we obtain
the $n$-th order TOP in the form
\begin{equation*}
T_n(x_1,\ldots,x_n)=T[\Lag(x_1)\ldots \Lag(x_n)]=\sum_{m=0}^{n+1}
t^0_m(x_1,\ldots ,x_n)\ :\h\psi \psi^{(m)}:,
\end{equation*}
where the distributions $t^0_m(x_1\ldots x_m)$ are known for all points with
the  exception of the diagonal $x_1=\ldots=x_m$. We need to extend it to the
diagonal, because the evolution operator $\S(g)$ smears the time-ordered
product $T_n(x_1,\ldots,x_n)$ with \emph{the same} test function $g(x)$ in
every variable, and hence the diagonal belongs to the support of the product
of the test functions.

If relative coordinates are employed, for instance $x_i^a=x_0^a+\xi_i^a$,
then the diagonal coincides with the origin of the $(\R^4)^{\times n}$ space,
where $\xi^i=0$. There is an elegant way to extend distributions singular at
the origin to that point, if only the distributions are homogeneous. If we
require that the scaling of the distribution must be preserved by the
extension, then the following possibilities arise:
\begin{itemize}
    \item the extension to the diagonal is unique (this is usually the
    case in the so-called tree Feynman diagrams) and may be obtained, for
    instance, by multiplying the Wick expansions for different
    configurations of points with an appropriate step
    (Heaviside) distributions;
    \item the extension to the diagonal is not unique (which is the case
    in the loop-containing Feynman diagrams) and is characterized by a
    finite number of free parameters, which correspond to the
    renormalization constants.
\end{itemize}

The truncated evolution operator $\S$, constructed with the help of the
time-ordered products of order not greater than $n$ and smeared with  test
functions $g(x)$ which specify the interaction region, describes the quantum
electrodynamics in finite-order perturbation theory. The results obtained in
this way correspond \cite{Scharf} to those obtained in the other
formulations of the perturbation theory (which in our opinion have a less
transparent structure).

\sp
\section{Algebra of Wick polynomials}\qquad Let from now on $\Alg(\O)$
denote the CAR algebra of the free Dirac field. The region $\O$ is an open
bounded region in spacetime which additionally is geodesically convex (a
double-cone). In order to incorporate the interacting field we shall
investigate algebras larger than $\Alg(\O)$. In particular, we will define
the algebra $\W(\O)$ which also contains the Wick polynomials\footnote{For
the purposes of quantum electrodynamics it is sufficient to restrict the
investigations to the Wick polynomials of undifferentiated field operators
with an equal number of field operators $\psi(x)$ and their adjoints
$\h\psi(x)$. This would, however, be insufficient if we wanted to define the
energy-momentum operator.} of the free fields. Before we define $\W(\O)$
consider the Wick square of the free Dirac field.

\begin{example}
The Wick squares are defined by the point-splitting limit:
\begin{subequations}\label{def:Wick_2}
\begin{align}
    : \psi^* \psi:(x)&=\lim_{y\ra x}\left[  \h \psi(x)\psi(y)-\T
    d(x,y)\right],\\
    : \psi \h\psi:(x)&=\lim_{y\ra x}\left[  \psi(x)\h \psi(y)-d(x,y)\right],
\end{align}
\end{subequations}
with appropriate distributions $d(x,y),\T d(x,y)$. The smeared Wick square
is a functional of one test function:
\begin{equation*}
: \psi^* \psi:(f)=\int d^4x\ d^4y \left[  \h \psi(x)\psi(y)-d(x,y)\right]\
\de(x-y) f(x).
\end{equation*}
The above formula looks as if we have smeared the ordinary operator-valued
distribution
\begin{equation*}
    \h\psi(x)\psi(y)-d(x,y),
\end{equation*}
(which when smeared with \emph{two test functions} becomes an operator
belonging to $\Alg$ and is well-defined for all $d(x,y)$) with the
\emph{distribution}\footnote{This is the nomenclature frequently employed in
\cite{HW2}.} $\de(x-y) f(x)$. The result is an OVD in the algebra of Wick
polynomials $\W$ irrespective of $d(x,y)$. However, in order to find a
representation of this algebra it will be necessary to restrict the class of
allowed $d(x,y)$.
\end{example}

\begin{definition}[Wick products of $n$-th order]
The Wick product of $n$-th order is a regularization of the product of $n$
field operators and $n$ adjoint filed operators (grouped in pairs). It is
defined inductively in terms of Wick products of order $n-1$ and lower. Let
$d(x,y)$ and $\T d(x,y)$ denote the two distributions which have \emph{the
same} wave front set:
\begin{equation*}
    WF'(d)=\{(x,\xi,y,\xi): x\sim y, \xi_0\geq 0\}.
\end{equation*}
 The $n$-th order product is given by
\begin{multline*}
:\h\psi(x_1)\psi(y_1)\dots \h\psi(x_n)\psi(y_n):\ \doteq
\h\psi(x_1)\psi(y_1)\dots \h\psi(x_n)\psi(y_n)-\\
-\sum_{k,l} A(x_k,y_l)\ :\h\psi(x_1)\psi(y_1)\dots
\h\psi(x_l)\del{\psi(y_l)}\dots \del{\h\psi(
x_k)}\psi(y_k)\dots\h\psi(x_n)\psi(y_n):-\\ \ (contraction\ of\ two\
pairs\ \psi  \h \psi) :(product\ of\ remaining\ operators): -\\-\ {
\sum_{k=3}^n (contractions\ of\ k\ pairs)\ :product\ of\ (n-k)\ pairs:},
\end{multline*}
where $A(x_k,y_l)$ denotes either $d(x_k,y_l)$ in case $l<k$ or $\T
d(x_k,y_l)$ in the other case. The tilde denotes the operators which do not
appear (have been contracted). The Wick product of an equal number of field
operators and their adjoints in arbitrary order is defined similarly, the
only difference is that in front of each contraction there is a factor
$(-1)^j$, where $j$ is the  number of permutations necessary in order to
bring the contracted operators together.
\end{definition}

\begin{remark}

The definition of Wick products makes use of only two distributions $d,\T
d$. It is assumed that they regularize the two-point functions of Hadamard
states, namely, the distributions
\begin{align*}
 \w&(\psi_A(x)\psi^*_B(z))-d_{AB}(x,z),\\
\w&(\psi^*_B(z)\psi_A(x))-\T d_{BA}(z,x)
\end{align*}
are assumed to be smooth for every Hadamard state $\w$. The pointwise
product of $d$ and $\T d$ is well-defined as a distribution, due to the
properties of their wave front sets. The multiplication enlarges the wave
front set; for instance, $d(x,y)\T d(x,y)$ has the wave front set
\begin{equation*}
    WF'(d\T d)=\{(x,\xi_1,y,\xi_2): \xi_i\in J^+
    \}.
\end{equation*}
This set is stable under further multiplications with
distributions having the Hadamard form of singularity \cite{BF}.
\end{remark}

\begin{lemma}\label{lema:OVD} Let $\Fou$ be the Hilbert space
constructed upon a Hadamard state $\w$. The Wick products defined above,
even upon multiplication with $\de(x_i-y_j)$, are operator-valued
distributions on $\Fou$.
\end{lemma}

\begin{remark}
We do not prove this lemma here, although for second-order Wick products it
is not difficult. The generalization to higher order products for the scalar
field has been given in \cite{BF}. In order to use these results here we
would need a characterization of the Wick products in terms of derivatives
of certain functionals which we do not have at the moment.
\end{remark}

In what follows we shall prove a lemma which greatly facilitates the Wick
ordering of the time-ordered products.

\begin{lemma} \label{lemma:anticommutation} If the regularizing distributions fulfill the condition
\begin{equation*}
d_{AB}(x,y)+\T d_{BA}(y,x)=S_{AB}(x,y)=\{\psi_A(x),\h\psi_B(y)\},
\end{equation*}
then the Wick products change sign upon commutation of adjacent field
operators:
\begin{multline*}
:\h\psi(x_1)\psi(y_1)\ldots  \h\psi(a)\psi(b) \ldots \h\psi(x_n)\psi(y_n) :\
=\\= - :\h\psi(x_1)\psi(y_1)\ldots \psi(b)\h \psi(a) \ldots
\h\psi(x_n)\psi(y_n):.\
\end{multline*}
\end{lemma}
\begin{proof}The proof is standard and can be completed with combinatoric methods.
We shall consider the case when the commutation occurs right at the end of
the Wick product\footnote{This makes the proof more evident. The general
case is more laborious.}. The LHS in that case is
\begin{equation*}
    LHS=\h\psi(x_1)\psi(y_1)\ldots \h\psi(x_n)\psi(y_n)
    \ \h\psi(a)\psi(b)+ contractions.
\end{equation*}
The RHS is similarly
\begin{multline*}
    RHS=\h\psi(x_1)\psi(y_1)\ldots \h\psi(x_n)\psi(y_n)
    \ \h\psi(a)\psi(b)+\\-S(b,a)\ \h\psi(x_1)\psi(y_1)\ldots
    \h\psi(x_n)\psi(y_n)+ \widetilde{contractions},
\end{multline*}
where "$contractions$" means all the due terms from the expansion of
\[:\h\psi(x_1)\psi(y_1)\ldots
\h\psi(x_n)\psi(y_n)\ \h\psi(a)\psi(b) :.\]
 Similarly, "$\widetilde{contractions}$" means all
the due terms from the expansion of
\begin{equation}\label{eq:expansion_tilde}
-:\h\psi(x_1)\psi(y_1)\ldots \h\psi(x_n)\psi(y_n)\ \psi(b)\h \psi(a) :.
\end{equation}
Evidently, the products of field operators are equal on both sides; what
remains to be done is to show that the terms with contractions are also
equal. Let us take the RHS and consider the two possible cases:
\begin{itemize}
    \item There is a contraction of $\h \psi(a)$ or $\psi(b)$ with the other
    points. Then in the Wick product there
    is a minus sign due to the even number of field operators
    between the contracted quantities. This minus compensates the minus of
    the whole expansion \eqref{eq:expansion_tilde}, and  what remains
    after the contraction corresponds precisely to the
    respective term of the expansion on the LHS.

    \item The operators $\psi(b) \h \psi(a)$ are contracted together. This
    leads to the term
    \begin{equation*}
    d(b,a)\ [all\ contractions\ of\
    :\h\psi(x_1)\psi(y_1)\ldots \h\psi(x_n)\psi(y_n):],
    \end{equation*}
    which is equal to
    \begin{equation*}
    d(b,a)\ [\h\psi(x_1)\psi(y_1)\ldots \h\psi(x_n)\psi(y_n)].
    \end{equation*}
    If we combine this with the other summand on the RHS (which involves $S(b,a)$) it yields
    \begin{equation*}
    -\T d(a,b)\ [\h\psi(x_1)\psi(y_1)\ldots \h\psi(x_n)\psi(y_n)],
    \end{equation*}
    which is precisely the sum of contractions on the LHS in the case of
    $\psi(a) \h \psi(b)$ contracted together.
\end{itemize}
\end{proof}

\begin{remark} As a simplest corollary we note that
\begin{equation*}
    :\psi(x)\h \psi(y):=-:\psi(y)\h\psi(x):.
\end{equation*}
Also
\begin{equation*}
    :\h \psi(x) \psi(x)\h \psi(y) \psi(y):=
    :\h \psi(y)\psi(y)\h \psi(x) \psi(x):.
\end{equation*}
Both the above expressions facilitate greatly the causal Wick
expansion of the second order TOP (see section
\ref{sec:second_order_TOP}).
\end{remark}

With this preparatory definitions we may now define the larger algebra
$\W(\O)$ which will be large enough to contain the non-linear field
quantities, in particular the interacting fields.

\begin{definition}
On the distributional level the algebra $\W(\O)$ is the algebra generated by
the Wick polynomials of an even number of elements
\begin{align*}
    W_2(x,y)&=\ :\h\psi(x)\psi(y):,\\
    W_4(x_1,x_2,y_1,y_2)&=\ :\h\psi(x_1)\psi(y_1) \h\psi(x_2)\psi(y_2):,\\
    \vdots\qquad&=\qquad\vdots,
\end{align*}
possibly multiplied with delta-distributions in order to define Wick powers.
The elements of the algebra $\W(\O)$ are formed, when the resulting quantity
is integrated with the test functions of compact support in $\O$:
\begin{equation*}
    \W(\O)\ni W(g_1,\dots,g_n)=\int d^4x_1\dots d^4x_n\
    g_1(x_1)\*\dots\*g_n(x_n)\ W(x_1,\dots,x_n),
\end{equation*}
where for brevity we have omitted the remaining arguments $y_1,\dots,y_n$
which should also be smeared with the test functions.
\end{definition}

\begin{remark}In applications the expectation values of the elements
of $\W$ will be studied. The expectation values of the Wick products
$W_n(x_1,\ldots,x_n)$ will become harmless smooth functions, but their
distributional coefficients $c_n$ will require investigation (see the
expression \eqref{wick_2_1} for a concrete example).
\end{remark}

\sp
\section{Locality in causal perturbation theory}
In this section we will introduce a new condition of locality for
quantum field theory on external backgrounds. This new locality
requirement will be stronger than the notion of locality usually
employed in quantum field theory without external field
backgrounds. In what follows we will define this new notion and
explain why it is necessary to incorporate it in the construction
of the quantum electrodynamics. We shall put less emphasis on the
mathematical rigor of the definition of locality\footnote{A
precise definition, which can be found in \cite{BFV}, uses the
theory of  categories.} in favor of a clarification of its
physical meaning.

The standard notion of locality in  quantum field theory means
that the observables in spatially separated regions should be
commensurate. More precisely, if we focus on the algebras of
observables $\Alg(\O)$ parameterized by spacetime regions, we have
the following

\begin{defi}[Quantum field theoretical locality]\label{def:QFT_local}
The net of algebras $\Alg(\O)$ is local, if each $A\in\Alg(\O)$
commutes with all the elements of the algebras of the regions
spatially related to $\O$, that is
\begin{equation*}
    [A,B]=0 \qquad \forall\ A\in\Alg(\O),\ B\in\Alg(\O'),\
    \O\  /\hspace{-2mm}\backslash \T\O.
\end{equation*}
The notation $\O\  /\hspace{-2mm}\backslash \T\O$ means, that the
regions $\O$ and $\T\O$ cannot be connected by a causal curve.
\end{defi}

\sp

As far as the standard notion of locality is concerned, there is
the following, simple

\begin{obs}\label{obs:lokalnosc} The Wick squares defined above are local in the
field-theoretical sense (def. \ref{def:QFT_local}), and this
property is independent of the choice of  the regularizing
distribution $d(x,y)$.
\end{obs}
\begin{proof}We investigate the commutator
\begin{equation*}
    [:\psi^*_A\psi_B:(x),:\psi^*_C\psi_D:(y)],
\end{equation*}
where the points $x$ and $y$ are spacelike related. If we undo the
point-splitting limit and order the expression appropriately we
obtain
\begin{equation*}
    [.,.]=S_{BC}(x',y)\psi^*_A(x)\psi(y')-
S_{DA}(y',x)\psi^*_C(y)\psi_B(x'),
\end{equation*}
which is independent of $d(x,y)$ and vanishes, due to the
properties of the anti-commuta\-tor distribution.
\end{proof}

\sp

\subsection{Local quantum field theory on
external field backgrounds}\label{subsec:LPI}\quad

At this point we may ask, whether it follows form observation
\ref{obs:lokalnosc} that any choice of the regularizing distribution
$d(x,y)$ leads to a local theory. If the definition \ref{def:QFT_local}
is all we ask for, then indeed the net of Wick polynomial algebras
$\W(\O)$ are local. However, there is a stronger locality requirement that
is in the spirit of general relativity. This requirement, called the
Local Position Invariance (LPI), lies at the foundation of the theory of
relativity; it is instructive to recall its definition\footnote{The
discussion of the equivalence principle (which has various ingredients)
as well as its current experimental status can be found in an excellent
review article by C.Will \cite{will}.} and the way it has been tested
experimentally. Local position invariance states that:

\sp \emph{The outcome of any non-gravitational experiment is independent
of where and when in the universe it is performed. The fundamental
constants of non-gravitational physics should be constants in space and
time.} \sp

\begin{figure}[ht]\centering
\includegraphics{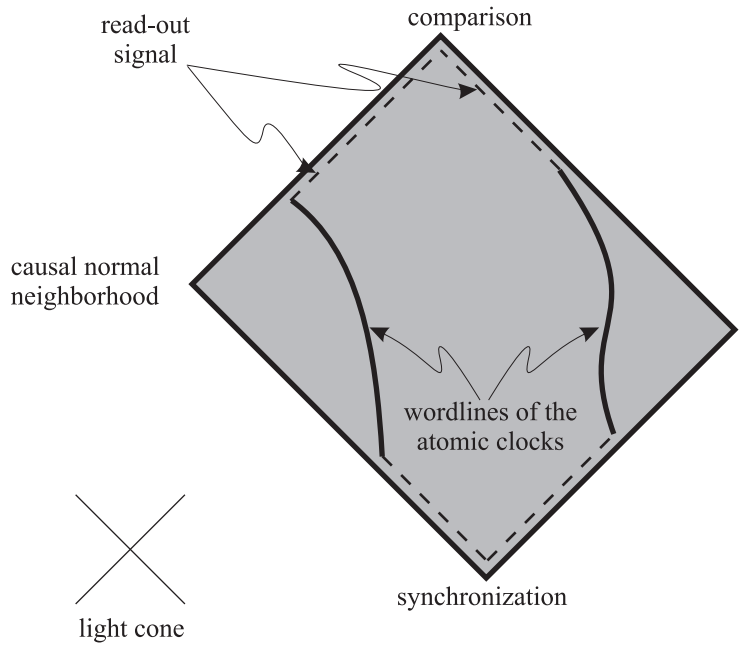}
\caption{GPS-like test of local position invariance. The experimental
result is allowed to depend only on the gravitational field in the causal
normal neighborhood containing the whole measurement apparatus (together
with the final signal read-out).} \label{gps}
\end{figure}

In experiments the LPI is measured with the help of the following
procedure (see figure \ref{gps}): two precise frequency standards (the
atomic clocks) are employed. They are synchronized with a light signal.
They follow their wordlines\footnote{ We can assume them to be geodesic in
order to rule out the influence of acceleration on the clocks.} and
continuously send the light signals which carry the information about
their states. Those signals are compared at a single event. The result is
scrutinized against general relativistic predictions, namely, one
calculates the geometric lengths of both wordlines. If the time lapse the
atomic clocks have measured is proportional to the length of their
wordlines, then indeed the non-gravitational experiments (here the quantum
optical experiments) are independent of the position in the universe.

Let us view the experiment described above from a different perspective
and ask what does the result of the measurement (of the comparison) depend
on? Clearly, if the LPI is fulfilled, then the result depends only on the
gravitational field in the region of spacetime which contains the
wordlines of the two detectors and all the geodesics which join them (the
clocks must be synchronized and their state must be compared). Therefore,
the LPI which at the moment is supported by strong experimental evidence
\cite{will} implies that the results of quantum optical experiments can
only depend on the external field in the causal neighborhood containing
the entire measurement setup.

We now turn back to the definition of local quantum field theory. Recall
that the quantum field theory consists of three main ingredients:
\begin{itemize}
\item an algebra of observables $\W(\O)$, on which

\item the time evolution acts as a group of automorphisms;

\item the states are functionals on $\W(\O)$; they describe the
expectation values of the observables.
\end{itemize}

\begin{figure}[h]\centering
\includegraphics{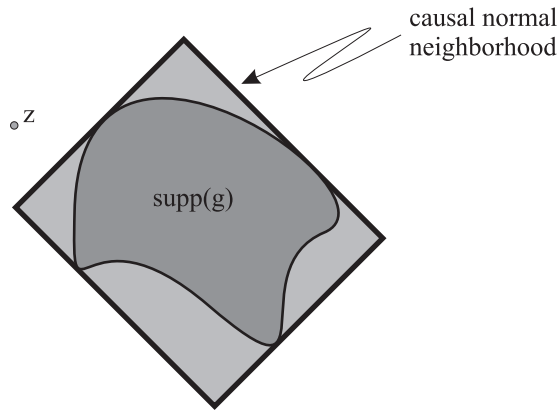}
\caption{Smallest causal neighborhood containing the support of the test
function $g$. The point $z$ lies outside of it.} \label{points}
\end{figure}

In this paper we have adapted the following way to incorporate the LPI
into quantum field theory. \index{Locality}
\begin{definition}[Strong locality]\label{def:locality}
The quantum field theory is local in a strong sense (LPI) if:
\begin{itemize}
    \item The bi-distribution employed in the definition of the Wick
    square ($d(x,y)$) and of all Wick polynomials
    (see Wick expansion, theorem \ref{thm:Wick_expansion}),
    is a functional of the external field (and its derivatives). In particular
    \begin{equation*}
    d(x,y)=d[A,\d A,\d^2 A,\ldots].
    \end{equation*}
    \item The distributional coefficients of the Wick powers,
    $c_n(x_1,\ldots,x_n)$, are also functionals of the external field.
    \item All the above distributions  depend functionally only on the
    external field in the causal normal neighborhood which contains
    all of their arguments. Specifically, if $g(x)$ denotes an arbitrary test
    function of compact support, then the functional derivatives of
    $\langle d,g\otimes g\rangle$ and $\langle
    c_n,g^{\otimes n}\rangle$, with respect to the external field $A(z)$, must
    necessarily vanish:
    \begin{align*}
    \frac{\de\,\langle d,g\otimes g\rangle}{\de A(z)}&=0,&
    \frac{\de\, \langle
    c_n,g^{\otimes n}\rangle}{\de A(z)}&=0,
    \end{align*}
    if the point $z$ does not belong to the smallest causal normal
    neighborhood containing the support of $g$ (see figure \ref{points}).
\end{itemize}
\end{definition}

\begin{remark}
As a consequence of the strong locality requirement we may compare two
quantum field theories on different external fields, say $A$ and $\T A$.
If those external fields match in a geodesically convex neighborhood
$\O$,\footnote{The causal normal neighborhood of such a region coincides
with itself. Typically, regions of this sort are diamonds of the form
$\O=J^+(y)\cap J^-(x)$ for some $x\in J^+(y)$.} the Wick algebras
$\W_A(\O)$ and $\W_{\T A}(\O)$ may be identified (are naturally
isomorphic). This leads to the picture of a local quantum field theory as
a functor between the categories of all possible external field
configurations and the category of unital $C^*$-algebras\footnote{The
morphisms of the respective categories are explained in \cite{BFV}.}. In
this framework it is possible to investigate the change of algebraic
elements under the compactly supported  variation of the external field
(relative Cauchy evolution) \cite{BFV}. It is important to stress,
however, that the notion of locality takes into account only the two first
ingredients of a quantum field theory: the observables and their time
evolution. There is an intrinsic non-locality in the construction of
states of the quantum field (this is discussed in the next section). We
should also note the new insight into the LPI of general relativity which
is gained from the quantum field theory on external field backgrounds,
namely, it is not true that there exist "non-gravitational experiments"
at all. Even the local influence of the external field may lead to the
alternation of non-gravitational phenomena such as the atomic dynamics.
This together with the non-locality of the two point function of the
radiation field, which in the view of the appendix \ref{app:spontaneous}
influences the atomic emission process, is a possible way to try to
explain the decrease of the fine structure constant at distant epochs
recently observed with the help of very modern methods by J.K.Webb and
collaborators \cite{webb}.
\end{remark}

\section{Non-locality of the two-point functions}

In this section we show that the two-point functions $G_\pm(x,y)$ contain
information about the external field not only in the causal neighborhood
of $x$ and $y$, and thus they do not fulfill the strong locality
requirement \ref{def:locality}.

We construct our argument for the massless, hermitian, scalar field, as in
that case the argument is transparent and simple.  Let the external
potential  be static and only dependent on one spatial variable, say
$x_3$, called "$x$" in the sequel. Investigate the classical scattering,
stationary wave functions in order to define the field operator with their
help.

\begin{figure}[h]\centering
\includegraphics{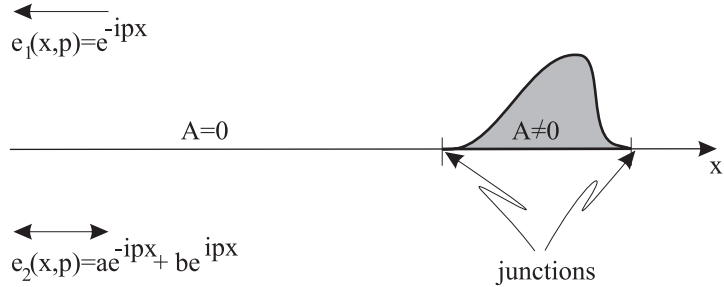}
\caption{Two independent solutions $e_1(x)$ and $e_2(x)$. The other
dimensions $(t,\v x_\bot)$ are suppressed.}
\end{figure}
The $x_1$ and $x_2$ (called "$\v x_\bot$" for brevity) dependence
factorizes:
\begin{equation*}
  u(t,\v x)=e^{iE(p_\bot,p)t-i\v p_\bot \v x_\bot}\cdot e(p,x).
\end{equation*}
As there is no external field for small $x$, both independent solutions
$e_{1,2}(x,p)$ will be linear combinations of the running waves: outgoing
$e^{-ipx}$ and incoming $e^{ipx}$. We may take as the first solution the
one for which $e_1=e^{-ipx}$ (which is a purely outgoing wave) for small
$x$. The second solution, orthogonal to $e_1$,  will have the form
\begin{equation*}
  e_2=ae^{-ipx}+be^{ipx},
\end{equation*}
where $a\neq 0$, because there is a potential scattering, $b\neq 0$
because it needs to be orthogonal to $w_1$. Now the field can be
quantized in a standard manner. We introduce the object (later, the
field-operator-valued-distribution)
\begin{multline*}
  \psi(t,x,x_\bot)=\int_{\R^2}d^2p_\bot\int_0^\infty
  \frac{dp}{\sqrt{2E(\v p_\bot,p)}}\ \cdot
  \left\{e^{iEt-ip_\bot x_\bot}[e_1(x,p)a_1(\v p)+e_2(x,p)a_2(\v
  p)]+H.c.\right\},
\end{multline*}
with the CCR for the $a$'s :
\begin{align*}
  [a_1(\v p) , a^*_1(\v k)]&=\de(\v p-\v k), &   [a_1(\v p) , a^*_2(\v k)]&=0, \\
  [a_2(\v p) , a^*_2(\v k)]&=\de(\v p-\v k), &   [a_.(\v p) , a_.(\v
  k)]&=0.
\end{align*}
Note that we have two sets of creation/annihilation operators, because we
have separated the positive and negative values of $p=p_3$.

The vacuum representation is defined by requiring
\begin{equation*}
  a_.(\v p)\Omega=0.
\end{equation*}
Now the two point function $(\Vac,\psi(\v x)\psi(\v y)\ \Vac)$ (for equal
times) is given by
\begin{equation*}
\int_{\R^2}d^2p_\bot\int_0^\infty
  \frac{dp}{2E}\ \cdot
  \left\{e^{-ip_\bot (x-y)_\bot}[e_1(x_3)\b e_1(y_3)+e_2(x_3)\b
  e_2(y_3)]\right\}.
\end{equation*}
The square bracket is evaluated to be
\begin{equation*}
[\ldots]=(1+|a|^2)e^{-ip(x-y)}+|b|^2e^{ip(x-y)}+ a\b b\ e^{ip(x+y)}+\b a b
\ e^{-ip(x+y)}.
\end{equation*}
On the other hand in the absence of the external field we have
\begin{equation*}
  [\ldots]_{Minkowski}=e^{-ip(x-y)}+e^{ip(x-y)}.
\end{equation*}
Both distributions are not equal.  $\quad \Box$

We infer from the above example that the two-point function contains
non-local information. We cannot implement the strong locality
requirement \ref{def:locality} by employing objects constructed with the
help of the two-point functions.

\section{Non-localities of the extensions of local distributions}
The construction of interacting field theory, as outlined in section
\ref{sec:CPT}, requires an extension of the time-ordered products to the
coincidence point, where the appropriate $c$-number-valued distributions
are singular. In this section, by means of an example, we discuss the
difficulties arising in the extension process.

Suppose that the appropriate distribution $t^0(x)$ away from the diagonal
already fulfills the strong locality requirement\footnote{For the sake of
simplicity, we investigate here distributions of one argument. The
variable $x$ in concrete applications has the meaning of the relative
coordinate of two points of the time-ordered product.}. Moreover, suppose
that the singularity of the distribution $t^0$ is such that its action is
well-defined for all test functions that vanish at the
origin\footnote{That means that the derivatives of the test functions do
not necessarily have to vanish there.} $x=0$. The following simple method
of extension of $t^0$ to the origin is generically employed in the usual
causal perturbation theory: define the subtraction $\mathfrak w$ on the
space of test functions
\begin{equation*}
    (\mathfrak w f)(x)=f(x)-w(x) f(0)\quad\forall f\in\dom(\R),
\end{equation*}
where $w(x)$ is an arbitrary $\dom$-function equal to  $1$ in the vicinity
of $x=0$. The extended distribution (denoted by $t$) is defined by
\begin{equation}\label{eq:extension_1}
\langle t,f\rangle=\langle t^0, \mathfrak w f \rangle.
\end{equation}
Evidently, the extension $t$ coincides with the distribution $t^0$, if the
test function vanishes at the origin. Moreover, the difference of two
different extensions, characterized by $w_1$ and $w_2$, is given by
\begin{equation*}
\langle t_1-t_2,f\rangle=\langle t_0,w_2-w_1\rangle f(0),
\end{equation*}
which is a distribution supported  only at the origin, and therefore
(which is a general property - see \cite{hilbert_courant2}) can be
expressed as a sum of the delta-distribution and its derivatives. If we
additionally require the extension to have the same scaling
behavior\footnote{A discussion of the extension of distributions (in the
context of interacting field theories on background manifolds) to a point
can be found in \cite{BF} section 5.2, where the construction of the
extended distribution with appropriate scaling is carried out explicitly.}
as $t_0$, we can say which derivatives of $\de$ appear. In the case of
$t^0(x)=1/x$ the result is
\begin{equation*}
    t_2(x)=t_1(x)+c\ \de(x)
\end{equation*}
with an arbitrary constant $c$.  For instance, the Cauchy principal value:
\begin{equation*}
  \langle P\tfrac{1}{x},g \rangle=\lim_{\e\ra 0} \int_{|x|\geq \e}
  \frac{g(x)}{x}
\end{equation*}
and the $i\e$ prescription
\begin{equation}
  t^0=\frac{1}{x+i\e}
\end{equation}
are precisely two different extensions of the distribution $t^0=1/x$
which differ by a $c\cdot \de(x)$ distribution (with $c=-i \pi$).

Let us  consider the difficulties with the strong locality
requirement (definition \ref{def:locality}) which arise in the
presence of external fields. Suppose the distribution $t^0$ was
local in the strong sense
 and had the same singularity
structure as $1/x$. An example of such a distribution is
\begin{equation*}
    t^0(x)=\frac{\int_0^1 A(xs)\ ds}{x}
\end{equation*}
which shares some similarity with the first term of the
parametrix of the Klein-Gordon operator on an external background
$A(x)$. The extension of this distribution by means of the
$\mathfrak w$ subtraction is
\begin{equation}
\langle t,f\rangle=\langle t^0, \mathfrak w f \rangle= \int dx\
ds\ \frac{1}{x} A(sx)\ [f(x)-w(x)f(0)].
\end{equation}
The conflict with the strong locality principle is now evident. The
general extension depends not only on the external field in the support
of $f(x)$, which would be the manifestation of strong locality, but also
on the external field in the region of support of $w(x)$, which spoils
the locality (if $\supp w\subset\!\!\!\!\!/\supp f $).

A further investigation is therefore needed in order to make the
renormalization method of Epstein and Glaser
\cite{dosch_mueller,BF} strongly local.

\section{Local causal perturbation theory in the lowest orders}\label{lowest_orders}

In this section we shall construct local Wick and time-ordered products of
first and second order. They are the building blocks of the evolution
operator $\S$ (when smeared out with test functions), and it is essential
to investigate them carefully, as any alternation of their form can
potentially lead to the modification of measurable quantities. The second
order time-ordered product already describes a number of important
phenomena.

We begin this section with some considerations of the lowest two orders of
Wick and time-ordered products of the standard approach to the external
field quantum electrodynamics. They are not local in the strong sense, as
has been explained before. We present them nonetheless, because this
discussion will be of help in the construction of the local Wick and
time-ordered products, and because it may help the reader to recognize the
more familiar objects and realize where the necessary modification will
be done, in order to make these objects local.

In quantum electrodynamics the first order time-ordered product is given
by the interaction term $\Lag_I$:
\begin{equation*}
    T_1(x)=\Lag_I(x)=ie :\b \psi_A(x) ?\g_\m ^AB? \psi_B(x): \ \A^\m(x).
\end{equation*}
The second order time-ordered product satisfies the causality relation
\begin{equation*}
    T_2(x,y)=T_1(x)T_1(y),
\end{equation*}
if only $y\nin J^+(x)$. Thus, for $x\neq y$ the fermionic part of the
second order TOP can be brought into the form:
\begin{equation*}
    T_2(x,y)=\sum_{n=0}^2\ t_n(x,y) \  :\h \psi \psi^{(n)}:
\end{equation*}
with the help of the Wick expansion \ref{thm:Wick_expansion}. The Wick
products on the RHS are independent of the relative position of the
points, but the distributions $t_n(x,y)$ are at first defined only for
non-coinciding points.

\subsection{Usual Wick product} Now we will recall some facts about
the Wick product of two Dirac field operators.  Let there be a fixed
representation of the free CAR algebra with the base state described by a
vector $\Vac$ in the Hilbert space. We introduce the commonly employed
abbreviations
\begin{align*}
    i ?G^+_AB?(x,y)&\doteq (\Vac, \psi_A(x)\b \psi_B(y) \Vac),\\
    i ?G^-_AB?(x,y)&\doteq (\Vac, \b \psi_A(x) \psi_B(y) \Vac).
\end{align*}
The (non-local) Wick products are defined by
\begin{subequations}
\begin{align}\label{Wick_psi_psibar}
    :\psi_A(x)\b \psi_B(y):&\doteq\psi_A(x)\b \psi_B(y)-iG^+_{AB}(x,y),\\
    :\b \psi_A(x) \psi_B(y):&\doteq\b \psi_A(x)
    \psi_B(y)-iG^-_{AB}(x,y).\label{Wick_psibar_psi}
\end{align}
\end{subequations}

Note that the Wick products are not linear and the definition
\eqref{Wick_psibar_psi} cannot be derived from \eqref{Wick_psi_psibar}
with the help of the anti-commutation relations\footnote{Indeed, consider
\begin{equation*}
    :\psi_A(x)\b\psi_B(y):+:\b\psi_B(y)\psi_A(x):=
    \{\psi_A(x),\b\psi_B(y)\}-iG^+_{AB}(x,y)-
    iG^-_{BA}(y,x)=0,
\end{equation*}which follows from the definition of the Wick products. On the other hand
linearity would give
\begin{equation*}
    \ldots=:\{\psi_A(x),\b\psi_B(y)\}:=:iG_{AB}(x,y):\neq 0.
\end{equation*}
Therefore, these definitions are logically independent. If we take (as was
done so far) $G^-$ and $G^+$  as the expectation values of the appropriate
field operators, then they are connected, as both of them are in
one-to-one correspondence with their projections, which however sum up to
an identity on $\H$, that is $P_++P_-=\id$. It follows from
\eqref{op_pola} that we have (in case $f,g\in \H $ are test functions)
\begin{align*}
    G^+(f,g)&=(\Vac,a(P_+f)a^*(P_+g)\Vac)=(f,P_+g),\\
    G^-(f,g)&=(\Vac,b(P_-f)b^*(P_-g)\Vac)=(f,P_-g).
\end{align*}
 }.
In the case of absent external field the distributions $G^\pm(x,y)$ are
well-known:
\begin{align*}
    i ?G^+_AB?(x,y)&=\frac{1}{(2\pi)^3}\int \frac{d^3p}{2p_0} (\dir{p}  +m)
    e^{-ip(x-y)},\\
    i ?G^-_AB?(x,y)&=\frac{1}{(2\pi)^3}\int \frac{d^3p}{2p_0} (\dir{p}  -m)
    e^{-ip(x-y)}.
\end{align*}
We may write
\begin{align*}
iG^+_{AB}(x,y)=(i\dir\d_x+m)_{AB} \* \De^+(x,y),\\
iG^-_{AB}(x,y)=(i\dir\d_x-m)_{BA} \* \De^+(x,y),
\end{align*}
where $\De^+(x,y)$ denotes the massive ($m$), scalar,  vacuum two-point
function.\footnote{As the free (from the external field) case is
extensively studied in \cite{Scharf}, it is perhaps valuable to establish
a correspondence with the notation used there. For the fermionic operators
we have
\begin{align*}
    -S^{(+)}_{ab}(x-y)&=G^+_{AB}(x,y),\\
    -S^{(-)}_{ab}(y-x)&=G^-_{AB}(x,y),
\end{align*}
whereas in the electromagnetic case
\begin{equation*}
iD_0^{(+)}(x-y)=\De^+_0(x,y)
\end{equation*}
(the expressions on the LHS are in the notation of G.Scharf
\cite{Scharf}).} It is important to stress that both $G^+$ and $G^-$ are
derivatives of \emph{the same} distribution. Consequently, the wave front
set of $\De^+$ contains the wave front sets of $G^\pm$.

 The anti-commutator distribution $G_{AB}(x,y)$ can be expressed in
 terms of $G^\pm$:
\begin{equation}\label{car_bar}
iG_{AB}(x,y)=\{\psi_A(x),\b \psi_B(y)\}=(\Vac,\{\psi_A(x),\b \psi_B(y)\}\
\Vac)=iG^+_{AB}(x,y)+iG^-_{BA}(y,x),
\end{equation}
which assures the validity of the lemma on the (anti)commutation of Wick
products (lemma \ref{lemma:anticommutation}). We also note that
\begin{equation*}
    i G_{AB}(x,y)=\{\psi_A(x),\psi_C^*(y)\}\ ?\g^{0C}_B?=S_{AC}(x,y)?\g^{0C}_B?,
\end{equation*}
and we find, in accordance with \ref{antykomutacja_pola_swobodnego}, the
equal-time anti-commutation relation
\begin{equation*}
iG_{AB}(\v x,\v y)=\g^0_{AB}\  \de(\v x - \v y).
\end{equation*}

The Wick product of the free electromagnetic fields is defined by
\begin{equation*}
    :\A_\m(x) \A_\n(y):\ =\A_\m(x) \A_\n(y)-\eta_{\m\n} \De^+_0(x,y),
\end{equation*}
where $\De^+_0(x,y)$ denotes the massless, scalar field two-point function
\eqref{eq:omega_0}.

\sp
\subsection{Second-order time-ordered product}\label{sec:second_order_TOP}\quad
Turning now towards the second order TOP, $T_2(x,y)$, in the case $y\nin
J^+(x)$, we will use the causal Wick expansion to order of the expression
\begin{equation*}
    T_2(x,y)=(ie)^2 :\b \psi_A (x) \g^{\mu AB}\psi_B(x):
    :\b \psi_C (y) \g^{\n CD}\psi_D(y): \ \A_\m(x) \A_\n(y).
\end{equation*}
As $T_1$ already involved a limiting procedure (in the case of the Dirac
field operators), in order to Wick-expand $T_2$ we will step back this
limit and consider
\begin{equation*}
    T_2(x,y)=(ie)^2 \left[:\b \psi_A (x) \psi_B(x'):
    :\b \psi_C (y) \psi_D(y'):\right] \g^{\mu AB} \ \g^{\n CD} \
    \A_\m(x) \A_\n(y),
\end{equation*}
where only the square-bracket term requires further attention. It is seen
to be equal to
\begin{multline*}
[\ldots]=\b \psi_A(x)\psi_B(x')\b\psi_C(y)\psi_D(y')-iG^-_{AB}(x,x'):\b
\psi_C(y)\psi(y'):-\\
-iG^-_{CD}(y,y'):\b\psi_A(x)\psi_B(x'):+G^{-}_{AB}(x,x')G^-_{CD}(y,y').
\end{multline*}
The first term should now be Wick-expanded. This leads to the expression
in which the limit $x'\rightarrow x$ and $y'\rightarrow y$ can be taken.
One obtains:

\begin{center}
\framebox{\parbox{6.0in}{
\begin{multline}\label{wick_2_1}
    T_2(x,y)=-e^2\left[:\b\psi_A(x)\psi_B(x)\b\psi_C(y)\psi_D(y):+iG^-_{AD}(x,y)
    :\psi_B(x) \b \psi_C(y):+\right. \\ \left.+iG^+_{BC}(x,y):\b\psi_A(x)\psi_D(y):
    -G^-_{AD}(x,y)G^+_{BC}(x,y)\right]\g^{\m AB}\g^{\n CD} \\
    \left[:\A_\m(x)\A_\n(y):+
    \eta_{\m\n}\De^+_0(x,y)\right].
\end{multline}
}}
\end{center}
The pointwise product of $G^-_{AD}(x,y)G^+_{BC}(x,y)$ is well-defined, and
its wave front set is not greater than that of $\De^+(x,y)^2$.

\sp

 In the other case, when $x\nin J^+(y)$, there is
\begin{equation*}
    T_2(x,y)=T_1(y)T_1(x),
\end{equation*}
and the similar Wick expansion can be performed. We only have to exchange
$x$ with $y$ (and also $A$ with $C$, $B$ with $D$ and $\mu$ with $\nu$).
The causal Wick expansion requires ,however, to order $T_1(y)T_1(x)$ in
such a way that the Wick products in $T_2(x,y)$ are the same as in
\eqref{wick_2_1}. In order to do that we employ the (anti)commutation
lemma \ref{lemma:anticommutation}. We arrive at
\begin{multline}\label{wick_2_2}
    T_1(y)T_1(x)=-e^2\left[:\b\psi_A(x)\psi_B(x)\b\psi_C(y)\psi_D(y):
    -iG^+_{DA}(y,x):\psi_B(x) \b \psi_C(y):+\right. \\
    \left.-iG^-_{CB}(y,x):\b\psi_A(x)\psi_D(y):
    -G^+_{DA}(y,x)G^-_{CB}(y,x)\right]\g^{\m AB}\g^{\n CD} \\
    \left[:\A_\m(x)\A_\n(y):+
    \eta_{\m\n}\De^+_0(y,x)\right].
\end{multline}
A comparison of the above expression with \eqref{wick_2_1} is presented
in the table 1.
\begin{table}[h]\label{tabelka}
\begin{center}
\begin{tabular}{c|c} \hline
 $y\nin J^+(x)$& $x\nin J^+(y)$\\ \hline
&\\
$G^-_{AD}(x,y)$& $-G^+_{DA}(y,x)$\\
&\\
$G^+_{BC}(x,y)$& $-G^-_{CB}(y,x)$\\
&\\
$G^-_{AD}(x,y)G^+_{BC}(x,y)$& $G^+_{DA}(y,x)G^-_{CB}(y,x)$ \\
&\\
$\De^+_0(x,y)$     &  $\De^+_0(y,x)$  \\
\hline
 \end{tabular}
\end{center}
\caption{Correspondence of factors of Wick products with different
relative position of the points $x$ and $y$.}
\end{table}

 What remains to be done in order to define the $T_2(x,y)$
 for all its arguments is to extend the distributional factors of Wick products
 to the diagonal $x=y$. Thus, in accordance with the general
 procedure (see section \ref{sec:CPT}), the definition of $T_2$ for all $x,y$ reduces to
the problem of the extension of number-valued distributions, which are
known for $x\neq y$, to the diagonal $x=y$. Some of the distributions
appearing in $T_2(x,y)$ can be extended to the diagonal uniquely; the
other, however, have a family of extensions parameterized by what is
known as the renormalization constants.

\sp
\subsection{Local definition of the Wick
product}\label{sec:local_Wick}\quad The Wick products defined with the
help of the $G^\pm$ are not local in the strong sense. We shall repair
this deficiency in what follows.

In order to make the Wick square an operator-valued distribution it is
necessary only to remedy the short-distance singularities of the product
of the field operators. We proceed therefore and give a local expression
for the Wick product, the uniqueness of which will be analyzed in a
separate section (see section \ref{sec:Wick_unique}). The leading
short-distance singularities of the expectation value of the product of
the Dirac-field operators in some Hadamard state are all contained in the
(local) Hadamard parametrix $H_{AB}(x,y)$ constructed in chapter
\ref{chapter:Local_Solutions}. We therefore define\index{Wick product}
\begin{equation*}
    :\psi_A(x)\b \psi_B(y):\doteq\psi_A(x)\b \psi_B(y)-H_{AB}(x,y)
\end{equation*}
and, similarly,
\begin{equation*}
    :\b \psi_A(x) \psi_B(y):\doteq\b \psi_A(x) \psi_B(y)-\T H_{AB}(x,y),
\end{equation*}
where the Hadamard parametrix $\T H$ must be separately
derived\footnote{The parametrix $\T H$ is not just the adjoint of $H$;
although $\g^0H^\dag \g^0$ would fulfill the appropriate equations, it
would not have the appropriate wave front set (this can be seen even on
the no-external field distributions $G^\pm$).} with the methods of chapter
\ref{chapter:Local_Solutions} for the case of the adjoint Dirac operator
acting on the variable $x$:
\begin{equation*}
\T H_{AB}(x,y)\ \overleftarrow{D^a_x}?\, ^A_C?=0,
\end{equation*}
with
\begin{equation*}
\overleftarrow{D^a_x}?\, ^A_C?=(-i\overleftarrow{\dir \d} -m +e\dir
A)?\,^A_C?.
\end{equation*}

We therefore establish the following substitution rule for the Wick
products:
\begin{align*}
H_{AB}(x,y)& \leftrightarrow iG^+_{AB}(x,y),\\
\T H_{AB}(x,y)& \leftrightarrow iG^-_{AB}(x,y).
\end{align*}

The definition of the Wick product with the help of the Hadamard
parametrix encounters, however, two obstacles: on the one hand the full
series defining $H_{AB}(x,y)$ is convergent, only if the external fields
are analytic functions\footnote{The coefficients $v_N(x,z)$ are
functionals of the external field dependent on the partial derivatives of
it up to  order $2(N+1)$ (and therefore require an appropriately high
differentiability of $A(z)$). What F.G.Friedlander claims in
\cite{friedlander} chapter 4.3 is that also smoothness of the external
field is not sufficient  for the convergence of the series $\sum_n^\infty
v_n(x,z) \G^n \Theta(\G)$ (where $\Theta(\G)$ is the analogue of $\ln
\G$).} (which is a narrow case), on the other hand the full $H_{AB}(x,y)$
is of no practical use, as it is unknown how to sum up the Hadamard
series.

If the coincidence limit $y\ra x$ is of interest (that is if only powers
of the field are of interest), then an observation of V.Moretti,
\cite{Moretti01}, is helpful, namely:

\begin{obs}The Wick product $:\psi_A(x)\b \psi_B(x):$ can
be defined with finitely many terms of the Hadamard expansion (in this
case all what is needed are $u,v_0 \t{ and } v_1$). The so-defined OVD
fulfills the wave equation (the Dirac equation) in its variable modulo a
constant:
\begin{equation*}
 \lim_{x\ra y} D_x :\psi_A(x)\b \psi_B(y):=3\, \OpD_x v_0(x)|_{x=y}
\end{equation*}
(with abbreviations of chapter \ref{chapter:Local_Solutions}, note that
$D_x$ and $\OpD_x$ are different).
\end{obs}
\begin{proof}As the field operator $\psi_A(x)$ fulfills the field
equation by definition, what remains to be done is the inspection of
\begin{equation*}
  \lim_{x\ra y} D_x H^{N}(x,y),
\end{equation*}
where $H^N(x,y)$ is the Hadamard parametrix \emph{without the smooth term}
$w(x,y)$ and cut at $v_N$. Not surprisingly the calculation which preceded
the derivation of the transport equations \eqref{rown_wn} will be helpful,
for
\begin{equation*}
    \lim_{x\ra y} D_x H^{N}(x,y)=  \lim_{x\ra y} \OpD_x
    \phi^{N}(x,y).
\end{equation*}
By inspection, we conclude that $\OpD_x \phi^N(x,y)$ consists of terms
proportional to $1/\G^2,\ 1/\G$, $ \ln \G$ as well as $\G^n$ and $\G^n\ln
\G$. All three first, singular terms will vanish as long as $N\geq 1$ due
to the transport equations fulfilled by $u,v_0$ and $v_1$. The
coefficients of the powers of $\G$ will have no effect in the limit $x\ra
y$. What will remain, however, is the coefficient of $\G^0$ which defined
the transport equation for $w_1(x,y)$ set to zero in our case. One finds
\begin{equation*}
  \lim_{x\ra y} \OpD_x \phi^{N}(x,y)=\lim_{x\ra y} \OpD_x
  \phi^{1}(x,y)=-v_1(12+B^a\d_a\G)-2\d^a\G\d_a v_1=
  -4v_1+\OpD_x v_0,
\end{equation*}
as a consequence of \eqref{rown_vn}. Now, from the recursion relation, we
have
\begin{equation*}
v_1(x,y)=-\T u(x,y)\int_0^1\tau d\tau\, \left.\frac{\OpD_w v_0(w,y)}{\T
u(w,y)}\right|_{w=y+\tau (x-y)},
\end{equation*}
in the limit $x\ra y$
\begin{equation*}
v_1(y,y)=-\frac{1}{2}\OpD_x v_0(x,y)|_{\, x=y}.
\end{equation*}
\end{proof}


\sp
\section{Local definition of the current operator; back-reaction effects}
In this section we formulate the problem of the back reaction of the
quantum fields on the background arena. Let us recall the field equations
of the theory developed in this paper (cf. chapter \ref{chapter:intro}):
\begin{subequations}
\begin{align}
        \left[i\g^a\d^x_a+eA_a^{class}(x)- m\right] \psi(x) &=-e\A_a(x)\psi(x),\\
         \d^b_x F^\A_{ab}(x)&=4\pi\ \b \psi(x)\g_a\psi(x),\\
        \d^b_xF_{ab}^{class}(x)&=4\pi J_{a}(x),
\end{align}
\end{subequations}
where $A^{class}\equiv A$ is the classical electromagnetic field which is
the background field, $\A$ is the quantum electromagnetic field, $\psi$ is
the quantum Dirac field and $J$ is the external classical current which is
the source of the $A$-field. The interaction of the quantum fields (the
RHS of the first two equations) is treated in a perturbative way.

The main assumption of the theory is that the external current $J$ is not
influenced by the state of the quantized radiation field $\A$. We propose
the following way to assess this influence: if the current produced by the
quantized Dirac field is comparable (locally) to $J$, then the external
field approximation is not reliable anymore.

Suppose that we focus on the current density $J^0(x)$ and the
current-density operator $\p(x)$ (to be defined later). There are three
issues arising, if we want to make the above criterion more precise.
Firstly, for a given state $|F\rangle$ of the Dirac field it is
insufficient to compare the expectation value $(F,\p(x) F)$ to $J^0(x)$.
The current-density operator is not positive, and therefore one should
also investigate higher momenta of it, for instance\footnote{All operators
appearing here are understood as follows: we choose a weighing function
$f$,
\begin{align*}
    f(x)&\geq0,&   \int f(x)\ d^4x&=1,
\end{align*}
in such a way that it is supported in a small neighborhood of the point
$x$. Then the current density and its fluctuation are defined as
\begin{align*}
(F,\p(x)F)&=(F,\p(f) F)& (F,\p(x)\p(x) F)&=(F,\p(f)\p(f) F).
\end{align*}
The need to investigate higher momenta of the charge distribution is
clearly visible in the following example: Suppose that we want to examine
the electric field of a photon. The expectation value of the
electric-field operator (which is also not a positive operator) in the
one-photon state is zero; on the other hand, the expectation value of the
square of the electric-field operator is non-zero, and precisely this
quantity produces an effect on a photodiode.} $(F,\p(x)\p(x) F)$.

The second issue is that we have only the statistical interpretation of
quantum theory, and thus we do not really know what happens in a single
experiment. This is related to the problem of measurement.  If the results
of quantum theory such as $\langle \p^2(x)\rangle$ do correspond only to
the average over many experimental realizations, why then should in a
single experiment the back-reaction effect be governed by such an
average? Let us put this question in a physical context: Imagine a weakly
localized electron which, upon an interaction with a measurement device,
for instance a CCD camera, produces a sharply localized, strong effect - a
macroscopically visible trace. The CCD camera is composed of various
charges and could have been regarded, for instance, as the source of the
classical current density $J^0$. Therefore, even arbitrarily low current
densities can produce intensive back-reaction effects. This shows that the
phenomenological criterion we have put forward in order to judge the
possible back-reaction effects can fail for some sources of the external
current $J$.

In quantum field theory there is an additional difficulty. The
current-density operator has to be defined as a coincidence limit of a
difference between the two-point function of the state of the quantum
field and a certain regularizing bi-distribution. This bi-distribution in
the no-external-field case is taken simply to be the vacuum expectation
value of the appropriate (bi-local) current-density operator. However, as
we have discussed in the chapter \ref{chapter:quantization} there is no
privileged state of the Dirac field on external field backgrounds. How
should the current-operator be defined?\footnote{We are not interested
here in a relative current density between two states, but rather in the
\emph{absolute} current density of a given state.}  It is precisely the
strong locality criterion (definition \ref{def:locality}) that allows for
a definition of the absolute current density. In other words, the strong
locality criterion allows us to separate the (state-independent)
singularity of the two-point function from the state-dependent information
which may cause the back reaction.

What follows in this section deals with the definition of the current
operator $j^\m(x)$. We shall define it  for the \emph{free} Dirac field by
the point-splitting procedure. The quantum version of $j^\m(x)$ should
correspond to the classical current
\begin{equation*}
  j_{class}^\m=e\ \b \psi(x)\g^\m\psi(x).
\end{equation*}
It should also posses the following properties:

\begin{itemize}

\item In the time-independent case, the integration of $j^0(x)$ over a
Cauchy surface,
\begin{equation*}
  Q=\int d^3x\ j^0(\v x),
\end{equation*}
should correspond to the total charge operator $Q=e(N^+-N^-)$, where
$N^\pm$ are electron-/positron-number operators.

\item The current operator should be conserved\footnote{With regard to the
current conservation we note that the energy-momentum operator is
covariantly conserved in the QFT on a curved spacetime as a consequence of
the diffeomorphism invariance (see \cite{BFV} theorems 4.2 and 4.3); the
conservation of the quantized current should follow from the gauge
invariance of the theory.}
\begin{equation*}
\d_\m j^\m(x)=0.
\end{equation*}
\end{itemize}

We recall that the classical current is conserved as a consequence of the
Dirac equation and its adjoint, namely the (co)-spinors $\b\psi(x)$,
$\psi(x)$ fulfill
\begin{align*}
  [i\dir \d+e\dir A(x)-m]\psi(x)&=0,  \\
  \b\psi(x)\ [-i\overleftarrow{\dir \d}+e\dir A-m] &=0,
\end{align*}
from which conservation of the current follows immediately.

As a first candidate for the current-density operator we investigate
\begin{equation}\label{j_first}
  j^\m(x)=:\b\psi(x)\g^\m\psi(x):_{\w}=
  \lim_{y\ra x}[\b\psi(y)\g^\m\psi(x)-iG^-_{AB}(y,x)\g^{\m AB}],
\end{equation}
where $iG^-_{AB}(y,x)=(\Vac,\b\psi(y)\psi(x)\ \Vac)$. The above operator
is not local in that it depends on the external field in the regions other
than the infinitesimal neighborhood of $x$; nonetheless we shall consider
its features. Let us look at the total charge. We find
\begin{multline*}
  \int d^3x\ j^0(x)=\int d^3x\ \int d\m_p\ d\m_k\
  \left[\con u(x,\v p)u(x,\v k)\h a(\v p) a(\v k)-
  \con v(x,\v p) v(x,\v k) \h b(\v k) b(\v p) \right]+\\
  +\int d^3x\ \int d\m_p\ d\m_k\ \left[\con u(x,\v p)v(x,\v k)\h a(\v p)\h
  b(\v k)+ \con v(x,\v p) u(x,\v k) b(\v k) a(\v p) \right].
\end{multline*}
Now, due to
\begin{align*}
  (u(\v k),u(\v p)) &=\de(\v k-\v p),& (v(\v k),v(\v p))&=\de(\v k-\v p), \\
  (u(\v k),v(\v p)) &=0,\\
\end{align*}
we obtain
\begin{equation*}
\int d^3x\ j^0(x)=\int d\m_p \left[\h a(\v p) a(\v p)-\h b(\v p ) b(\v
p)\right],
\end{equation*}
which is just the total current $Q=e(N^+-N^-)$.

\subsection{Charge conservation, local definition of the current density}
The non-locality of \eqref{j_first} was not the only shortcoming of this
expression. It is easy to see that the so-defined current density is not
conserved. Indeed, as the differentiation is performed before the
coincidence limit $y\ra x$ is taken, the $\b\psi$-terms are not
differentiated and, consequently, the current is not conserved.

With regard to the point-splitting procedure of defining $j^\m$ there are
therefore two issues:
\begin{itemize}
\item The conservation of the current density.
\item The appropriate behavior of the coincidence limit.
\end{itemize}

Various procedures have been devised in the literature to remedy both of
the above problems\footnote{See for instance \cite{dosch_mueller}, where
the ideas originally due to J.Schwinger together with their possible
generalizations are discussed.}. The main theme of those procedures is:
\begin{itemize}
\item to remedy the non-conservation by means of the symmetrization of
$j^\m(x,y)$ w.r.t. its variables;

\item to assure the finiteness of the $x\ra y$ limit by  a subtraction of a certain local
bi-distribution.
\end{itemize}
A current density which fulfills all our requirements and which is
constructed as indicated above is\index{Current operator}:
\begin{equation}\label{eq:LCD}
  j^\m(x)=\lim_{y\ra x}[\b\psi(y)\g^\m\psi(x)-\t H_{AB}(y,x)\g^{\m AB}+
  \b\psi(x)\g^\m\psi(y)-\t H_{AB}(x,y)\g^{\m AB}].
\end{equation}

\sp
\subsection{Uniqueness of the current operator}\label{sec:Wick_unique}
\quad The uniqueness of the current operator is connected with the
uniqueness of the Wick square of the field operators. Suppose that we
alter the definition of the Wick product by terms of lower order in the
field operators:
\begin{equation*}
    :\b\psi\psi:=\b\psi\psi-\T d(x,y)\g^0+c(x) \psi+d(x)\h \psi+f(x,y)\id.
\end{equation*}
If we require the current to be selfadjoint and to fulfill the standard
locality requirement \ref{def:QFT_local}, then we must have $c=0=d$. In
order to further restrict the remaining ambiguity we use the locality
requirement, namely, the function $f(x,y)$ can depend functionally only on
the external field in the smallest causal neighborhood of $x$ and $y$. In
the point splitting limit this should become a functional of the external
field at the point $x$. Such functionals are known to be functions of the
external field $A(x)$ and its derivatives $\d_x A(x)$. If we further
require these functions to be analytic in the external field and the mass
$m$ of the Dirac field, we can expand $f(x,x)$ as a power series in $A$,
$\d^n A$ for all $n$, and in the mass $m$. Furthermore, the Wick product
must have a dimension of $1/cm^3$ and both the external field $A$ and the
mass $m$ have dimension $1/cm$. Due to this and the Lorentz condition $\d
A=0$, we obtain the result

\begin{lemma}The local current-density operator is unique up four
arbitrary numbers $a_1\ldots a_4$. Any two local definitions differ at
most by
\begin{equation*}
\De:j^\mu:(x)=a_1\ A^\n\d_\n A^\m+a_2\ m^2 A^\m+a_3\ A_\n A^\n A^\m +a_4\
\Box A^\m
\end{equation*}
\end{lemma}


\sp
\section{Scaling transformations for local
observables}\label{section_scaling} In this section the (re)scaled
non-linear observables will be investigated. The purpose of the scaling is
to approach the short scale-limit of the theory.

\subsection{Scaling transformation in the scalar case} In the following we
shall investigate the behavior of local observables costructed out of the
scalar massive fields under scaling.

In analogy to the extraction of the long-range part of the electromagnetic
field introduced by Gervais and Zwanziger \cite{gervais},
\begin{equation*}
    A_a^{long}(x)=\lim_{\la\ra\infty}\ \la A_a(\la x),
\end{equation*}
we shall associate with a (classical) scalar massive field $\vp(x)$ a
field $\T\vp(x)$ rescaled by $\la$:
\begin{equation*}
    \T\vp(x)=\la \vp(\la x).
\end{equation*}
We define the scaling transformation for classical quantities:
\begin{defi}[classical rescaling]
A scaling transformation $\s$ is a prescription to obtain various
quantities at a different scale $\la$, where $\la\ra 0$ corresponds to the
short-distance limit. In order to investigate the rescaled theory we
\begin{itemize}
\item rescale all quantities, that is
\begin{equation*}
\s: f(x,y,\ldots)\ra f(\la x,\la y,\ldots);
\end{equation*}
\item interpret the result in terms of the rescaled field
\begin{equation*}
\T\vp(x)=\la \vp(\la x).
\end{equation*}
\end{itemize}
\end{defi}

As an example, let us find the field equation fulfilled by $\T\vp$. We
rescale
\begin{equation*}
    0=\s\left[(\Box_x+m^2)\vp(x)\right]=(\la^{-2}\Box_x+m^2)\vp(\la x)=
    \la^{-3}(\Box_x +\la^2m^2)\T \vp(x)=0,
\end{equation*}
and therefore $\T\vp(x)$ is a massive scalar field with the smaller
mass\footnote{In the interesting case $\la<1$.} $\la m$.

\sp Now we shall define the scaling of the (massive, hermitean, scalar)
quantum field

\begin{defi}[quantum rescaling]\index{Rescaling transformations}
The scaling transformation is a prescription which transforms
operator-valued distributions constructed with the help of the (linear)
quantum field $\vp(x)$ on the GNS Hilbert space of the state $\w$ into an
operator-valued distributions of the scaled quantum field $\T \vp(x)$ on
the GNS Hilbert space of the scaled state $\T \w$. The two-point functions
of the respective states are related by:
\begin{equation*}
    \T \w(x,y)=\la^{2}\w(\la x,\la y).
\end{equation*}
\end{defi}

\begin{remark}The scaling transformation relates the quantities of one
quantum dynamical system to the other. The dynamical laws in both systems
are different, as the field operators $\vp (x)$ and $\T\vp (x)$ fulfill
different field equations. Both systems (on a single Cauchy surface) are
representations of the same CCR algebra:
\begin{align*}
    [\vp(\v x),\vp(\v y)]&=\de(\v x-\v y),
    & [\T \vp(\v x),\T \vp(\v y)]&=\de(\v x-\v y).
\end{align*}
Even on this single surface, however, the states $\w$ and $\T \w$ are not
even locally equivalent; as we shall see, their short distance singularity
structure is different.
\end{remark}

As an example let us investigate the scaling of the Wick product which is
given by
\begin{equation*}
    :\vp(x)\vp(y):\doteq \vp(x)\vp(y)-H(x,y),
\end{equation*}
where $H(x,y)$ is a certain bi-distribution, for instance
\begin{equation*}
    H(x,y)=\frac{1}{\G}+m^2\ln(\G m^2).
\end{equation*}
Let us now perform the scaling $\s$ on the Wick product. We have
\begin{equation*}
    \s_\la[:\vp(x)\vp(y):]=\frac{1}{\la^2}[\la^2 \vp(\la x)\vp(\la y)
    -\la^2H(\la x,\la y)]=\frac{1}{\la^2}[\T \vp(x)\T \vp(y)-H_\la(x,y)],
\end{equation*}
where we have denoted the rescaled regularizing bi-distribution by
$H_\la$:
\begin{equation*}
    H_\la(x,y)=\la^2 H(\la x,\la y)=\frac{1}{\G}+(\la m)^2\ln
    \left[\G(\la m)^2\right].
\end{equation*}
The singularity structure of $H_\la$ clearly differs form $H$ (as long as
the field is massive). It is also evident that in the short distance limit
$\la\rightarrow 0$ the less singular term in $H_\la$ will disappear,
\begin{equation*}
    \lim_{\la\rightarrow 0} H_\la(x,y)=\frac{1}{\G}=\De_0^+(x,y),
\end{equation*}
where $\De_0^+(x,y)$ is the vacuum, massless two-point function.

\section{Scaling transformation for the Dirac field in external potentials}

The action for the classical Dirac field on a background external
potential $A_a(x)$ is given by
\begin{equation*}
  S=\int \t{d$^4$x } \left\{[i\hbar \ \b \psi\g^a\d_a \psi-mc\ \b\psi\psi]-
  [\b\psi\g^a\psi \ \frac{e}{c}A_a]\right\}
\end{equation*}
Here the action is dimensionless and the Dirac field $\psi(x)$ has a
dimension of cm$^{-3/2}$. Analogously to the scalar case, the scaling is
performed by a substitution $x\ra \la x$. Here, however, the rescaled
field (classical and quantum) is defined via
\begin{equation*}
    \T\psi(x)=\la^{3/2}\psi(\la x).
\end{equation*}
Again the scaling transformation transforms operator-valued distributions
on the GNS Hilbert space $\Fou$ into the operator-valued distributions on
the GNS Hilbert space $\T \Fou$ constructed with the help of
\begin{equation*}
    \T\w(x,y)=\la^{3}\w(\la x,\la y).
\end{equation*}
\begin{remark}
Similarly to the scalar-field case, we note that on a fixed Cauchy surface
the OVD $\psi(x)$ and $\T \psi(x)$ fulfill the same anti-commutation
relations. However,
\begin{itemize}
\item they obey a different dynamical law, and \item the state $\T\w$ is
not locally equivalent to $\w$.
\end{itemize}
\end{remark}

It is interesting to note\footnote{This (in part) will be the conclusion
of the next subsection.} that if we introduce the rescaled external field
$\T A$ and the rescaled mass $\T m$,
\begin{align*}
    \T m&=\la m, & \T A_a(x)&=\la A_a(\la x),
\end{align*}
then for all operator-valued distributions $F[x,y,\ldots\psi(x),A,m]$ there
is
\begin{equation*}
    \s\left\{F[x,y,\ldots,\psi(x),A,m]\right\}=
    F[x,y,\ldots,\T \psi(x),\T A,\T m].
\end{equation*}

An investigation of the Wick product
\begin{equation*}
    :\psi_A(x)\b{\psi}_B(y):=\psi_A(x)\b \psi_B(y)-H_{AB}(x,y),
\end{equation*}
with $H_{AB}$ chosen to be a parametrix (with certain $w_0(x,y)$)
dependent on $m$ and the external field $A_a(x)$ as explained in chapter
\ref{lowest_orders}, reveals
\begin{equation*}
\s_\la[:\psi_A(x)\b \psi_B(y):]=\frac{1}{\la^3}[:\psi_A(x)\b
\psi_B(y):-H_{\la\ AB}(x,y)]
\end{equation*}
with
\begin{equation*}
    H_{\la\ AB}(x,y)=\la^3 H_{AB}(\la x,\la y),
\end{equation*}
being the scaled Hadamard parametrix. In what follows $H_{\la\ AB}$ will
be investigated in greater detail.

\subsection{Scaling of the Hadamard parametrix for the Dirac field}
In chapter \ref{chapter:Local_Solutions} we have found the Hadamard
parametrix $H_{AB}(x,z)$ of the Dirac field. It was given by
\begin{equation*}
H_{AB}(x,z)=(i\g^a\d_a+e\g^aA_a(x)+m?)_A^C? \phi_{CB} (x,z),
\end{equation*}
where $\phi_{CB}(x,z)$ is the Hadamard parametrix of the operator
\begin{equation*}
    \OpD=[\Box -2ieA^a(x)\d_a+m^2+e^2A^2(x)]-\frac{e}{2}?\s^abA_B? F_{ab}(x),
\end{equation*}
which was determined with the help of the progressing wave expansion:
\begin{equation*}
    \phi_{CB}(x,z)=\frac{u_{CB}(x,z)}{\G}+v_{0CB}(x,z)\ln \G +
    \sum_1^\infty v_{nCB}(x,z) \G^n\ln
    \G+\sum_1^\infty w_{nCB}(x,z) \G^n
\end{equation*}
where the spinor indices have been restored in order to emphasize the
bi-spinorial character of the smooth coefficients $u,v_n,w_n$.
\begin{remark}
The term $\ln\G$ cannot stand as it is, because $\G=(x-z)^2$ has a
dimension. This dimension can be cancelled with the only length scale
available in the theory - the Compton wavelength of the electron.
Therefore, everywhere $\ln \G$ should be understood as $\ln(m^2\G)$. It
is another matter that a different scale (if present) would give a
parametrix differing from the above by a smooth function.
\end{remark}

Because the parametrix (or certain parts of it ) have been used to define
nonlinear quantities, it is important to investigate the scaling
properties of the diagonal ($x=z$) values of the smooth coefficients under
the transformation $\s$. We find the following

\begin{lemma}The diagonal values $v_n(z,z), w_n(z,z)$ of the
coefficients of the Hadamard parametrix scale under the transformation
$\s_\la$ as:
\begin{align*}
    \s[v_n(z)]&=\frac{1}{\la^{2(n+1)}}\ \T v_n(z),\\
    \s[w_n(z)]&=\frac{1}{\la^{2(n+1)}}\ \T w_n(z),
\end{align*}
where $n=0,1,\ldots$ for the $v$'s and $n=1,2,\ldots$ for the $w$'s. The
$\T v$ and $\T w$ denote $v$ and $w$ with $A,m$ replaced by $\T A,\T m$.
The coefficient of the highest singularity $u(z,z)=1$ by assumption for
all $\la$.
\end{lemma}
\begin{proof}Before considering the case of arbitrary $n$, let us first illustrate
the reasoning in case of $n=0$ and $n=1$. Everywhere we consider $z$ as a
fixed point of reference so that all the differential operators are being
taken with respect to the variables other than $z$. We use the
abbreviation
\begin{equation*}
    u(x,z)=\exp\left[ie\chi(x,z)\right]=\exp\left[ie(x-z)^a\int_0^1
    A_a(y)|_{y=z+\tau(x-z)} \ d\tau\right].
\end{equation*}
One easily verifies that in the limit $x\rightarrow z$ there is
\begin{align*}
    (\d_a\chi)&=A_a(z),\\
    \Box \chi&=0,
\end{align*}
with the last equation being an indirect consequence of the Lorentz gauge
of the external field.

The coefficient $v_0$ is found from\footnote{In the following
considerations, for brevity, we drop the tilde above the scalar part of
$u$.}
\begin{equation*}
    v_0(x,z)=-\frac{u}{4}\int_0^1 \frac{\OpD_y u}{u}|_{y_0=z+\tau_0(x-z)}\ d\tau_0.
\end{equation*}
Clearly, in the limit of interest the distributions under the integral
become $\tau_0$-independent and can be pulled out of the integral, which
then assumes the value $1$. Therefore,
\begin{equation*}
    v_0(z,z)=-\frac{1}{4} \OpD_y u(y,z)|_{y=z}.
\end{equation*}
Now the differentiation of $u(y,z)$ results in terms of the sort $\Box_x
\chi(x,z)$, $A^a(x)\d_a\chi(x,z)$, $[m^2+e^2A^2(x)]\* \chi(x,z)$, as e.g.
explained in and above the formula \eqref{nice_v0}. The scaling means a
replacement of $(x,z)$ by $(\la x,\la z)$. We shall also make use of the
rescaled mass and the external field
\begin{equation*}
    \T m=\la m \qquad \T A_a(x)=\la A_a(\la x).
\end{equation*}
By the explicit formulas for $\chi(x,z)$ and all its derivatives, it is
evident that every term of $\OpD u$ transforms under $\s_\la$
homogeneously with degree $-2$. Therefore, we have
\begin{equation*}
\s_\la[v_0(z,z)]=\frac{1}{\la^2}\T v_0(z,z),
\end{equation*}
where $\T v_0$ contains the rescaled $\T A$ and $\T m$.

 Let us consider then the next coefficient, $v_1(x,z)$, which is
the prototype of a general situation. According to the formulas of the
chapter \ref{chapter:Local_Solutions} we have:
\begin{equation*}
    v_1(x,z)=\frac{1}{16}u(x,z)\int_0^1 \tau_1\ d\tau_1 \ u^{-1}(y_1,z)
    \OpD_{y_1}\left[u(y_1,z)\int_0^1 u^{-1}(y_0,z)\OpD_{y_0} u(y_0,z)\
    d\tau_0\right],
\end{equation*}
where $y_1=z+\tau_1(x-z)$ and $y_0=z+\tau_0(y_1-z)$. Now the differential
operator $\OpD_{y_1}$ either acts on $u(y_1,z)$ in front of the integral
or can be pulled under the integral, where (as all the expressions there
depend only on $y_0$) a chain rule has to be invoked which e.g. transforms
$\d_{y_1}$ into $\d_{y_0}\* \tau_0$. In the first case there will be
derivatives of $\chi(y_1,z)$ appearing in front of the $\tau_0$-integral,
and in the second the derivatives of $\chi(y_0,z)$ and of $A_a(y_0)$ or
$c(y_0)$. In the limit $x\rightarrow z$ all those factors will become
$\tau_0$- and $\tau_1$-independent (which also assures that they will be
functions of the external field at $z$ - and thus - local expressions) and
can be pulled out of all the $\tau$-integrals. All the derivatives of
$\chi$ are easily seen to have the homogeneous scaling, namely,
\begin{equation*}
    \s_\la[\d'_{a_1}\ldots\d'_{a_n} \chi(y,z')|_{y=z'}=
    \frac{1}{\la^n}\chi(z,z).
\end{equation*}
As each differential operator $\OpD$ contains two such differentiations
(or a factor with identical scaling), it follows in particular that
\begin{equation*}
\s_\la[v'_1(z',z')]=\frac{1}{\la^4}v_1(z,z),
\end{equation*}
and in general
\begin{align*}
    \s_\la[v'_n(z',z')]&=\frac{1}{\la^{2(n+1)}}v_n(z,z),\\
    \s_\la[w'_n(z',z')]&=\frac{1}{\la^{2(n+1)}}w_n(z,z).
\end{align*}
 The general expressions follow from the recursion relations of
chapter \ref{chapter:Local_Solutions}.
\end{proof}

\chapter{Physical applications}
The considerable modification and generalization of existing quantum
electrodynamics in strong external fields described in the previous
chapters raises two important issues. The first is whether the framework
presented in this thesis is concrete enough to describe physical
situations. One may ask how to apply the general construction of states
of chapter \ref{chapter:quantization} together with the local causal
perturbation theory of chapter \ref{non_linear}. The other issue is
whether our improvements lead to predictions which differ from the
conventional ones. In this chapter we will address the first issue while
the second must be deferred to future investigations.

The local causal quantum electrodynamics constructs the local $\S$-matrix
which describes the evolution of  observables. This $\S$-matrix in each
order of perturbation is composed  of local Wick and time-ordered
products, defined in previous chapters. When smeared out with a test
function $g(x)$, all these elements become operators belonging to the
algebra $\W$. In order to be able to investigate the physical properties
of matter and radiation it is necessary to specify the representation of
this algebra and identify in some way the states on this algebra with
physical situations. In a concrete representation the  non-linear
quantities will become operators on a Hilbert space.

The non-linear observables were constructed in such a way that the GNS
representation based upon a state $\w$ of the free field algebra $\Alg$
can be extended to the enlarged algebra $\W$. This extension is possible
only, if the two-point function of the state $\w$ is of Hadamard form
(chapter \ref{chapter_hadamard}).

The "base state" $\w$ carries with itself the concept of excitations.  If
$\w$ is prescribed by a projection operator $P_+$, the states of the form
$\psi(f_-)\Vac$, $\h \psi(f_+)\Vac$ with $f_\pm=P_\pm f\ni \H$ describe
excitations which would be conventionally called particles. It is the aim
of interacting quantum electrodynamics to investigate the evolution of
such excitations.

The identification of mathematical objects  with physical reality always
has its limits; they are particularly exposed in the interacting theory.
Suppose, for instance, that the external field is static. Then the
excitations can describe  particles in bound states (if such states are
allowed by the external field). But what are these states? The quantum
electrodynamic interaction changes the dynamics of the Dirac field, and
what was stationary in the free theory is not stationary  any more in the
interacting theory. We would like to know the energy of the excited state
2P of an electron in a hydrogen atom. We describe this state by
$\psi^*(f_{2P})\Vac$, where $f_{2P}$ is the wave function of the 2P state
calculated with the help of the Dirac equation. However, because this
state is not stationary, we will never be sure what physical situation it
corresponds to or how to produce such a state in experiment.

With this remark in mind we begin to address the first of the issues
named at the beginning; in what follows we shall show, how the theory
developed in this thesis is applied to describe the simplest non-trivial
physical effects.

\section{Electrodynamics in the presence of a static background}
Static backgrounds are important, mainly because they provide a
sufficiently well-characterized and isolated regime for a study of the
effects of the QED developed in this thesis. Not only the fully ionized
heavy elements (see \cite{uranium}) but also fine atomic/ionic traps (see
\cite{dehmelt} and many others) allow for a study of the electrodynamic
effects with great precision. In many of these experiments a single
electron probes the external field effects. Perhaps for those experimental
reasons the conventional literature on the external field QED has focused
on static external fields\footnote{Another reason might be the sense of
uniqueness that the ground state vacuum carries with itself.}
\cite{mohr,shabajev}.

\subsection{Vacuum representation, static background}
The construction of the vacuum representation in the case of static
external backgrounds has been treated in chapter \ref{section_static}.
Here we review it from a more heuristic point of view and restore the
appropriate spinor indices.

The free-field operator in the static external fields  "at a point $x$" is
an operator-valued distribution which can be decomposed according to:
\begin{align*}
 \psi^A(x)&=\int d^3p\ \left\{u^A_s(\v x,\v p)\  e^{-iE^+(\v p)t}
 \ a^s(\v p)+
  v^A_s(\v x,\v p)\ e^{-iE^-(p)t} \ b^{*s}(\v p)\right\}+\\
 &+\sum_{n=0}^N u_n^A(\v x)\ e^{-iE(n)t}\ a_n.
\end{align*}
Here we have separated the bound states from the scattering states. The
$u(\v x,\v p)$, $v(\v x,\v p)$ denote the positive/negative-frequency
scattering states (the index $\v p$ is continuous). The $u_n(\v x)$ denote
the spinorial wave functions of the bound states, which we have assumed to
have positive energies (for simplicity); there may exist an infinite
number of them ($N\ra \infty$). The symbols $E^\pm(\v p)$ denote the
energies (the eigenvalues of the time-independent Hamilton operator) of
the positive/negative-frequency scattering solutions, respectively. The
creaton/annihilation operators obey the standard CAR relations. The vacuum
$\Vac$ is the state annihilated by all of the annihilation operators:
\begin{align*}
  a(\v p)\Vac &=0, & a_n\Vac&=0, \\
  b(\v p)\Vac &=0.
\end{align*}

\subsection{First-order processes, creation of the electron-positron pair}
In quantum electrodynamics with the external field treated as a
perturbation it is known that one photon (with sufficiently large energy)
can create an electron-positron pair \cite{Ber_QED}. In what follows we
will investigate this process within the framework of local QED.  In the
zero-th and first order the evolution operator $\S$ is given by
\begin{equation*}
  \S=\id+\int d^4x\ g(x)\ :\b \psi(x)\g^\m \psi(x):\A_\m (x),
\end{equation*}
where the test function $g(x)$ equals one in the interaction
region\footnote{Later we shall take $g=1$ on the strip $t\in (0,T)$ (i.e.
we shall abandon the spatial smearing) and assume $g$ to vanish just
outside of this strip.}. In order to investigate the process in which a
one-photon state evolves into a state with fermionic content ("creation
of an electron-positron pair") we will calculate the matrix elements of
the $\S$-operator with respect to the following states of the free fields:
\begin{align*}
  \t{the initial state}:\  & I=\Vac\otimes |f\rangle, \\
  \t{the final state}:\  & F=|u,v\rangle\otimes \Vac,
\end{align*}
where $f$ denotes the photon's wave packet and $u,v$ denote the wave
packets of the electron and positron respectively (they may, for
instance,  be wave functions of some bound states). We obtain
\begin{equation*}
  (I,\S \ F)=\int d^4x \ g(x)\
  (\Vac, :\b \psi(x) \g^\m \psi(x): |u,v\rangle)\cdot
  (|f\rangle,\A_\m(x)\Vac)
\end{equation*}
with
\begin{equation*}
(|f\rangle,\A_\m(x)\Vac)=\frac{1}{\sqrt{2\pi}^3}
    \int \frac{d^3k}{\sqrt{2k^0}}\ \v e^\a_\m(\v k)
  \b f_\a (\v k) e^{-ikx}.
\end{equation*}
What remains  to be evaluated is the matrix element of the Wick product of
Dirac field operators:
\begin{equation*}
  (Dirac)\equiv (\Vac,:\b \psi_A(x)\psi_B(x):\ |u,v\rangle)\ \g^{AB\m},
\end{equation*}
where we will have to employ the local version of the Wick product. The
final state of the Dirac field is orthogonal to the initial state (the
vacuum), and therefore the distribution which has been chosen to
regularize $\b\psi(x)\psi(y)$ will have no influence on the result.
Consequently, the amplitude will turn out to be equal to the standard one
calculated with the help of usual formulations of external field QED
\cite{akh_ber}.
    We get
\begin{equation*}
 (Dirac)= \b{ v(t,\v x)}\
  \g^\m \ u(t,\v x),
\end{equation*}
where the classical Dirac Hamiltonian governs the time-evolution of the
wave packets. Finally,
\begin{equation*}
  (I,\S \ F)=\frac{1}{\sqrt{2\pi}^3} \int d^4x \ g(x)\ \int
  \frac{d^3k}{\sqrt{2k^0}}
  \  \b{ v(t,\v x)}\
   \dir{\v e}^\a(\v k) u(t,\v x)
  \ \b f_\a (\v k) e^{-ikx}.
\end{equation*}
Now we can explain, why it is possible to abandon the spatial smearing of
$g(x)$. The electron and positron wave packets are intended to be regular
wave packets\footnote{We may also take as $u$ or $v$ the bound state wave
functions (if such bound states exist in a given external field
background).}. Wave functions of such packets vanish faster than any
power of $1/|\v x|$ at spatial infinity. There is therefore no need to
further strengthen the decay behavior at spatial infinity. We take
$g(t,\v x)=1$ for each $t\in(0,T)$.

\subsection{Second-order processes, consequences of the redefinition of the
Wick product}

In free quantum electrodynamics on external backgrounds the electrons may
reside in bound states. Such states correspond to some discrete
eigenvalues of the free Dirac Hamiltonian which lie in the interval
$(-mc^2,mc^2)$. We shall investigate the process of annihilation of a
bound electron with a scattering state of a positron in a process, where
two photons are emitted.

We shall use the local version of the second-order time-ordered product
\eqref{wick_2_1}. As there will be two incoming fermions and two outgoing
photons, only the summands with the Wick product of two Dirac and two
electromagnetic field operators will have a non-vanishing expectation
value. In the case $y\nin J^+(x)$ they are
\begin{equation*}
   -e^2\left[\T H_{AD}(x,y) :\psi_B(x) \b \psi_C(y):+
   H_{BC}(x,y) :\b\psi_A(x)\psi_D(y):\right]\g^{\m AB}\g^{\n CD} \\
    :\A_\m(x)\A_\n(y):,
\end{equation*}
where we have already utilized the local version of the Wick product. The
initial state is defined to be a two-particle state: the ground electron
state and some scattering state of the positron:

\begin{equation*}
  |I\rangle=a^*_0 \ b^*(f) \Vac,
\end{equation*}
where
\begin{equation*}
  b^*(f)=\int d^3p\ f^\a(\v p)\ b^*_\a(\v p), \quad \a=1,2,\quad f\in
  P_-\H,
\end{equation*}
and the two polarization functions $f^\a$ are normalized by
\begin{equation*}
\int d^3p\ \b{f^\a}(\v p)  f_\a(\v p)=1
\end{equation*}
The outgoing state contains two photons of with certain polarizations:
\begin{equation*}
|F\rangle=|f_1f_2\rangle=\int d^3p\ d^3k \ f^\a_1(\v p) f^\be_2(\v
k)a^*_\a(\v p)a^*_\be(\v k)\Vac,
\end{equation*}
where $a^*_\a(\v k)$ denotes the creation operator for the $\a$-th
polarization.

The amplitude of the process at hand requires us to calculate the
following expectation value\footnote{Here we denote the electromagnetic
and the Dirac vacuum with the same letter, as this should cause no
confusion.}:
\begin{align}
  A_1&=(|I\rangle\otimes \Vac,T_2(x,y)\ \Vac\otimes |F\rangle)=\\
  &=\bigl(|I\rangle, :\b\psi^A(x) (\g^\m H_F \g^\n)_{AB}\psi^B(y):\ \Vac
  \bigr)\cdot (\Vac,:\A_\m(x)\A_\n(y):|F\rangle)
\end{align}
(the other summand of $T_2$ will be considered later; here $H_F$ denotes
the Feynman parametrix which is the Hadamard parametrix with $\G_F$). The
electromagnetic expectation value clearly gives
\begin{multline*}
(\Vac,:\A_\m(x)\A_\n(y):|F\rangle)=\frac{1}{2(2\pi)^3}\int \frac{d^3q_1\
d^3q_2}{\sqrt{|\v q_1||\v q_2|}}\ [\v e^\a_\m(\v q_1)\v e^\be_\n(\v q_2)\
e^{-i(q_1x+q_2y)}+x\leftrightarrow y]\*\\ \*f_{1,\a}(\v q_1) f_{2,\be}(\v
q_2).
\end{multline*}
The Dirac-field expectation value,
\begin{equation*}
(Dirac)\equiv(|I\rangle, :\b\psi^A(x) \psi^B(y):\ \Vac),
\end{equation*}
turns out again to be indifferent to the modification of the Wick product
(this is because the initial state is orthogonal to the final state). We
obtain
\begin{equation*}
  (|I\rangle,:\b \psi^A(x) \psi^B(y):\Vac)=\b u^A(x) \int d^3p
  \ f^\a(\v p) v^B(y,\v p),
\end{equation*}
where the time dependence of the wave functions again is governed by the
Dirac equation.

The amplitude of the process differs from that of usual formulations of
QED in that the Feynman parametrix $H_F(x,y)$ stays in place of the
Feynman propagator $S_F(x,y)$. The former is strongly local, the latter
is not. In order to calculate the amplitude one should use the expansion
of $H_F$ in powers of $\G=(x-y)(x-y)+i\e$, similar to that of the
parametrix described in chapter \ref{chapter:Local_Solutions}.

\section{Outlook}
The investigations of this thesis point into three directions which are
in our opinion good candidates for future research. In what follows we
shall describe them briefly.
\begin{enumerate}
\item The local definition of the current operator, eq. \ref{eq:LCD},
allows for a calculation of the expectation value of the current in some
given state of the free Dirac field. We expect a non-trivial result, if
this quantity is computed for a ground state of some external field
configuration. This bears some resemblance to the investigations of
Casimir effect in the framework of strongly local quantum field theory
\cite{fewster}.

\item The topic of (perturbative) physical properties of
interacting matter has only been touched by the investigations of this
chapter. It is important to understand, how the usual predictions of
external field quantum electrodynamics, derived for instance in chapter V
of the textbook of Akhiezer and Berestetski \cite{akh_ber}, are altered,
if we use the local theory developed in this thesis. Furthermore, a
detailed investigation of the dynamics of the excited states of an
electron in a hydrogen atom appears possible. Issues like the shift of
energies of the states or the time dependence of the excited state's
amplitude and phase  with today's methods are of direct experimental
significance.

\item One of the improvements our thesis has brought is the
possibility to handle time-varying external fields. They can be
investigated, because the states of the Dirac field on the surfaces of
constant time are not assumed to be globally equivalent to one another
anymore. Thus, if a deep potential well is turned on extremely quickly,
the state of the free Dirac field, which was the vacuum state in the past,
evolves into some state in the future\footnote{It would be premature to
call such a state excited, because in time-dependent situations there is
no preferred reference state.}. It would be important to know in what way
this picture is altered by quantum electrodynamic interaction. The
interaction of the radiation field with the Dirac field presumably
changes the final state of the Dirac field, but it is impossible to tell
without  calculation whether this change is significant. It is our hope
that models of this type can shed some new light on the process of
collapse of matter in the formation of black holes in the theory of
general relativity.
\end{enumerate}

\newpage
\begin{center}
{\bf Acknowledgements}
\end{center}

At this point I would like to express my gratitude towards my supervisor
Prof. K.Fredenhagen. He not only proposed the topic of my thesis (which
harmonized well with my previous experience and coincided with my
interests), but also guided my studies of quantum field theory. I would
also like to thank all those who, by pointing out my ignorance helped in
the development of this thesis. Specifically, the constant interest of my
friend N. Szpak is gratefully acknowledged. I am indebted to M.Porrmann,
who read the manuscript of this thesis carefully, and pointed out
numerous flaws of it. Particular thanks also to my wife Monika for her
support and the day-to-day sacrifices she made in order to help me.
Financial support of the DFG is also gratefully acknowledged.

\appendix
\chapter{The electromagnetic units}\label{units}

As is well known, it is possible to construct a dimensionless quantity out
of the unit of electric charge, $\hbar$ and the velocity of light:
\begin{equation}\label{137}
  \frac{e^2}{\hbar c}=\frac{1}{137.036}.
\end{equation}
Consequently, it is possible to eliminate all electrical artificial units
like Coulomb or Volt and express them in terms of mechanical "cgs" units
only. In Gauss units the attraction force of two charges is
\begin{displaymath}
  F=\frac{q_1q_2}{r^2},
\end{displaymath}
where the distance is measured in "cm" and charge in charge units i.e.
one electron possesses the charge of
\begin{displaymath}
  e=\sqrt{\frac{\hbar c}{137}}=4.803 \cdot 10^{-10} \sqrt{\frac{\text{g
  cm}^3}{\text{s}^2}}.
\end{displaymath}
The electromagnetic field potential, typically expressed in Volts, now has
the unit of
\begin{displaymath}
  [V]=\frac{\sqrt{\text{g cm}}}{s}.
\end{displaymath}

\section{Action of the Maxwell-Dirac electrodynamics}

The unit of action is the same as the unit of Planck's constant:
\begin{displaymath}
  [S]=[\hbar]=\text{erg s}=\frac{\text{g cm}^2}{s}.
\end{displaymath}
The Lagrange density $\Lag$
\[S=\int \text{dx$^0$ d$^3$x } \Lag,\]
where $x^0=c\cdot t$,  has the dimension of
\begin{displaymath}
  [\Lag]=\frac{\t{g}}{\t{cm$^2$ s}},
\end{displaymath}
which happens to be the dimension of the energy density. The Dirac field
has the dimension of
\begin{displaymath}
  [\psi]=\frac{1}{\sqrt{\text{cm}^3}},
\end{displaymath}
so that the probability density is measured in 1/cm$^3$:
\begin{displaymath}
  \p=\h{\psi}\psi.
\end{displaymath}

The action of the classical electrodynamics coupled to the classical Dirac
field, which
\begin{itemize}
\item contains the appropriate combination of physical constants, \item
has all the appropriate signs, \item leads to the Maxwell-Diraca system,
\end{itemize}
is
\begin{displaymath}
  S=\int \t{d$^4$x } \left\{[i\hbar \ \b \psi\g^a\d_a \psi-mc\ \b\psi\psi]
  +[\b\psi\g^a\psi \ \frac{e}{c}A_a]-
  \left[\frac{1}{16\pi \ c}F_{ab}F^{ab}\right]\right\}.
\end{displaymath}
It leads to the Maxwell-Dirac system
\begin{align*}
  \d_b F^{ba}(x) &=4\pi j^a(x), \\
  \g^a\left[i\hbar \d_a-\frac{e}{c}A_a(x)\right]\ \psi(x) &=mc\ \psi(x),
\end{align*}
where the electromagnetic current is
\begin{displaymath}
  j^a(x)=e \b \psi \g^a \psi.
\end{displaymath}

\section{$\hbar=1=c$, particular combinations of electromagnetic
quantities}

In order to use Plank's units, $\hbar=1=c$ we note that setting $c=1$ is
equivalent to $s=cm$. Subsequently, $\hbar=1$ means $g=1/cm$. In such a
way the electric charge unit becomes dimensionless. The interaction term
in the action and, therefore, the first order time-ordered product has the
dimension of
\begin{equation*}
[T_1]=\left[\b\psi\g^a\psi \ \frac{e}{c}A_a\right]=[1/cm^4],
\end{equation*}
due to
\begin{equation*}
  [A_a]=1/cm, \qquad [\psi]=1/cm^{3/2}.
\end{equation*}
The two-point function has the same dimension as the current density
\begin{equation*}
  [(\Vac,\b \psi \psi \ \Vac)]=1/cm^3.
\end{equation*}
Not surprisingly, all the dimensions above correspond to the rescaling
powers in the transition to the rescaled quantities discussed in chapter
\ref{section_scaling}.

\chapter{Microlocal analysis}\label{appendix:microlocal_analysis}
Microlocal analysis deals with pseudo-differential operators (\PDO),
distributions and their singularities. In this appendix we summarize the
most important definitions and results. The purpose of our presentation
is to exhibit the beauty and usefulness of microlocal analysis in a
readable way. We shall thus not put much emphasis on the correctness of
our notation; neither shall we present the most general versions of the
theorems (which sometimes obscure their immediate value).

\begin{defi} A wave front set at $x$, $WF_x(u)$, of a distribution $u$ is the closed,
conic set of directions in the tangent space at $x$ in which the Fourier
transform of $u$, localized at $x$, does not decay rapidly. In other words
\begin{equation*}
  k\nin WF_x(u) \Leftrightarrow \widehat{(\vp u)}(\la k) \t{ decays rapidly for
  }\la\ra \infty.
\end{equation*}
Here $\vp(y)$ is a $\dom$-function with support in a small neighborhood
around $x$. The wave front set $WF(u)$ is simply
\begin{equation*}
  WF(u)=\{(x,k):\ k\in WF_x(u)\}.
\end{equation*}
\end{defi}

\begin{defi}A pseudo-differential operator $A$ is a linear operator
defined via
\begin{equation*}
  (A\psi)(x)=\int dk\ a(x,k)\ e^{ikx}\ \widehat{\psi(k)},
\end{equation*}
the smooth function $a(x,k)$ is called a symbol of $A$. The \PDO \ is of
order $m$, if its symbol fulfills
\begin{equation*}
  \left(\frac{\d}{\d k}\right)^\be a(x,k)\leq C (1+|k|)^{m-\be}
\end{equation*}
for some constant $C=C(\be)$ and all $\be$. The derivatives with respect
to $x$ do not affect the decay behavior in $|k|$.
\end{defi}

\begin{remark}It follows from the above definition that the symbols
have to be $\smooth$ in both of their variables. Quite often, however, one
would like to use operators the symbols of which possess a discontinuity
at certain points (eg. symbols of the sort $|k|$ or $\sqrt{|\v
k|^2+m^2}$).
\end{remark}

\begin{defi} An asymptotic expansion of a symbol $a(x,k)$ is a sum of
symbols of the form
\begin{equation*}
p_m(x,k)+p_{m-1}(x,k)+p_{m-2}(x,k)+\ldots
\end{equation*}
such that
\begin{equation*}
a(x,k)-\sum_{n=m}^{-\infty} p_n(x,k)
\end{equation*}
is a symbol of order $-\infty$. Here $p_n(x,k)$ are symbols of order $n$
homogeneous in $k$. The first of those symbols, $p_m(x,k)$, is called the
principal symbol of $A$.
\end{defi}

\begin{defi}The \PDO\ $A$ is properly supported, if its distribution
kernel\footnote{From the Schwarz kernel theorem every \PDO\  may be
represented as \[(A\psi)(x)=\int dy\ A(x,y) \psi(y),\] where $A(x,y)$ is
called the distribution kernel of $A$.} $A(x,y)$ is such that $A(K,y)$ as
well as $A(x,K)$ for a compact $K$ have compact support\footnote{The
notation is obvious: $A(K,y)$ has compact support, if $A(x,y)$ has
compact support for all $x\in K$.}.
\end{defi}

The following theorems are of great practical importance:

\begin{Thm}[Pseudolocal property]\label{pseudolocal_property}
Let $A$ be a \PDO\ of any order $m$ and let $u$ be a distribution. Then
\begin{equation*}
  WF(Au)\subset WF(u).
\end{equation*}
\end{Thm}
\begin{Thm}[Propagation of singularities] \label{thm:propagation} Let $A$ be a \PDO\ of order $m$ which is properly
supported and  has a real principal symbol $p_m(x,k)$. If $u$ is a
distribution which solves the (in)homogeneous equation
\begin{equation*}
  Au=f
\end{equation*}for $f\in\dom$, then
\begin{enumerate}
\item \begin{equation*} WF(u)\subset a^{-1}(0)\setminus\{0\}=\{(x,k):
a(x,k)=0\}.
\end{equation*}
\item If a certain point $(x_0,k_0)$ is in the wave front set of $u$, then
also the whole Hamiltonian trajectory of this point lies in $WF(u)$. Such
a trajectory is derived from the Hamilton equations of motion with
$p_m(x,k)$ taken as the Hamiltonian.
\end{enumerate}
\end{Thm}

Quite often the wave front set of distributions has a form $(x,p,y,-p)$;
it is common to introduce a primed wave front set in order to get rid of
the minus sign:

\begin{defi}The primed wave front set of a bi-distribution is defined
via
\begin{equation*}
  WF'(x,k,y,p)=WF(x,k,y,-p).
\end{equation*}
\end{defi}

With that definition we have: $WF'[\de(x,y)]=(x,k,x,k)\subset
M\times M$.

\begin{defi} Let $u(x,y)$ be a bi-distribution, $x\in X,\ y\in Y$. Then we
introduce the notation
\begin{align*}
  WF_X(u) &=\{(x,k): (x,k,y,0)\in WF(u) \t{ for some }y\in Y\},\\
  WF_Y(u) &=\{(y,k): (x,0,y,k)\in WF(u) \t{ for some }x\in X\}.
\end{align*}
\end{defi}

\begin{Thm}[Composition of bi-distributions] Let $u,v$ be properly
supported bi-distributions on $X\times X$. The composition of them
\begin{equation*}
  (u\circ v)(x,z)=\int dy\ u(x,y)v(y,z),
\end{equation*}
exists if\footnote{Here $WF'_2(u)$ denotes $WF'_X(u)$ in the second
variable.}
\begin{equation*}
  WF'_{2}(u)\cap WF_{1}(v)=\emptyset.
\end{equation*}
In such a case we have
\begin{align*}
  WF'(u\circ v)&\subset WF'(u)\circ WF'(v)\ \cup\\
  &\cup \ \left[WF_1(u)\times(X\times\{0\})\right]\ \cup \ \left[
  (X\times \{0\})\times WF'_2(v)\right],
\end{align*}
where the composition of wave front sets above means:
\begin{equation*}
(x,k,y,p)\in WF'(u)\circ WF'(v)\Leftrightarrow \exists (z,q):\
(x,k,z,q)\in WF'(u) \t{ and } (z,q,y,p)\in WF'(v).
\end{equation*}

\end{Thm}

\chapter{Quantum Dirac field in the absence of any external potentials}
\label{swobodne pole diraca}
In the present appendix the free Dirac field in the absence of any external
potentials will be quantized. The quantization procedure contains two steps:
\begin{itemize}
    \item Defining the CAR algebra.
    \item Constructing a representation of the CAR.
\end{itemize}
We will present both of them in a somewhat heuristic manner.

\section{CAR Algebra}
Consider the mode decomposition of the classical Dirac field:
\begin{equation}\label{operator_pola_Diraca}
  \psi^A(x)=\frac{1}{\sqrt{2\pi}^3} \int d^3p\ \left\{u^A_s(\v p)\ a^s(\v p)\  e^{-ipx}+
  v^A_s(\v p)\ b^{*s}(\v p)\ e^{ipx}\right\}.
\end{equation}
 Here $s=1,2$  are connected with the orientation of
the spin of the electron/positron (they enumerate the basis spinors), and
$u^A(\v p)$, $v^A(\v p)$ are the basis spinors\footnote{Capitular letters
denote spinor indices.}; $a^s(\v p)$ and $b^{*s}(\v p)$ are ordinary
functions. There holds
\begin{align*}
  \sum_s  |u_s(\v p)\rangle\langle  u_s(\v p)|&= P^+(\v p),\\
  \sum_s  |v_s(\v p) \rangle\langle v_s(\v p)|&= P^-(-\v p),
\end{align*}
which in spinorial indices means
\begin{align}\label{operatory_rzutowe}
  \sum_s  u_s^A(\v p)  u^{+B}_s(\v p)&= P^{+AB}(\v p),\\
  \sum_s  v_s^A(\v p)  v^{+B}_s(\v p)&= P^{-AB}(-\v p).
\end{align}
The operators $P^\pm(\v p)$ project onto the
positive/negative-frequency\footnote{The frequency is the eigenvalue of
the Hamiltonian
\begin{equation*}
    H=\g^0(i\g^i\d_i+m).
\end{equation*}}
subspaces of $\H$. They can therefore also be written as
\begin{equation*}
    P^\pm(\v p)=\frac{1}{2p_0}(p_0\pm H(\v p)),
\end{equation*}
where $p_0=|H(\v p)|$. The basis spinors supplied with the Fourier factor
\begin{equation*}
u(x,\v p)=e^{-ipx}u(\v p), \qquad\qquad v(x,\v p)=e^{ipx}v(\v p),
\end{equation*}
are solutions of the classical Dirac equation:
\begin{equation*}
  (i\g^a\d_a -m)\psi=0.
\end{equation*}
From this equation we infer
\begin{equation*}
  \g^0 (2p_0) \ P_-(\v p)\ u(\v p)=0
\end{equation*}
as well as
\begin{equation*}
  \g^0 (2p_0) \ P_+(-\v p)\ v(\v p)=0.
\end{equation*}
Thus
\begin{align*}
  P_+(\v p) v(-\v p)&=0,\\
  P_-(\v p) u(\v p)&=0,
\end{align*}
which is in accordance with $P^\pm$ being orthogonal projections. The
basis spinors fulfill also:
\begin{align*}
(u^s,u^r)&=\de^{sr},\\
(v^s,v^r)&=\de^{sr}.
\end{align*}
$(.,.)$ denotes here the classical scalar product
\begin{equation*}
  (\psi,\psi)=\int d^3x \ \b \psi \g^0 \psi=\int d^3x \  \psi^* \psi.
\end{equation*}

Thus far we have used the mathematical structure of the classical Dirac
field. There is a Hilbert space $\H$ of four-component spinors, which are
square-integrable with respect to the scalar product $(.,.)$ given above.
The basis spinors are eigendistributions which correspond to the
continuous spectrum of the free Dirac operator.

The expression \eqref{operator_pola_Diraca} is a basis for a more common
quantization of the Dirac field. One introduces the ("sharp")
creation/annihilation operators in place of the ordinary functions
(Fourier components) $a(\v p)$ and $b(\v p)$. They are objects which
fulfill the following anti-commutation relations,
\begin{align*}
  \{ a_s(\v p), a^*_r(\v q)\}=\de_{rs}\de(\v p - \v q),\\
  \{ b^*_s(\v p), b_r(\v q)\}=\de_{rs}\de(\v p - \v q).
\end{align*}
When smeared out with square-integrable test functions
\begin{align*}
    a_s(f)&=\int d^3p \ P_+(\v p) f(\v p) a_s(\v p),\\
    b_s(f)&=\int d^3p \ P_-(\v p) f(\v p) b_s(\v p),
\end{align*}
they fulfill
\begin{align*}
  \{ a_s(f), a^*_r(g)\}=\de_{rs}(f_+,g_+),\\
  \{ b^*_s(f), b_r(g)\}=\de_{rs}(g_-,f_-).
\end{align*}

The anti-commutators  between $a$ and $b$ are assumed to vanish. The Fock
space is constructed in that one assumes that there is a vacuum $\Vac$:
\begin{equation*}
    a(f)\Vac=0,\qquad b(f)\Vac=0,
\end{equation*}
and one constructs the many-particle subspaces by a successive application
of the creation operators on $\Vac$.

At this point the representation of the CAR algebra $\Alg$, that is the
algebra of the polynomials of the smeared-out field operators
\eqref{operator_pola_Diraca}, has been constructed. We may verify the CAR
property explicitly
\begin{multline*}
  \{\psi(\v x), \psi^* (\v y)\}=\frac{1}{(2\pi)^3} \int d^3p \sum_{s}
  \left\{u^s(\v p) u^{*}_s(\v p) e^{i\v p(\v x-\v y)}+
  v^s(\v p) v^{*}_s(\v p) e^{i\v p(-\v x+\v y)}\right\}=\\=\frac{1}{(2\pi)^3} \int d^3p
  \left\{P_+(\v p) + P_-(\v p)\right\} e^{i\v p(\v x-\v y)}=\de(\v x - \v y),
\end{multline*}
where in the positron part the change of variables $\v p\ra  -\v p$ was
performed. Therefore,
\begin{equation}\label{antykomutacja_pola_swobodnego}
  \{\psi(\v x), \psi^* (\v y)\}= \de(\v x - \v y),
\end{equation}
which can be written as
\begin{equation*}
  \{\psi(f), \psi^* (g)\}= (f,g),
\end{equation*}
where as before $(.,.)$ denotes the scalar product in the classical
Hilbert space $\H$ (this is the analogue of \eqref{CAR_t}).

\chapter{Model of the spontaneous atomic emission of
light}\label{app:spontaneous}

This appendix contains a field-theoretical model of the emission of
radiation from an excited state of an atom. Although this problem has been
attacked by many authors in the past we believe that our model describes
 this fundamental process in a better way. A huge part of what has been done on
this subject relies on the perturbation theory w.r.t. the interaction of
the electromagnetic field with the electron under consideration. Such a
treatment can only give reliable information about the beginning of the
decay process, but that is not what we are aiming at. On the other hand, a
different approach due to V.Weisskopf and E.Wigner establishes a closed
integro-differential equation for the time-dependent amplitude of the
excited-state part of the electron's wave function. Our model is a
refinement of a version of this approach which has been described in the
textbook of M.Scully and M.Zubairy\footnote{The original paper of
V.Weisskopf and E.Wigner is not easily accessible (published in 1930). It
makes use of the methods of the early days of quantum mechanics. The
modern expositions of that approach often use dubious mathematical tricks
as in \cite{SZ}.} \cite{SZ}. Our model goes further because:
\begin{itemize}
\item it is constructed with the help of the field-theoretical methods,
describes a unitary time evolution of the system;

\item in the derivation it is possible to keep a freedom of the initial
state of the radiation field. It can be recognized at the end that the
spontaneous decay  is caused directly by the two-point (autocorrelation)
function of this state.
\end{itemize}
In particular, we regard the second point to be very important. In recent
years many attempts to experimentally alter the spontaneous emission
process have been devised  (eg. the presence of a squeezed state
\cite{kimble} or a light reflecting cavity). Those experimental setups
influence directly the autocorrelation function of the electromagnetic
field, and thus can be easily investigated in our framework.

The spontaneous emission of radiation is important for the general context
of this thesis for the following reasons. Firstly, it provides \emph{a
testing device, a detector} of the state of the quantum field. Indeed,
with a help of a single isolated atom it is possible, for instance, to
investigate the properties of the KMS (thermal) states of the radiation
field. With an ensemble of such atoms it is even possible to measure its
temperature \cite{pi_krak}. The model can be of help in the  investigation
the sub-vacuum fluctuations of certain exotic states of the radiation
field. The other issue is that our model of the spontaneous emission
gives evidence of what happens in an interacting field theory. Apart from
the ground state (whose dynamics are also altered) all other "excited
states" of the free Dirac field are unstable in the interacting theory.
Accordingly, one should exercise caution with the claims about the
properties of the excited states and their energy shifts \cite{mohr,bach}.

The problem of an interaction of non-relativistic bound-states with the
radiation field has been investigated with great rigor by V.Bach and
collaborators (see for instance \cite{bach})\footnote{We are grateful to
K.Fredenhagen for pointing out this reference to us.}. In comparison to
those investigations our model is a modest attempt - a special case at
most. It provides, however, a relatively simple evolution equation which
gives an insight into the time dependence of the system. In our opinion
such an insight complements the investigations of V.Bach.

\section{Hilbert space and the interaction}
Consider a system consisting of an electron with two bound states
(non-relativistic, described by quantum mechanics)  coupled to the quantum
radiation field. The natural Hilbert space for such a system is
\begin{equation*}
    \H=L^2(\R^3)\otimes \Fou,
\end{equation*}
where $\Fou$ denotes the physical (transversal) Fock space of the Maxwell
field. Later we will restrict this Hilbert space to its subspace $\H_1$
spanned by the vectors
\begin{align*}
\psi_1(\v x)&\otimes \Vac, & \psi_0(\v x)\otimes |f\rangle,
\end{align*}
where $\psi_{0/1}$ denote the wave functions of the two bound states,
$\Vac$ is the electromagnetic vacuum, and $|f\rangle$ denotes the
one-photon state
\begin{equation*}
|f\rangle=a^*(f)\Vac=\int d^3p\ f^\a(\v p)a_\a^*(\v p)\ \Vac.
\end{equation*}
Here $f_\a(\v p)\in\mathcal{S}$ is a test function which describes the
$\a$-th polarization component of the photon's wave packet.

In order to describe the interaction we introduce the interaction
Hamiltonian\footnote{In what follows we shall use the units $\hbar=1=\v
c$ (the speed of light in this chapter is denoted by a bold $\v c$). We
shall also omit the factor $-e/m\v c$ for brevity. We note that in this
chapter $p$ means $p_0=\v c |\v p|$. The correct factors as well as the
atomic units will be adapted at the end of the calculation.},
\begin{equation}\label{equ:interaction}
  V(t,\v x)=-\frac{e}{m\v c}\ \v p_i\otimes A^i(t,\v x),
\end{equation}
where $\v p_i=-i\hbar \d_i$ and $A^i(t,\v x)$ denotes the electromagnetic
field operator in the radiation gauge with the free time evolution already
implemented\footnote{Which means that the time dependence of $A^i$ is
generated by the purely electromagnetic Hamiltonian $\int d^3p\ p_0 \h a
(\v p) a(\v p)$.}. This is a standard interaction Hamiltonian describing
the interaction of the non-relativistic systems with the electromagnetic
field under the assumption that  $A^2$ can be neglected, what we hereby
also assume. The electromagnetic field operator is expressed in terms of
the creation/annihilation operators (cf. chapter \ref{chapter_EM}):
\begin{equation*}
     A_i(x)=\frac{1}{\sqrt{2\pi}^3} \int \frac{d^3k}{\sqrt{2k^0}}
  \  e^\a_i(\v k)
  \left\{ a^*_\a (\v k)\ e^{ikx}+a_\a (\v k)\ e^{-ikx}\right\}.
\end{equation*}

We adopt the interaction picture w.r.t. the free electronic Hamiltonian
which is given by $\v p^2/2m$, the unperturbed energies of the states
$\psi_{0,1}$ will be denoted by $E_{0,1}$. In order to develop the model
we now restrict the Hilbert space and the Hamiltonian to the subspace
$\H_1$. By doing so, we shall obtain a closed quantum system with a
bounded, selfadjoint Hamiltonian. It is another matter to what extent our
restriction describes the physical situation well. We expect that, as long
as $\psi_{0,1}$ are the lowest two bound states which differ by $\pm 1$
in angular momentum, our assumptions are reasonable.

We start by writing the time-dependent state vector $S(t)$, which is an
element of $\H_1$, with the help of $c(t)$ and $f_t(\v p)$:
\begin{equation*}
S(t)=c(t)\ \psi_1(t,\v x)\otimes \Vac+\psi_0(t,\v x)\otimes |f_t\rangle.
\end{equation*}
The interaction \eqref{equ:interaction} leads to the following evolution
equation
\begin{align*}
  i&\dot c(t)\ e^{-iE_1t}\ \psi_1\otimes \Vac+i e^{-iE_0t}\ \psi_0\otimes
  |\dot f_t\rangle=\\&
  c(t) \ e^{-iE_1t}\  \v  p_i\ \psi_1\otimes
  |A^i(t,\v x)\Vac\rangle+e^{-iE_0t}\ \v p_i\ \psi_0\otimes
  |A^i(t,\v x)f_t\rangle.
\end{align*}
Here, however, what stays on the RHS does not belong to $\H_1$. We
restrict the RHS to $\H_1$ in order to obtain a closed system. We do it by
contracting (taking the scalar product) the above equation with the
vectors $\psi_1\otimes \Vac$ and $\psi_0\otimes |a_\be(\v k)\Vac\rangle$
separately\footnote{The operator $a(\v k)$ is taken sharp here; however,
this is allowed, as the use of $a(\v k)$ still leads to expressions which
make sense as distributions.}. We obtain
\begin{equation*}
  i\dot c(t)=e^{i\w t}\ \big(\psi_1, a^{i}_f(t,\v x)\ \v p_i\psi_0 \big),
\end{equation*}
where\footnote{This comes from the contraction of the negative-frequency
part of the electromagnetic field operator with the $f_t(\v p)$-smeared
creation operator.}
\begin{equation*}
  a^i_f(t,\v x)=(\Vac,A^i(t,\v x)a^*(f)\Vac)=\int \frac{d^3p}{\sqrt{2p}}
  \ e^{-ipt+i\v p\v x}\
  f^\a_t(\v p) e^i_\a(\v p),
\end{equation*}
and $\w=E_1-E_0$. In the calculation we have used the fact that
$(\Vac,A^i\ \Vac)=0$, which is true for any quasi-free state $\Vac$, in
particular also for the vacuum.

The second equation reads:
\begin{align*}
i\dot f_{\be\ t}(\v k)&= c(t)e^{-i\w t}\ \bigl(\psi_0,(\Vac,a_\be(\v
k)A^{i}(t,\v x)\ \Vac)\
\v p_i\psi_0\bigr)\\
&=c(t)e^{-i\w t}\left(\psi_0,\frac{e^{ikt-i\v k\v x}\ e^i_\be(\v
k)}{\sqrt{2k}}\ \v p_i\psi_0\right),
\end{align*}
here we have again utilized the assumption that $\Vac$ is a quasi-free
state: the expectation value of an odd number of creation/annihilation
operators vanishes for such states. The particular form of the operator
in the bracket above,
\begin{equation*}
\frac{e^{ikt-i\v k\v x}\ e^i_\be(\v k)}{\sqrt{2k}},
\end{equation*}
is true only, if $(\Vac,a_\be(\v k)\h a_\a(\v p)\ \Vac)=\de(\v p -\v
k)\de_{\a\be}$, i.e. if $\Vac$ is the vacuum. For other states $\Vac$ we
would get some other expression, but the analysis which follows proceeds
analogously.

We introduce some convenient abbreviations:
\begin{align*}
\chi_i(\v p)&=\int d^3x\ e^{-i\v p\v x}\ \b \psi_1(\v x)\v p_i \psi_0(\v x),\\
\vp_i(\v p)&=\int d^3x\ e^{i\v p\v x}\ \b \psi_0(\v x)\v p_i \psi_1(\v x).
\end{align*}
If the external field which binds the electrons is smooth, then it follows
from the elliptic regularity that $\chi_i(\v p)$ is a smooth function of
rapid decay. The vector index $i$ of $\chi_i,\vp_i$ is always contracted
with polarization vectors $e^i_\a(\v p)$. If only the transversal photon
polarizations are emitted (thus $\a=1,2$), then all the polarization
vectors are orthogonal to the vector $\v p_i$:
\[\v p_i\ e^i_\a(\v p)=0.\]
Then
\begin{equation*}
  \vp_i(\v p)=\b {\chi_i(\v p)}.
\end{equation*}

The system of differential equations that needs to be solved is
\begin{align*}
  i\dot c(t) &=e^{i\w t} \int \frac{d^3p}{\sqrt{2p}} \ e^{-ipt}\
  \chi_i(\v p)\ e^i_\be(\v p) \ f^\be_t(\v p), \\
  i\dot f_{\be,t}(\v p) &=c(t)e^{-i\w t}\frac{e^{ipt}}{\sqrt{2p}}\
  \b{\chi_i(\v p)}\ e^i_\be(\v p).
\end{align*}
It is not difficult to see that it is the Schr\"odinger equation on $\H_1$
with the restricted interaction Hamiltonian
\begin{equation*}
  H\upharpoonright_{\H_1}=P V P,
\end{equation*}
where $P$ is the projection from $\Fou$ onto $\H_1$. Moreover, the
restricted Hamiltonian is bounded (because the function $\chi(\v p)$
decays rapidly for large $\v p$) and symmetric, therefore selfadjoint.

The initial value condition (no one-photon component at $t=0$) is
\begin{equation*}
  f_{t=0}(\v p)=0, \quad c(0)=1.
\end{equation*}
The equation for $f_{\be,t}(\v p)$ may be integrated:
\begin{equation*}
f_{\be\ t}(\v p)=-i\int_0^tds\ c(s)e^{-i\w s }\frac{e^{ips}}{\sqrt{2p}}\
\vp_i(\v p)\ e^i_\be(\v p).
\end{equation*}
Upon insertion into the first equation of the system, one obtains
\begin{equation}\label{eq}
  \dot c(t)=-e^{i\w t}\int_0^tds\ c(s) e^{-i\w s}S(t-s),
\end{equation}
where
\begin{equation*}
  S(t-s)=\int\frac{d^3p}{2p}e^{-ip(t-s)}\left(\de^{ij}-\frac{\v p^i\v p^j}{\v p^2}\right)\
  \b{\chi_i(\v p)} \chi_j(\v p)
\end{equation*}
is the $\chi_i$-smeared transversal two-point function\footnote{We have
made use of $\b \chi =\vp$. The term $(\de^{ij}-\v p^i\v p^j/\v p^2)$
comes from the contraction of two transversal polarization vectors:
$\sum_\a e^i_a(\v p) e^j_a(\v p)$. The assumption on the transversality of
the emitted radiation leads to this factor.}. The equation \eqref{eq}
together with the initial condition $c(0)=1$ describes the atomic emission
of light. The equation \eqref{eq} is well-defined. The function $S(t-s)$
in the limit $s\rightarrow t$ gives a finite number, as $\chi$ is a smooth
function\footnote{Thus the infrared point $p=0$ is an integrable
singularity.} of rapid decay\footnote{The decay property makes the
ultraviolet part $p\ra\infty$ harmless.}.

We also note, that $S(t-s)$ decreases for large $t-s$; this is important
in the investigation of the behavior of $c(t)$ for large $t$.

The equation \eqref{eq} may also be brought to a more familiar form. We
may integrate it, namely, from zero to $T$ and (with an appropriate change
of variables) obtain
\begin{equation*}
    c(T)=1-\int_0^T Z(T-s) c(s)\ ds, \qquad
\end{equation*}
where
\begin{equation*}
    Z(\tau)=\int_0^\tau dt\ S(t) e^{-i \w t}.
\end{equation*}
The above function is a smooth function of its argument. Therefore we
remain with a task of solving the Volterra type integral equation of
second type with a smooth kernel.

\section{Atomic units, comments, outlook}
\begin{figure}\centering
\includegraphics{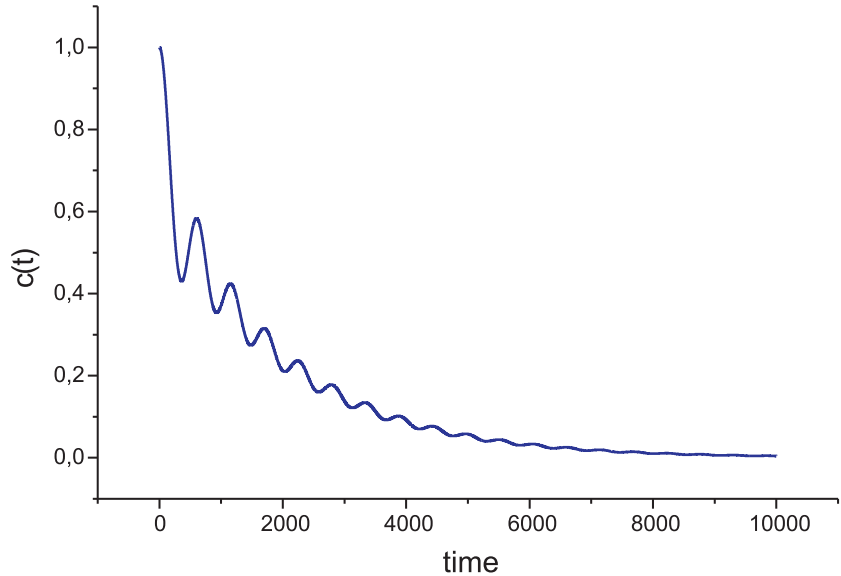}
\caption{Spontaneous emission of light from an atom. The figure presents a
numerical solution for some artificial values of various parameters. The
presence of oscillations at short times accompanied by decay for large
times appears to be a general feature of the model (the number of visible
oscillations and the decay rate are
parameter-dependent).}\label{fig:emisja}
\end{figure}

It is difficult to say anything general about the solutions of equation
\eqref{eq}. The numerical experience with it shows that the solutions
seem to oscillate on a short time scale and then later decay to zero.
Figure \ref{fig:emisja} shows one of the solutions (for some artificial
data, however).

In order to investigate the concrete atomic emission processes we
introduce atomic units. Formally this means that we set $e,m,\hbar$ equal
to one and note that

\begin{itemize}
\item $r=1$ means $r=0.53\*10^{-8}\ cm$ (N.Bohr radius),

\item $E=1$ means $E=27.2\ eV$ (typical atomic ionization energies),

\item $t=1$ means $t=2.4\*10^{-17}\ s$ (typical atomic oscillation times),

\item $c=137$ (result of the relation $e^2/\hbar c=1/137$).
\end{itemize}

With those units it is possible to investigate realistic systems. One
realizes that the only non-trivial constant is the speed of light, $\v c$
which appears only in the expression of the smeared two point function
$S(t-s)$. We have
\begin{equation*}
  S(t-s)=\int\frac{d^3p}{2\v c p}\ e^{-ip\v c (t-s)}\left(\de^{ij}-\frac{\v p^i\v p^j}{\v p^2}\right)\
  \b{\chi_i(\v p)} \chi_j(\v p).
\end{equation*}
It is not difficult to find the explicit expressions for $\chi_j(\v p)$ in
case of the harmonic oscillator or the hydrogen atom wave functions. The
numerical analysis, however, stumbles upon the following difficulties:
\begin{itemize}
\item The typical, experimental, decay times in case of the $2P\ra 1S$
transition are of the order of nanoseconds. Such a time corresponds to
$10^{8}$ in atomic units. This raises the question of the propagation of
the numerical errors.

\item The appearance of the speed of light in the function $S(t-s)$ makes
it both small and supported only for very short $|t-s|$. This even
further worsens the problem, as the time step must be accordingly small.
\end{itemize}

Those difficulties, however, could have been anticipated from the very
beginning: the model constructed here describes equally well the emission
of X-rays from the atomic nuclei and the emission of {rf} radiation from
the oscillatory levels of molecules. In terms of the decay times those
processes are separated by a vast gap of more than ten orders of
magnitude.

In the case of the hydrogen atom bound states $1S$ and $2P$ the smeared
electromagnetic two-point function is:
\begin{equation*}
 S(d)=\frac{16\pi}{3\v c}\int_0^\infty p\ dp\ e^{-ip\ d\v c}\left[\frac{A^2-p^2}
    {(A^2+p^2)^3}\right]^2,
\end{equation*}
where $A=3/2$, all in atomic units.

 The main merit of the model
constructed above (apart from its well-defined mathematical status) lies
in the appearance of the two-point function of the "base" state of the
radiation field on the right-hand side of equation \eqref{eq}. Thus, it is
clear that the "vacuum fluctuations" cause the decay of the atomic energy
levels. Indeed it is relatively easy in the presented framework to
investigate the spontaneous emission in the presence of a thermal state or
a squeezed state of the radiation field.

For the latter type of states such an investigation may have important
consequences, namely, it could clarify the issue why the squeezed states
produce atomic spectra with lines narrower than the natural line
width\footnote{This arises presumably, because the squeezed states exhibit
a sub-vacuum level of fluctuations of the electromagnetic field. However,
the periods of reduced fluctuations seem to last no longer than $10^{-14}
s$ (they are bounded by quantum inequalities \cite{pi}), which is far less
than the usual spontaneous emission time of the energy levels.}
\cite{kimble}.

\chapter{GNS construction and thermo-field dynamics}\label{Thermo_field}

In the chapter devoted to the construction of  representations of the CAR
algebra $\Alg$ we have indicated that interesting things occur, if the
integral kernel $B$ of the two-point distribution $\w$ is not a
projection.

    We shall consider a simple example of this here. Let $\H=\mathbb C^2$ and
\begin{equation*}
B=\begin{pmatrix}
  1 & 0 \\
  0 & 1/2 \\
\end{pmatrix}.
\end{equation*}
If we look for the Gelfand ideal in the one-field operator subalgebra of
$\Alg$, we will find that
\begin{equation*}
    \psi\left[(\a,0)\right]
\end{equation*}
is the only (one-field) annihilation operator present there.

In order to obtain a representation of $\Alg$ we may pretend that we have
the creation/annihilation operators, as usual, with the vacuum:
\begin{align*}
a(\sqrt B f)\ \Vac&=0,\\
b(\sqrt{1-B} f)\ \Vac&=0,
\end{align*}
but those operators  will in general no longer be elements of $\Alg$. In
the standard Fock space built with the help of $\h a,\h b$'s we may
construct the representation of $\Alg$ via
\begin{equation*}
    \psi(f)=a(\sqrt B f)+b^*(\sqrt{1-B} f).
\end{equation*}
The representation defined as above turns out, however, to be reducible;
there are operators different from the identity which act in the
representation space and commute with the representation of $\Alg$ given
above.

Much better insight into the nature of the problem is gained, if we extend
$\H$ to $\H_2=\H\otimes \H$ and choose
\begin{equation*}
    B_2=\begin{pmatrix}
      B & \sqrt B \sqrt{1-B}\\
      B \sqrt{1-B} & 1-B \\
    \end{pmatrix},
\end{equation*}
which is a projection operator on $\H_2$ irrespective of $B$. We shall
now search for a representation of a doubled algebra generated by the
elements $\Psi(f)$, $f\in \H_2$, and their adjoints. In our example
\begin{equation*}
    B_2=\begin{pmatrix}
      1 & 0 & 0 & 0 \\
      0 & 1/2 & 0 & 1/2 \\
      0 & 0 & 0 & 0 \\
      0 & 1/2 & 0 & 1/2 \\
    \end{pmatrix}.
\end{equation*}
If we introduce an orthonormal basis of $\H_2$:
\begin{align*}
    e&=(1,0,0,0),\\
    \T e&=(0,0,1,0),\\
    u&=(0,1,0,1)/\sqrt{2},\\
    v&=(0,1,0,-1)/\sqrt{2}.
\end{align*}
then
\begin{equation*}
    B_2=P_e+P_u,
\end{equation*}
where $P$'s denote the respective 1-dimensional projections. The Gelfand
ideal reveals four annihilation operators:
\begin{align*}
a_1&=\Psi(e), & b_1&=\h\Psi(\T e),\\
a_1&=\Psi(u), & b_1&=\h\Psi(v),
\end{align*}
which obey the standard anti-commutation relations among themselves. An
irreducible representation of the doubled algebra is thus given on the
Fock space constructed with the help of the adjoints of the above
operators. What is interesting is that on such a Hilbert space there acts
also the representation $\pi(\Alg)$ of the original algebra $\Alg$ with
\begin{equation*}
    \psi(f)\equiv\Psi[(f_+,f_-,0,0)]=
    f_+\ a_1+\frac{f_-}{\sqrt{2}}\ (a_2+\h b_2).
\end{equation*}
But in the representation space there act also the operators of the form
\begin{equation*}
    \T \psi(g)\equiv\Psi[(0,0,g_+,g_-]=
    \frac{g_-}{\sqrt{2}}\ (a_2-\h b_2)+g_+\ \h b_1.
\end{equation*}
This field and also its adjoint anti-commute with the representation of
$\Alg$. Thus, the observables constructed with their help\footnote{The
observable algebra is the algebra of even (i.e. containing even number of
field operators) elements of $\Alg$.} all commute with the elements of
$\pi(\Alg)$.

The doubling of $\H$ (and in turn of $\Alg$) is frequently done in the
literature on "thermo field dynamics". Further discussion of the
representation theory constructed on a thermal two-point function can be
found in \cite{Haag} chapter V.1.4 , or \cite{fre_cst}.

\bibliographystyle{amsalpha}
\bibliography{bibl}

\providecommand{\bysame}{\leavevmode\hbox to3em{\hrulefill}\thinspace}
\providecommand{\MR}{\relax\ifhmode\unskip\space\fi MR }
\providecommand{\MRhref}[2]{%
  \href{http://www.ams.org/mathscinet-getitem?mr=#1}{#2}
}
\providecommand{\href}[2]{#2}
\begin{thebibliography}{BLP82}

\bibitem[AB65]{akh_ber}
A.I. Akhiezer, and V.B. Berestetski, \emph{Quantum
  electrodynamics}, Interscience Publishers, 1965.


\bibitem[Ara71]{araki_CAR}
H.~Araki, \emph{On quasifree states of CAR and Bogoliubov automorphisms}, Publ.
  RIMS Kyoto Univ. \textbf{6} (1970/71), 385--442.


\bibitem[BFS99]{bach}
V.~Bach, J.Fr\"ohlich and I.M.~Sigal, \emph{Spectral analysis for systems
of atoms and molecules coupled to the quantized radiation field}, Commun. 
Math. Phys. \textbf{207} (1999), 249 



\bibitem[Ble50]{bleuler}
K.~Bleuler, \emph{Eine neue Methode zur Behandlung der longitudinalen und
  skalaren Photonen}, Helv. Phys. Acta \textbf{23} (1950),  567.

\bibitem[BLP82]{Ber_QED}
V.B. Berestetski, E.M. Lifshitz, and L.P. Pitaevskii, \emph{Quantum
  electrodynamics, vol 4. of Landau and Lifshitz's course of theoretical
  physics}, Pergamon Press, 1982.



\bibitem[BF00]{BF}
R.~Brunetti, and K.~Fredenhagen, \emph{Microlocal analysis and interacting
quantum field theories: Renormalization on physical backgrounds}, Commun. 
Math. Phys. \textbf{208} (2000), 623 and {\tt www.arXiv.org} \textbf{math-ph} 
(2000), 9903028.




\bibitem[BFV01]{BFV}
R.~Brunetti, K.~Fredenhagen, and R.Verch, \emph{The generally covariant locality principle - a new 
paradigm for local quantum field theory}, {\tt www.arXiv.org} \textbf{math-ph} 
(2001), 0112041.




\bibitem[Deh90]{dehmelt}
H.~Dehmelt, \emph{Experiments with an isolated subatomic particle at rest},
Rev. Mod. Phys. \textbf{62} (1990), 525.




\bibitem[DB60]{DeWittBrehme}
B.~DeWitt and R.~Brehme, \emph{Radiation damping in gravitational field}, Ann.
  Phys. \textbf{9} (1960), 220.

\bibitem[DM75]{dosch_mueller}
H.G. Dosch, and V.F. M\"uller, \emph{Renormalization of QED in an arbitrary
  strong time independent external field}, Fort. Phys. \textbf{23} (1975), 661.

\bibitem[FV01]{FV}
C.~Fewster, and R.Verch \emph{A quantum weak energy inequality
for Dirac fields in curved spacetime}, 
Commun. Math. Phys. \textbf{225}, 331 (2002) 
  and {\tt www.arXiv.org} \textbf{math-ph} (2001),
  0105027.


\bibitem[Fe03]{fewster}
C.~Fewster, \emph{Quantum energy inequalities and the Casimir effect}, Notes from
a talk given at DESY, June (2003).



\bibitem[Fre99]{fre_cst}
K.~Fredenhagen, \emph{Quantenfeldtheorie in gekr\"ummter Raumzeit}, Notes from
a lecture held at  Hamburg University, available at {\tt www.desy.de/uni-th/lqp},
  SS 1999.

\bibitem[Fre94]{fre_super}
K.~Fredenhagen, \emph{Superselection sectors}, Notes from
a lecture held at  Hamburg University, available at {\tt www.desy.de/uni-th/lqp},
  WS 1994.



\bibitem[Fri74]{friedlander}
F.~G. Friedlander, \emph{The wave equation on a curved spacetime}, Cambridge
  University Press, 1974.

\bibitem[GZ80]{gervais}
J.-L. Gervais, and D.Zwanziger, \emph{Derivation of the long range QED from the first
principles}, Phys. Lett. \textbf{B94} (1980),  389.


\bibitem[Gup50]{gupta}
S.N. Gupta, \emph{The theory of longitudinal photons in quantum
  electrodynamics}, Proc. Phys. Soc. \textbf{A63} (1950), 681.

\bibitem[Ha96]{Haag}
R.~Haag, \emph{Local quantum physics: fields, particles, algebras}, 2-nd ed., Springer, 1996.

\bibitem[HC66]{hilbert_courant2}
D.~Hilbert and R.~Courant, \emph{Methods of mathematical physics}, vol.~2,
  Interscience Publishers, 1966.

\bibitem[Hol99]{hollands_dirac}
S.~Hollands, \emph{The Hadamard condition for Dirac fields and adiabatic states
  on Robertson-Walker spacetimes}, Commun. Math. Phys. \textbf{216}, 635 (2001) 
  and {\tt www.arXiv.org} \textbf{gr-qc} (1999),
  9906076.

\bibitem[HW1]{HW1}
S.~Hollands, and R.M. Wald, \emph{Local Wick polynomials and time ordered products
of quantum fields in curved spacetime}, Commun. Math. Phys. 
\textbf{223}, 289
 {\tt www.arXiv.org} \textbf{gr-qc} (2001),
  0103074.

\bibitem[HW2]{HW2}
S.~Hollands, and R.M. Wald, \emph{Existence of local covariant time ordered products 
of quantum fields in curved spacetime}, {\tt www.arXiv.org} \textbf{gr-qc} (2001),
  0111108.



\bibitem[IZ78]{itzykson_zuber}
C.~Itzykson and J.-B.~Zuber, \emph{Quantum field theory}, 
  McGraw-Hill, 1978.


\bibitem[Ju02]{junker}
W.~Junker, \emph{Erratum}, Rev. Math. Phys. \textbf{14} no. 5 (2002),
  511.


\bibitem[Ka00]{KK}
K. Kartzert, \emph{Singularity structure of the two-point function
of the free Dirac field on a globally hyperbolic spacetime }, 
Annalen  Phys. \textbf{9} (2000), 475, available at 
{\tt www.desy.de/uni-th/lqp}.


\bibitem[KRW97]{RKW}
 B.~Kay, M.J.Radzikowski, and R.M. Wald, 
\emph{QFT on spacetimes with a compactly generated Cauchy horizon},
 Commun. Math. Phys. 
\textbf{183} (1997), 533, 
 {\tt www.arXiv.org} \textbf{gr-qc} ,
  9603012.



\bibitem[Ma02]{pi}
P.~Marecki, \emph{On the application of quantum inequalities 
to quantum optics}, Phys. Rev. A
\textbf{66}, 053801,
 {\tt www.arXiv.org} \textbf{quant-ph} (2002),
  0203027.

\bibitem[Ma01]{pi_krak}
P.~Marecki, \emph{Electrons at TESLA will experience thermal radiation}, Notes from
a talk given at the GrK meating in Cracow, (2001).


\bibitem[Mi01]{FEL}
S.V.~Milton et al., \emph{Exponential gain and saturation of a Self-Amplified 
Spontaneous Emission Free-Electron Laser}, Science  \textbf{292} (2001),
  2037.


\bibitem[MPS98]{mohr}
P.~Mohr, G.Plunien, and G.Soff \emph{QED corrections in heavy atoms},
 Phys. Rep. \textbf{293} (1998), 228.


\bibitem[Mor01]{Moretti01}
V.~Moretti, \emph{Comments on the stress-energy tensor operator in curved
  spacetime}, {\tt www.arXiv.org} \textbf{gr-qc} (2001), 0109048.

\bibitem[PCK92]{kimble}
E.S. Polzik, J.Carri, and H.J. Kimble,
\emph{Spectroscopy with squeezed light},
 Phys. Rev. Lett. \textbf{68(20)} (1992), 3020.



\bibitem[PS70]{pow_stroe}
R.T.~ Powers and E.~St\"ormer, \emph{Free states of the canonical
  anticommutation relations}, Commun. Math. Phys. \textbf{16} (1970), 1.


\bibitem[Ra96]{Rad}
M.J.~Radzikowski, \emph{Micro-local approach to the Hadamard condition in quantum
field theory on curved space-time},
 Commun. Math. Phys. 
\textbf{179} (1996), 529.




\bibitem[RS75]{RS2}
M.~Reed, and B.~Simon \emph{Methods of modern mathematical physics II},
Academic Press, 1975.


\bibitem[Ru77]{Ru}
S.N.M.~Ruijsenaars, \emph{Charged particles in external fields. I. Classical 
theory},
 J. Math. Phys. 
\textbf{18} (1977), 720.


\bibitem[SV1]{SV1}
H.~Sahlmann, and R. Verch, \emph{Passivity and microlocal spectrum condition},
 Commun. Math. Phys. 
\textbf{214}, 705, 
 {\tt www.arXiv.org} \textbf{math-ph} (2000),
  0002021.

\bibitem[SV2]{SV2}
H.~Sahlmann, and R. Verch, \emph{Microlocal spectrum condition and Hadamard
form for vector valued quantum fields in curved spacetime},
 Rev. Math. Phys. 
\textbf{13}, 1203, 
 {\tt www.arXiv.org} \textbf{math-ph} (2001),
  0008029.



\bibitem[Sch96]{Scharf}
G.~Scharf, \emph{Finite quantum electrodynamics}, Springer, 1996.



\bibitem[SZ97]{SZ}
M.O.~Scully, and M.S.Zubairy \emph{Quantum optics}, Cambridge, 1997.

\bibitem[Sha02]{shabajev}
V.M.~Shabaev, \emph{Two-time Green's function method in QED of high-Z 
few-electrons atoms},
Phys. Rep. \textbf{356} (2002), 121.


\bibitem[SS65]{SS}
D.~Shale and W.F.~Stinespring, \emph{Spinor representations of infinite 
orthogonal groups}, J. Math. Mech. \textbf{14} (1965), 315.

\bibitem[St\"o98]{uranium}
Th.St\"ohlker, \emph{et. al.}, \emph{Charge-exchange cross section and beam
lifetimes for stored and decelerated bare uranium ions},
Phys. Rev.  \textbf{A58} (1998), 2043.



\bibitem[Tay95]{tay2}
M.E. Taylor, \emph{Partial differential equations}, vol.~2, Springer, 1995.

\bibitem[Tha91]{Thaller}
B.~Thaller, \emph{The Dirac equation}, Springer, 1991.

\bibitem[Ver94]{verch}
R.~Verch, \emph{Local definiteness, primarity and quasiequivalence of quasifree
Hadamard quantum states in curved spacetime}, 
Commun. Math. Phys. \textbf{160} (1994), 507

\bibitem[Wa94]{Wald}
R.M.~Wald, \emph{Quantum field theory in curved spacetime and black hole
thermodynamics}, University of Chicago Press, 1994.

\bibitem[Web02]{webb}
J.K.~Webb, \emph{et. al.}, \emph{Further evidence for cosmological evolution
of the fine structure constant},
Phys. Rev. Lett. \textbf{87} (2002), 091301.


\bibitem[Wi01]{will}
C.M.~Will, \emph{The confrontation between General Relativity and experiment},
Living Rev. Relativity \textbf{4} (2001), 4;  
{\tt www.livingreviews.org/Articles/Volume4/2001-4will/}. 



\end{thebibliography}

\printindex

\end{document}